\documentclass[reprint,groupedaddress,showpacs,amsmath,amssymb,aps,prc,floatfix,nofootinbib]{revtex4-1}

\usepackage{graphicx} 
\usepackage{dcolumn} 
\usepackage{bm} 
\usepackage{appendix}
\usepackage{xcolor}
\usepackage{setspace}

\newcommand{\ee}{$e^{+}e^{-}$}

\newcommand{\auau}{$^{197}$Au+$^{197}$Au}
\newcommand{\auc}{$^{197}$Au+$^{12}$C}
\newcommand{\agag}{$^{107}$Ag+Ag}

\newcommand{\agev}{$A$~GeV}
\newcommand{\gevcc}{GeV/$c^{2}$}

\newcommand{\gevc}{GeV/$c$}
\newcommand{\mevc}{MeV/$c$}
\newcommand{\gev}{GeV}
\newcommand{\mev}{MeV}

\newcommand{\sqrtsNN}{$\sqrt{s_{NN}}$}

\begin{document}


\title{Proton number fluctuations in $\mathbf{\sqrt{s_{NN}}}$ = 2.4~\gev\ Au+Au collisions studied
with the high-acceptance dielectron spectrometer HADES}

\author{J.~Adamczewski-Musch,$^{5}$ O.~Arnold,$^{11,10}$ C.~Behnke,$^{9}$ A.~Belounnas,$^{17}$
A.~Belyaev,$^{8}$ J.C.~Berger-Chen,$^{11,10}$ A.~Blanco,$^{2}$ C.~Blume,$^{9}$ M.~B\"{o}hmer,$^{11}$
P.~Bordalo,$^{2}$
% ...
\\
(HADES Collaboration$^d$)}


\affiliation{
\mbox{$^{1}$Institute of Physics, Slovak Academy of Sciences, 84228~Bratislava, Slovakia}\\
\mbox{$^{2}$LIP-Laborat\'{o}rio de Instrumenta\c{c}\~{a}o e F\'{\i}sica Experimental de Part\'{\i}culas, 3004-516~Coimbra, Portugal}\\
\mbox{$^{3}$Institute of Nuclear Physics, Polish Academy of Sciences, 31-342~Krak\'{o}w, Poland}\\
\mbox{$^{4}$Smoluchowski Institute of Physics, Jagiellonian University of Cracow, 30-059~Krak\'{o}w, Poland}\\
\mbox{$^{5}$GSI Helmholtzzentrum f\"{u}r Schwerionenforschung GmbH, 64291~Darmstadt, Germany}\\
\mbox{$^{6}$Technische Universit\"{a}t Darmstadt, 64289~Darmstadt, Germany}\\
\mbox{$^{7}$Institut f\"{u}r Strahlenphysik, Helmholtz-Zentrum Dresden-Rossendorf, 01314~Dresden, Germany}\\
\mbox{$^{8}$Joint Institute of Nuclear Research, 141980~Dubna, Russia}\\
\mbox{$^{9}$Institut f\"{u}r Kernphysik, J. W. Goethe-Universit\"{a}t, 60438 ~Frankfurt, Germany}\\
\mbox{$^{10}$Excellence Cluster 'Origin and Structure of the Universe' , 85748~Garching, Germany}\\
\mbox{$^{11}$Physik Department E62, Technische Universit\"{a}t M\"{u}nchen, 85748~Garching, Germany}\\
\mbox{$^{12}$II.Physikalisches Institut, Justus Liebig Universit\"{a}t Giessen, 35392~Giessen, Germany}\\
\mbox{$^{13}$Institute for Nuclear Research, Russian Academy of Science, 117312~Moscow, Russia}\\
\mbox{$^{14}$Institute of Theoretical and Experimental Physics, 117218~Moscow, Russia}\\
\mbox{$^{15}$National Research Nuclear University MEPhI (Moscow Engineering Physics Institute), 115409~Moscow, Russia}\\
\mbox{$^{16}$Department of Physics, University of Cyprus, 1678~Nicosia, Cyprus}\\
\mbox{$^{17}$Laboratoire de Physique des 2 infinis Ir\`{e}ne Joliot-Curie, CNRS-IN2P3, Universit\'{e} Paris-Saclay, F-91400~Orsay Cedex, France}\\
\mbox{$^{18}$Nuclear Physics Institute, The Czech Academy of Sciences, 25068~Rez, Czech Republic}\\
\mbox{$^{19}$LabCAF. F. F\'{\i}sica, Universidad de Santiago de Compostela, 15706~Santiago de Compostela, Spain}\\
\mbox{$^{20}$Uniwersytet Warszawski, Wydzia\l ~Fizyki, Instytut Fizyki Do\'{s}wiadczalnej, 02-093 Warszawa, Poland}\\
\mbox{$^{21}$Dipartimento di Fisica and INFN, Universit\`{a} di Torino, 10125~Torino, Italy}\\
\\
\mbox{$^{a}$ also at Coimbra Polytechnic - ISEC, ~Coimbra, Portugal}\\
\mbox{$^{b}$ also at Technische Universit\"{a}t Dresden, 01062~Dresden, Germany}\\
\mbox{$^{c}$ also at Frederick University, 1036~Nicosia, Cyprus}\\
\mbox{$^{d}$e-mail: hades-info@gsi.de}\\
\mbox{$^{\dagger}$ deceased}
}

\date{\today}

\begin{abstract}
We present an analysis of proton number fluctuations in \sqrtsNN\ = 2.4~\gev\ \auau\ collisions measured with the High-Acceptance DiElectron
Spectrometer (HADES) at GSI.  With the help of extensive detector simulations done with IQMD transport model events including nuclear clusters,
various nuisance effects influencing the observed proton cumulants have been investigated.  Acceptance and efficiency corrections have been
applied as a function of fine grained rapidity and transverse momentum bins, as well as considering local track density dependencies.
Next, the effects of volume changes within particular centrality selections have been considered and beyond-leading-order 
corrections have been applied to the data.  The efficiency and volume corrected proton number moments and cumulants $K_n$ of
orders $n =$ 1, 2, 3, and 4 have been obtained as a function of centrality and phase-space bin, as well as the corresponding correlators $C_n$.
We find that the observed correlators show a power-law scaling with the mean number of protons, i.e.\ $C_n \propto \langle N \rangle^n$,
indicative of mostly long-range multi-particle correlations in momentum space.  We also present a comparison of our results
with Au+Au collision data obtained at RHIC at similar centralities, but higher \sqrtsNN.
\end{abstract}


\maketitle

\tableofcontents

\section{Introduction}
\label{Sec:Intro}


Lattice QCD calculations sustain that, at vanishing baryochemical potential $\mu_B$ and a temperature of order $T = 156$~\mev,
the boundary between hadron matter and a plasma of deconfined quarks and gluons is a smooth crossover \cite{Aoki2006,Bazavov2014}
whereas, at finite $\mu_b$ and small $T$, various models based on chiral dynamics clearly favor a 1\textsuperscript{st}-order
phase transition \cite{Stephanov2004}, suggesting the existence of a critical endpoint (CEP).  Although the QCD critical endpoint
is a very distinct feature of the phase diagram, it can presently not be located from first-principle calculations, and experimental
observations are needed to constrain its position.  Mapping the QCD phase diagram is therefore one of the fundamental goals of
present-day heavy-ion collision experiments.

Observables expected to be sensitive to the CEP are fluctuations of overall conserved quantities -- like the net electric charge,
the net baryon number, or the net strangeness -- measured within a limited part of phase space \cite{Hatta2003,Stephanov2009,Asakawa2009}.
Phase space has to be restricted to allow for fluctuations in the first place, yet remain large enough to avoid the regime of
small-number Poisson statistics \cite{Begun2004}.  Tallying the baryon number event by event is very challenging experimentally,
as e.g.\ neutrons are typically not reconstructed in the large multipurpose charged-particle detectors in operation.  It has been argued,
however, that net-proton fluctuations too should be sensitive to the proximity of the CEP: first, on principal grounds because
of the overall isospin blindness of the sigma field  \cite{Hatta2003} and, second, because of the expected equilibration of isospin in
the bath of copiously produced pions in relativistic heavy-ion collisions \cite{Kitazawa2012}.  Using the net proton number as a proxy
of net-baryon fluctuations and by studying its $\mu_B$ dependence, one may therefore hope to constrain the location of the CEP in the
phase diagram.  This is best achieved with a beam-energy scan, the characteristic signature being a non-monotonic evolution as a
function of \sqrtsNN\ of any experimental observable sensitive to critical behavior.

Ultimately, the characteristic feature of a CEP is an increase and even divergence of spatial fluctuations of the order parameter.
Most fluctuation measures originally proposed were related to variances of event-by-event observables such as particle multiplicities
(net electric charge, baryon number, strangeness), particle ratios, or mean transverse momentum.  Typically, the critical contribution
to variances, i.e.\ 2\textsuperscript{nd}-order cumulants, is approximately proportional to $\xi^2$, where $\xi$ is the spatial
correlation length which would ideally diverge at the CEP.  The magnitude of $\xi$ is limited trivially by the system size but much
more so by finite-time effects, due to critical slowing down, to an estimated 2 -- 3~fm \cite{Stephanov1999,Berdnikov2000,Bluhm2019}.
In addition, the fluctuating quantities are obtained at the chemical freeze-out point only which may be situated some distance away
from the actual endpoint.  This makes discovering a non-monotonic behavior of any critical contribution to fluctuation observables a
challenging task, particularly if those measures depend on $\xi$ too weakly.  To increase sensitivity, it was therefore
proposed \cite{Asakawa2009,Stephanov2009,Athanasiou2010} to exploit the higher, i.e.\ non-Gaussian moments or cumulants of the
multiplicity distribution as the latter are expected to scale like $K_n \sim \xi^{5n/2 - 3}$.  In particular, $K_4$ is
considered \cite{Stephanov2011} to be universally negative when approaching the CEP from the low-$\mu_B$ region, i.e.\ by lowering \sqrtsNN.
For a more complete review of this theoretical background see e.g.\ Refs.~\cite{Koch2010,Asakawa2016,Bzdak2020}.

It is also important to keep in mind that other sources can produce non-Gaussian moments: remnants of initial-state fluctuations,
reaction volume fluctuations, flow, etc.  A quantitative study of such effects is necessary to unambiguously identify the critical
signal.  It is clear that an energy scan of the QCD phase diagram is mandatory to understand and separate such non-dynamical
contributions from the genuine CEP effect, the latter being a non-monotonic function of the initial collision energy \sqrtsNN\
as the CEP is approached and passed over.  The fact that non-Gaussian moments have a stronger than quadratic dependence
on $\xi$ causes them to be much more sensitive signatures of the CEP.  Because of the increased sensitivity, they are, however,
also more strongly affected by the nuisance effects mentioned above and this must be investigated very carefully.

Various fluctuation observables have been scrutinized in heavy-ion collisions from SPS to LHC energies:  at the SPS in particular,
balance functions and scaled variances of charged particles \cite{Alt2007,Alt2008}, dynamical fluctuations of particle
ratios \cite{Anticic2013,Anticic2014}, as well as proton intermittencies \cite{Anticic2010,Anticic2015,Mackowiak2019,Grebieszkow2019};
at the RHIC and LHC, net-proton number fluctuations \cite{Aggarwal2010,Adamczyk2014a,Luo2015b,Rustamov2017b}, net-charge
fluctuations \cite{Adare2008,Adamczyk2014b,Adare2016}, and net-kaon fluctuations \cite{Adamczyk2018}.  In this context,
the first RHIC beam-energy scan, covering center-of-mass energies of \sqrtsNN\ = 7.7 -- 200~\gev, provided indications
of a non-monotonic trend with decreasing energy of the net-proton fluctuations \cite{Luo2015b,Adam2020a}.  Unfortunately,
the limited statistical accuracy of these data as well as their low-energy cutoff do not yet allow for firm conclusions.
A second approved scan aims, however, at greatly improving the statistical quality and at extending the measurements down
to \sqrtsNN\ = 3~\gev\ by complementing the standard collider mode with a fixed-target arrangement \cite{Yang2017}. 

Here we present results from a high-statistics measurement of proton number fluctuations in the reaction system \auau\
at \sqrtsNN\ = 2.4~\gev\ done with the charged-particle spectrometer HADES at SIS18.  In Sec.~II of this article a brief description
of the experiment, as well as of the analysis procedures, is given.  In Sec.~III the reconstructed proton multiplicity distributions
are presented and their relevant features in terms of moments, cumulants and factorial cumulants are discussed.  In Sec.~IV we show
and discuss the centrality and acceptance dependencies of (net-) proton number fluctuations.  Finally, Sec.~V summarizes and
concludes the paper.

\section{The Au+Au experiment}
\label{Sec:Exp}

\begin{figure*}[htb]
  \begin{center}
     \resizebox{0.7\linewidth}{!} {
       \includegraphics{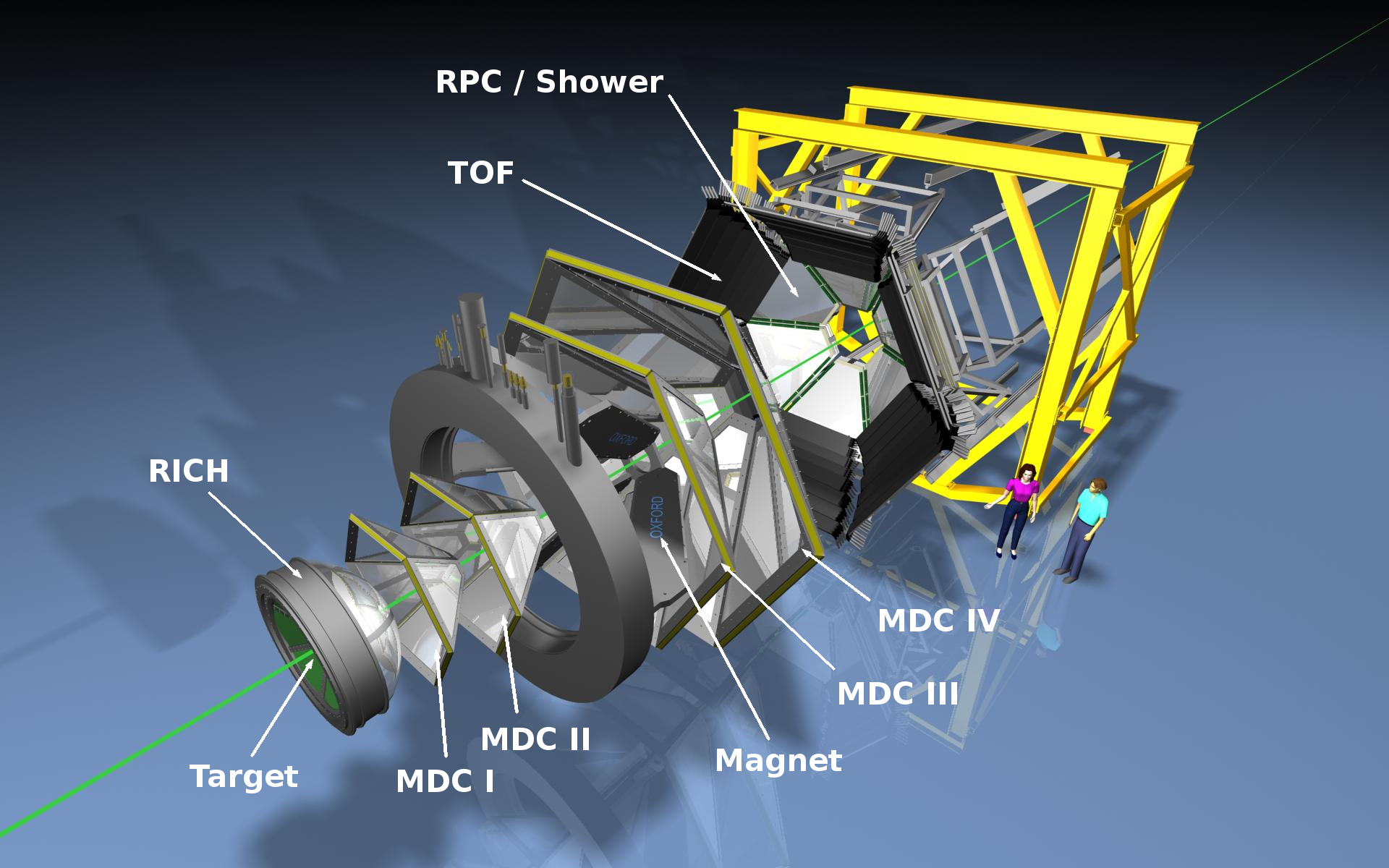} 
     }
  \end{center}
  \vspace*{-0.2cm}
  \caption[]{(Color online) Schematic explosion view of the HADES experiment.
       The forward hodoscope FWALL used for centrality selection (see text) is not shown on this picture.
    }
  \label{fig:hades}
\end{figure*}

\subsection{The HADES setup}
The six-sector high-acceptance spectrometer HADES operates at the heavy-ion synchrotron
SIS18 of GSI Helmholtzzentrum f\"{u}r Schwerionenforschung in Darmstadt, Germany.
Although its original design was optimized for dielectron spectroscopy, HADES is in fact
a versatile charged-particle detector with large efficiency, good momentum resolution, and high
trigger rate capability.  The HADES setup consists of an iron-less, six-coil toroidal
magnet centered on the beam axis and six identical detector sectors located between the coils.
With a nearly complete azimuthal coverage and spanning polar angles $\theta = 18^{\circ} - 85^{\circ}$,
this geometry results in a laboratory rapidity acceptance for protons of $y \simeq 0.1 - 1.8$.
In the configuration used to measure the data discussed here, each sector was equipped with a central
hadron-blind Ring-Imaging Cherenkov (RICH) detector, four layers of Multi-Wire Drift Chambers (MDC) used for tracking
-- two in front of and two behind the magnetic field volume, a time-of-flight detector made of plastic scintillator
bars (TOF) at angles $\theta > 44^{\circ}$ and of Resistive-Plate Chambers (RPC) for $\theta < 45^{\circ}$, 
and a pre-shower detector.  Figure \ref{fig:hades} shows a schematic view of the setup; more detailed technical
information can be found in \cite{Agakishiev2009tech}.  Hadron identification in HADES is based mainly on particle velocity,
obtained from the measured time of flight, and on momentum, reconstructed by tracking the particle through the
magnetic field using the position data from the MDC.  Energy-loss information from the TOF as well as from
the MDC tracking chambers can be used to augment the overall particle identification power.  The RICH detector,
specifically designed for electron and positron candidate identification, was not used in the present analysis.

The event timing was provided by a 60~$\mu$m thick monocrystalline diamond detector (START) positioned in the beam pipe
25~mm upstream of the first target segment.  The diamond detector material \cite{Pietraszko2014} is radiation hard, has high
count rate capability, large efficiency ($\epsilon \geq 0.9$), and very good time resolution ($\sigma_t \simeq 60$~ps).
Through a 16-fold segmentation in $x$ direction (horizontal) and in $y$ direction (vertical), of its double-sided metallization,
START also provided position information on the incoming beam particles, essential for beam focusing and position monitoring
during the experiment.  Combined with a multi-hit capable TDC, the fast START signal can be used to recognize and largely
suppress event pileup within a time slice of $\pm$0.5 $\mu$s centered on an event of interest.  Furthermore, as discussed below,
the easily identifiable \auc\ reactions in the diamond material of the START can be used to set a limit on background
from Au reactions on light nuclei (H, C, N, and O) in the target holder material.

A forward hodoscope (FWALL), positioned 6.9~m downstream of the target and covering polar angles of
$\theta = 0.33^{\circ} - 7.2^{\circ}$ was used to determine the reaction plane angle and the event centrality.
This device comprises 288 square tiles made of 25.4~mm thick plastic scintillator and each read out with a
photomultiplier tube.  The FWALL has limited particle identification capability based on the measured time of flight
and energy loss in the scintillator.

The \auau\ reactions investigated took place in a stack of fifteen gold pellets of 25~$\mu$m thickness,
adding up to 0.375~mm and corresponding to a nuclear interaction probability of 1.35\%.  Each of the 2.2~mm
diameter gold pellets was glued onto the central 1.7~mm eyelet of a 7~$\mu$m thick Kapton holding strip.
These strips were in turn supported by a carbon fiber tube of inner diameter 19~mm and wall thickness 0.5~mm,
realizing an inter-pellet spacing (pitch) of 3.7~mm.  All target holder parts were laser-cut with a tolerance
of 0.1~mm in all dimensions (see \cite{Kindler2011} for more details).  The total length of the segmented target
assembly was 55~mm.  This design (target segmentation, low-Z and low-thickness holder) was optimized to minimize
both multiple scattering of charged particles and conversion of photons into \ee\ pairs in the target.
It also helped to minimize production of spallation protons, that is emission of protons from any material
in the target region through secondary knockout reactions induced by primary particles.

\subsection{Online trigger and event selection}
In the present experiment, a gold beam with a kinetic energy of $E_{\textrm{kin}}$ = 1.23\agev\ and an average intensity
of $1 - 2 \times 10^6$ particles per second impinged onto the segmented gold target.  Several physics
triggers (PT) were implemented to start the data read-out:   Based on hardware thresholds set on the analog multiplicity
signal corresponding to at least 5 (PT2) or 20 (PT3) hits in the TOF detector, and coincident with a signal in the in-beam diamond
START detector.  The PT2 trigger was down-scaled ($\div \;8$) and PT3 was the main event trigger covering the 43\% most
central collisions\footnote{The centrality range selected by the HADES trigger was determined with Glauber Monte Carlo calculations \cite{Adamczewski2018}.}.
In total, $2.1 \times 10^9$ high-quality PT3 events were recorded of which, for performance reasons mostly, we used only a subset
of $1.6 \times 10^8$ events in the current analysis.

\subsection{Event pileup}

\begin{figure}[!htb]
  \begin{center}
     \resizebox{0.8\linewidth}{!} {
       \includegraphics{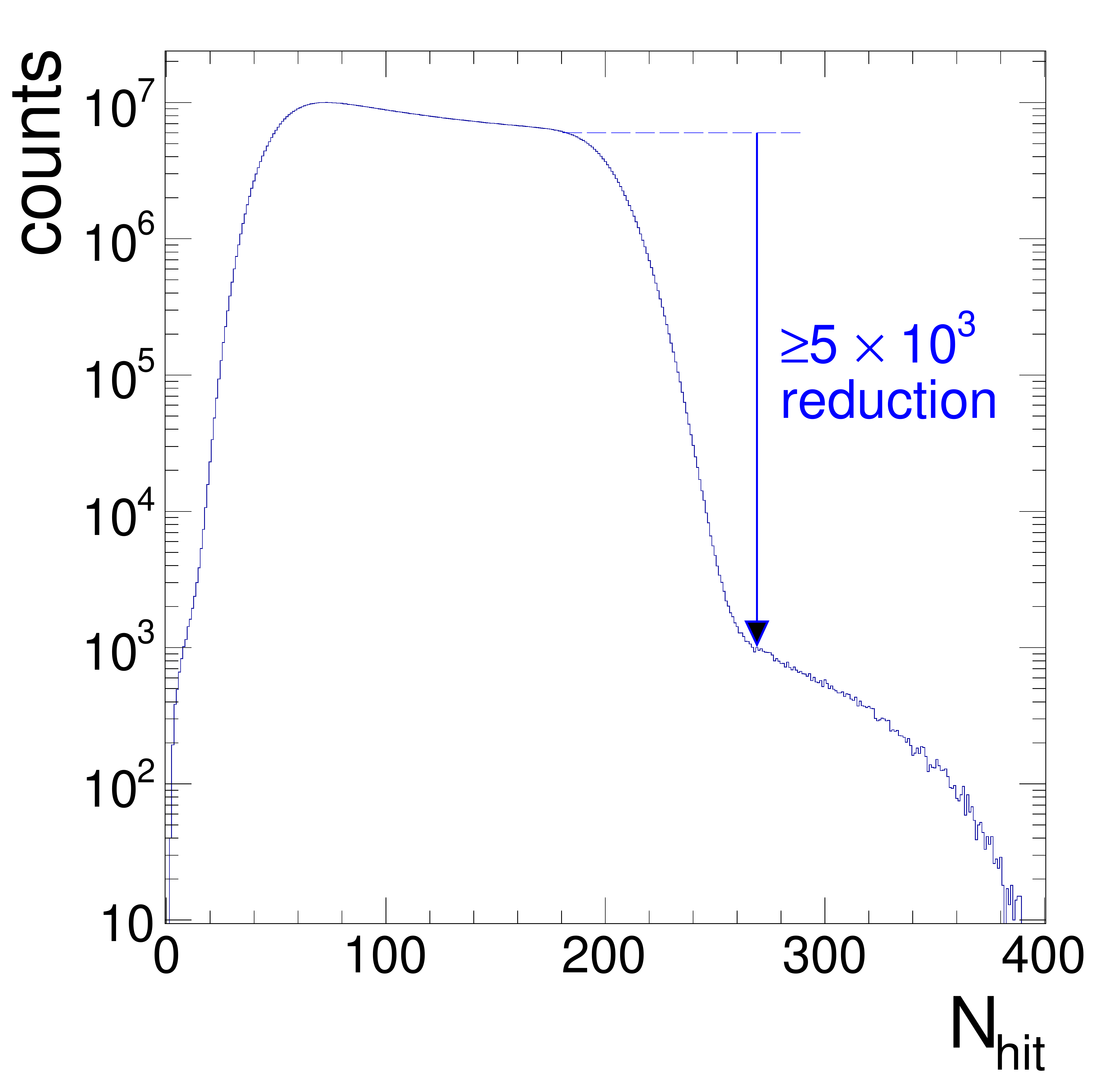} 
     }
  \end{center}
  \vspace*{-0.2cm}
  \caption[] {(Color online) Distribution of the total number of hits in the HADES time-of-flight
              detectors TOF and RPC in the 43\% most central Au+Au events accepted by the PT3 trigger.
              The tiny contribution of order $\leq 2 \times 10^{-4}$, visible for $N_{\textrm{hit}} > 260$, is attributed to event pileup. \\
             }
  \label{fig:nhits}
\end{figure}

Running the HADES experiment with high beam currents bears the danger of event pileup, that means, of having a sizable
chance that two or even more consecutive (minimum bias) beam-target interactions take place within the readout window
opened by a trigger.  Such events appear to have higher than average track multiplicity and, if their fraction becomes
sufficiently large, they will have a noticeable impact on the observed event-by-event particle number fluctuations.
The multi-hit TDC of the START detector provided, however, the possibility to reject piled up events by counting the number
of registered incident beam particles within a $\pm 0.5\; \mu$s time window centered on any accepted trigger.  However,
because of the finite efficiency of the START (determined to be $\geq 90$\%) a remaining small contribution of pileup
events is still visible in Fig.~\ref{fig:nhits} as a shoulder at large $N_{\textrm{hit}}$ values.  From the size of the resulting
``step'', one can estimate the overall pileup probability in our event sample to be $\leq 2 \times 10^{-4}$.

In fact, the contamination of identified proton yields by pileup turns out to be even much smaller. 
Because of its excellent timing properties, the START allowed to continuously monitor the instantaneous beam rate,
typically of order 4 -- 8 ions/$\mu$s, i.e.\ a factor four larger than the rate averaged over beam spills.
With this rate and a total beam interaction probability of 1.7\% (on START, gold targets, and Kapton combined),
and realizing that particle identification via momentum-velocity correlation implicitly puts a tight constraint
on the flight time of tracks (with $|\Delta t| < 1.5$~ns) for them to fulfill the required $p - \beta$ congruence,
one can estimate the pileup effect on identified particles to be $\leq 3 \times 10^{-5}$.  As discussed further below,
this value is low enough for pileup to be of no concern to our fluctuation analysis.

\subsection{Offline event selection and centrality definition}

\begin{figure}[!htb]
  \begin{center}
     \resizebox{0.9\linewidth}{!} {
       \includegraphics{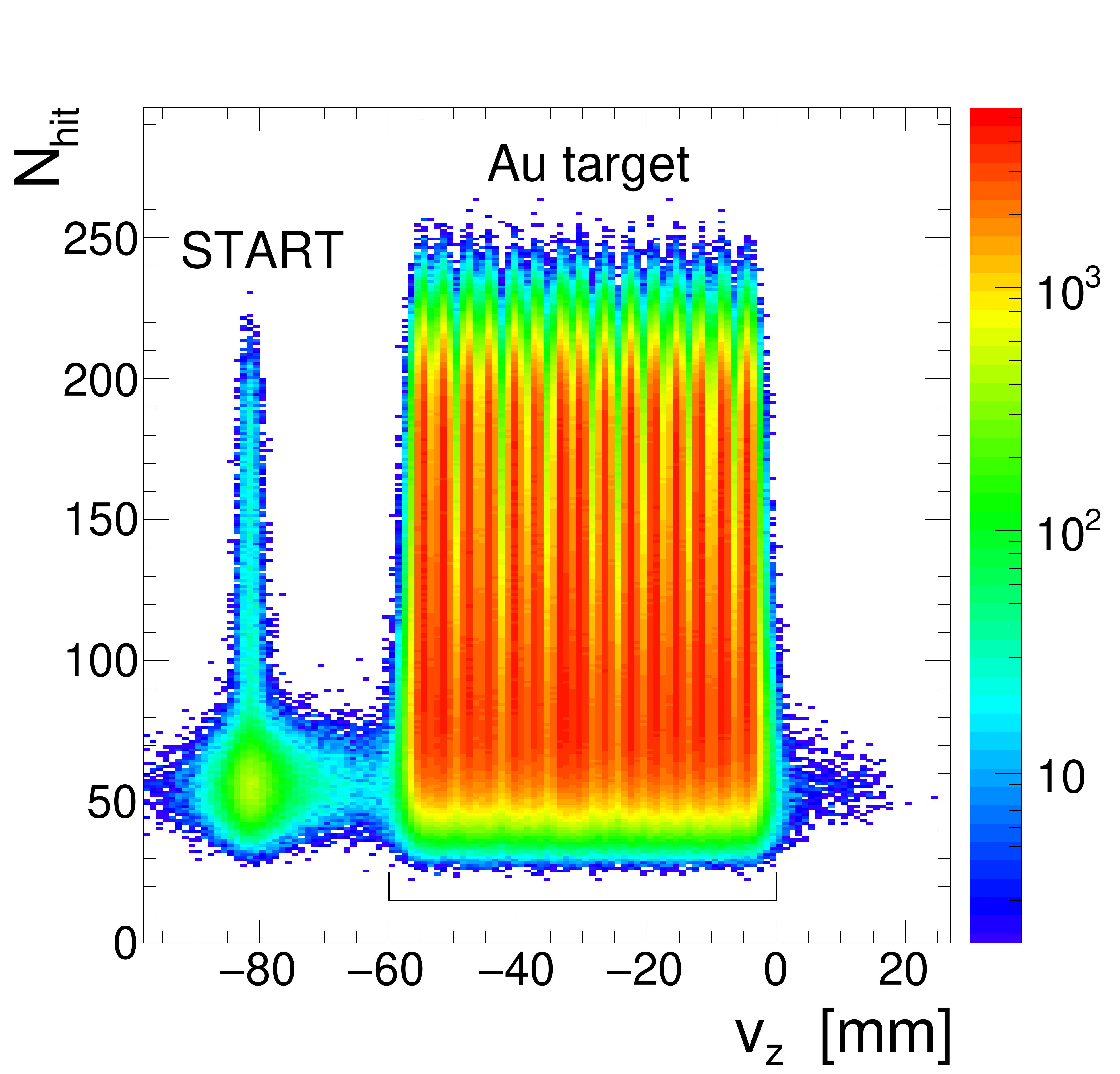} 
     }
  \end{center}
  \vspace*{-0.2cm}
  \caption[] {(Color online) Total number of charged hits in the time-of-flight detectors (TOF and RPC) as a function of the 
              reconstructed event vertex $v_z$ along the beam axis $z$. The fifteen gold target segments are clearly separated
              from the diamond START detector as well as from one another; the horizontal solid line indicates the vertex
              cut used to select reactions in the targets.  Note that the faint high-$N_{\textrm{hit}}$ tail ($N_{\textrm{hit}} > 100$)
              visible for the START detector is due to Au+Au reactions in its thin gold platings.}
  \label{fig:zvert}
\end{figure}

\begin{figure}[!htb]
  \begin{center}
     \resizebox{0.9\linewidth}{!} {
       \includegraphics{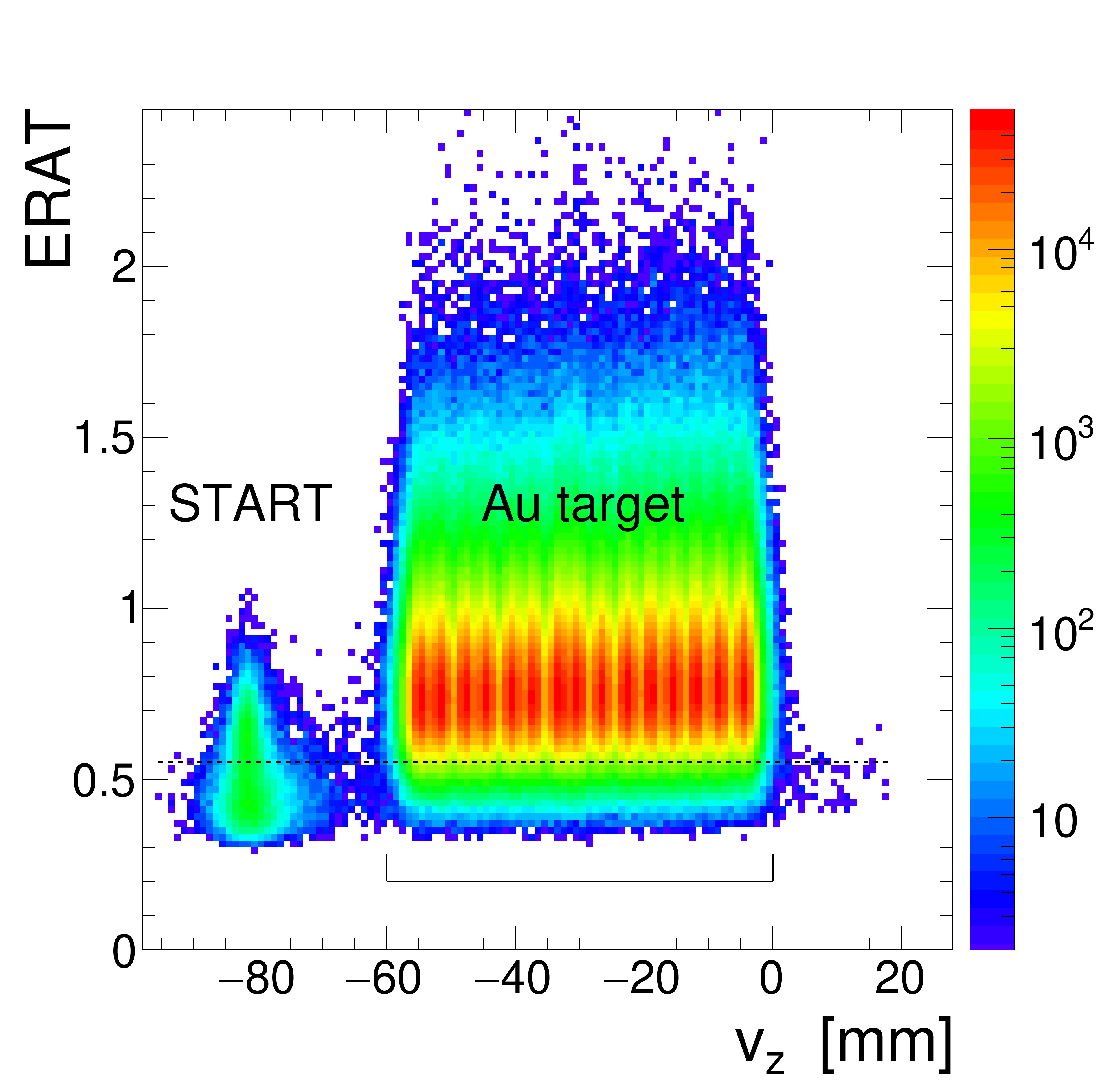} 
     }
  \end{center}
  \vspace*{-0.2cm}
  \caption[] {(Color online) Ratio of transverse to longitudinal energy $E_{RAT}$ as a function of the 
              reconstructed event vertex along the beam axis $z$.  The horizontal dashed line
              indicates the $E_{RAT} > 0.55$ cut used to further suppress Au+C reactions in the Kapton
              strips of the target holder.  The solid line delimits the vertex cut applied on the target region.
             }
  \label{fig:erat}
\end{figure}

Offline, events were selected by requiring that the global event vertex, determined from reconstructed tracks with
a resolution of $\sigma_x = \sigma_y \simeq$ 0.7~mm and $\sigma_z \simeq$ 0.9~mm, was within the $\approx60$~mm extension
of the segmented target.  Figure~\ref{fig:zvert} shows the combined number of hits observed in the RPC and TOF
as a function of the reconstructed event vertex $v_z$ along the beam axis.  The diamond START detector and the
segmented gold targets are clearly distinguishable along $z$, and the difference in hit multiplicity between Au+Au
reactions on the target pellets and Au+C reactions on the diamond detector is very evident as well.  The amount
of reactions on the START detector accepted by the PT3 trigger can furthermore be used to put an upper limit on a possible
contamination from reactions on the Kapton holding strips (containing H, C, N, and O) by comparing the effective thickness
and resulting interaction probability in Kapton and START, respectively.  Considering the intricate crisscross geometry
of those strips \cite{Kindler2011}, the diameter of their central eyelet, and the tight focus of the gold beam, we estimate
that reactions on Kapton can contaminate the recorded rate of semi-central Au+Au events at most on the level of $10^{-4}$.
This is also supported by control data taken with the Au beam vertically offset by 3.5~mm such as to miss the gold targets
altogether and hit only Kapton.  Note finally that the Au + C contamination affects mostly the peripheral event classes
whereas central Au+Au events are basically free from it because of their much higher average hit multiplicity.

\textcolor{black}{As the lateral event vertex resolution of $\sigma_{\perp} \simeq 1.1$~mm was not sufficient to fully avoid reactions
on the Kapton strips}, in the analysis, this background has been reduced further by applying a cut on the quantity
$E_{RAT}$ defined as the ratio of total detected transverse to total detected longitudinal energy in the
laboratory\footnote{Note that our definition of $E_{RAT}$ differs from the one used for centrality determination in \cite{Reisdorf1997}.}, 
namely

\begin{equation}
  E_{RAT} = \frac{E_{\perp}^{\textrm{tot}}}{E_{\parallel}^{\textrm{tot}}} = \frac{\sum_i E_{\perp, i}}{\sum_i E_{\parallel, i}} = \frac{\sum_i E_i \sin \theta_i}{\sum_i E_i \cos \theta_i} \:,
\label{eq:ERAT}
\end{equation}

\noindent
where $E_i$ and $\theta_i$ are the particle's total energy and polar angle, respectively, and the index $i$ runs over all detected
and identified particles.  The $E_{RAT} > 0.55$ cut applied, shown in Fig.~\ref{fig:erat} as horizontal dashed line, suppresses Au+C
reactions by an additional factor 4, while losing less than 2\% of the Au+Au events.  The remaining Au+C contamination is thus
at most $2.5 \times 10^{-5}$ for all centralities.  Our schematic simulations show that this level of contamination is of no concern
for our proton cumulant analyses, in agreement also with conclusions resulting from similar investigations discussed
in \cite{Bzdak2018,Garg2017}.  Finally, by monitoring the mean charged particle multiplicity per each
HADES sector,\footnote{Averaged over sets of 50 -- 100 thousand PT3 events, corresponding to about 2 -- 3 minutes of run time.}
we have made sure that in the data runs selected for the present analysis no hardware conditions occurred that could have caused
substantial drifts or even jumps of the proton yield.  All estimates of various background contributions potentially affecting
the proton multiplicities measured in our experiment are summarized in Table~\ref{tab:nuisance}.


\begin{figure}[!htb]
  \begin{center}
     \resizebox{0.9\linewidth}{!} {
       \includegraphics{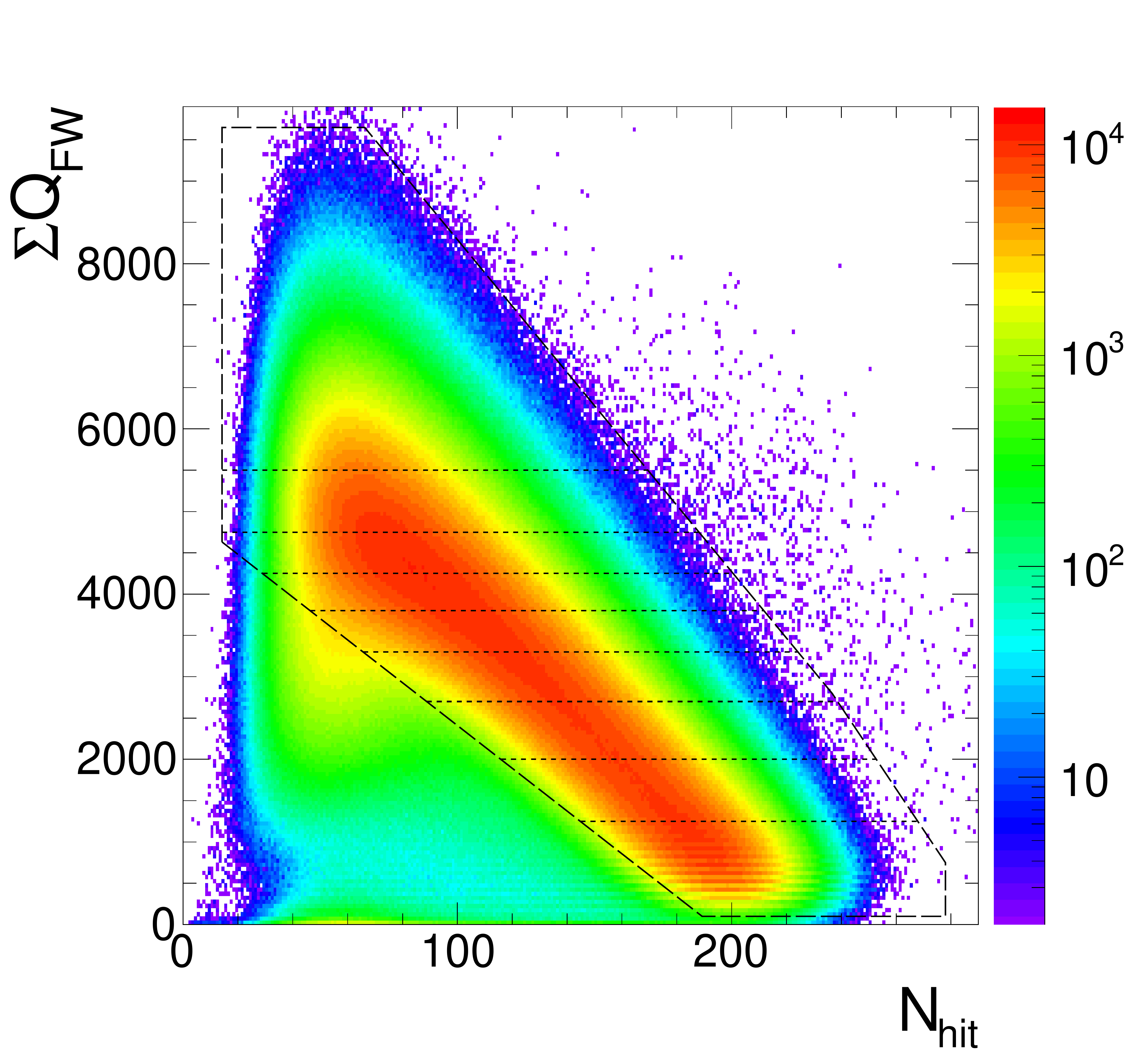} 
     }
  \end{center}
  \vspace*{-0.2cm}
  \caption[] {(Color online) Sum signal $\Sigma Q_{FW}$ measured in the forward wall vs. number of hits $N_{\textrm{hit}}$
              in the TOF and RPC.  The black long-dashed line indicates a loose 2D cleaning cut applied to remove
              the contamination from peripheral events.  Centrality selections, e.g.\ in steps of 5\%, were done by
              applying additional, more restrictive cuts on $\Sigma Q_{\textrm{FW}}$ as indicated by the set of short-dashed
              horizontal lines.
             }
  \label{fig:sumfwnhit}
\end{figure}

In the HADES experiment, event centrality determination is usually done by putting a selection on either the total number
of hits $N_{\textrm{hit}}$ in the time-of-flight detectors TOF and RPC, or on the total number of tracks $N_{\textrm{trk}}$ reconstructed in
the MDC \cite{Adamczewski2018}.  It is however important to realize that in our investigation of particle number fluctuations,
correlations exist between the observable of interest, that is the number of protons $N_{\textrm{prot}}$ emitted into a given phase space
and the $N_{\textrm{hit}}$ or $N_{\textrm{trk}}$ observable used to constrain the event centrality.  At a bombarding energy of 1.23\agev, protons
constitute by far the most dominant particle species and, as every detected proton will produce at least one hit and also
one reconstructed track, we expect indeed very strong autocorrelations when using $N_{\textrm{hit}}$ or $N_{\textrm{trk}}$ as centrality selectors.
A systematic simulation study \cite{Luo2013} done with UrQMD transport model events has already demonstrated the disturbing effect
of autocorrelations on fluctuation observables.  To avoid or, at least, minimize such effects, we have instead used for centrality
determination the cumulated charge $\Sigma Q_{\textrm{FW}}$ of all particles observed in the FWALL detector.  Figure~\ref{fig:sumfwnhit}
shows this measured quantity as a function of the number of hits in the HADES time-of-flight detectors.  The two observables
anticorrelate, as expected, demonstrating that $\Sigma Q_{\textrm{FW}}$ is also a useful measure of centrality.  As discussed in more detail
in Sec.~\ref{Sec:Npart}, a weak anticorrelation between $N_{\textrm{prot}}$ and $\Sigma Q_{\textrm{FW}}$ does exist as well, but its influence on the
proton fluctuation observables can be corrected for.  The coverage of the FWALL being restricted to a range of $0.33^{\circ} - 7.2^{\circ}$
in polar angle results in the loss of the most peripheral events where the projectile fragment passes undetected through the
central hole left open around the beam pipe.  For such events, $\Sigma Q_{\textrm{FW}}$ tends to decrease again, visible as a down-bending
in Fig.~\ref{fig:sumfwnhit}, which contaminates the most central selections with peripheral events.  This can be cured, i.e.\ the
monoticity of $\Sigma Q_{\textrm{FW}}$ with centrality can be restored, by applying a rather loose 2D cut on $\Sigma Q_{\textrm{FW}}$ vs. $N_{\textrm{hit}}$.
Centrality selections are realized, as indicated in Fig.~\ref{fig:sumfwnhit}, by additional dedicated cuts on $\Sigma Q_{\textrm{FW}}$.

From Glauber Monte Carlo calculations and also a direct comparison with UrQMD model calculations (version 3.4) \cite{Bass1998} we determined
that the PT3 hardware trigger selected only the 43\% most central events \cite{Adamczewski2018}.  For the fluctuation analysis,
a finer centrality binning was realized by applying on the measured $\Sigma Q_{\textrm{FW}}$ signal a sequence of 5\% cuts or,
in some instances, 10\% cuts.  The behavior of the various cuts was studied in detailed detector simulations using the GEANT3
software package \cite{GEANT3}.  For that purpose, Au+Au events were generated with the Isospin Quantum Molecular Dynamics (IQMD) transport model
(version c8) \cite{Hartnack1998,Gossiaux1997} supplemented with a minimum spanning tree (MST) clusterizing algorithm
in coordinate space \cite{Leifels2018} which allowed to obtain events including bound nuclear clusters like d, t, $^{3}$He,
$^{4}$He, etc.   At the bombarding energies where HADES takes data, the clusters contribute substantially to the track density;
in our Au+Au data they correspond indeed to about 40\% of the charged baryons detected \cite{Szala2019}.

\subsection{Proton reconstruction and identification}

Charged-particle trajectories in HADES were reconstructed using the MDC hit information \cite{Agakishiev2009tech,Pechenova2015align};
in this procedure,  the trajectories were constrained to start from the vicinity of the global event vertex.
The resulting tracks were subjected to several quality selections provided by the hit matching and Runge-Kutta track fitting algorithms.
Finally, the retained tracks were spatially correlated with time-of-flight information from TOF or RPC, and -- for lepton candidates --
also with ring patterns found in the RICH as well as electromagnetic shower signatures from the pre-shower detector. 

\begin{figure}[!htb]
  \begin{center}
     \resizebox{0.9\linewidth}{!} {
       \includegraphics{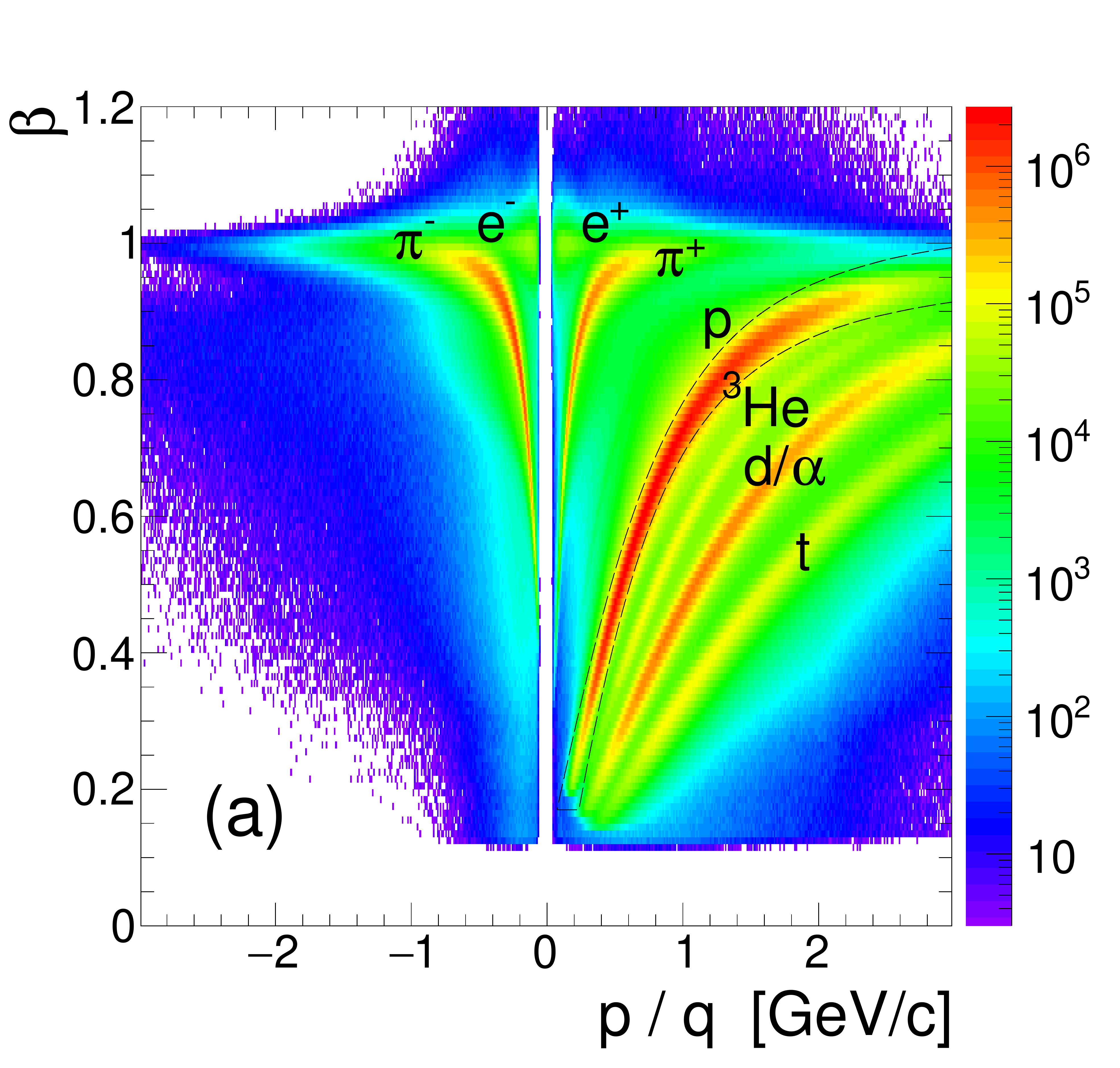}
     }
     \resizebox{0.9\linewidth}{!} {
       \includegraphics{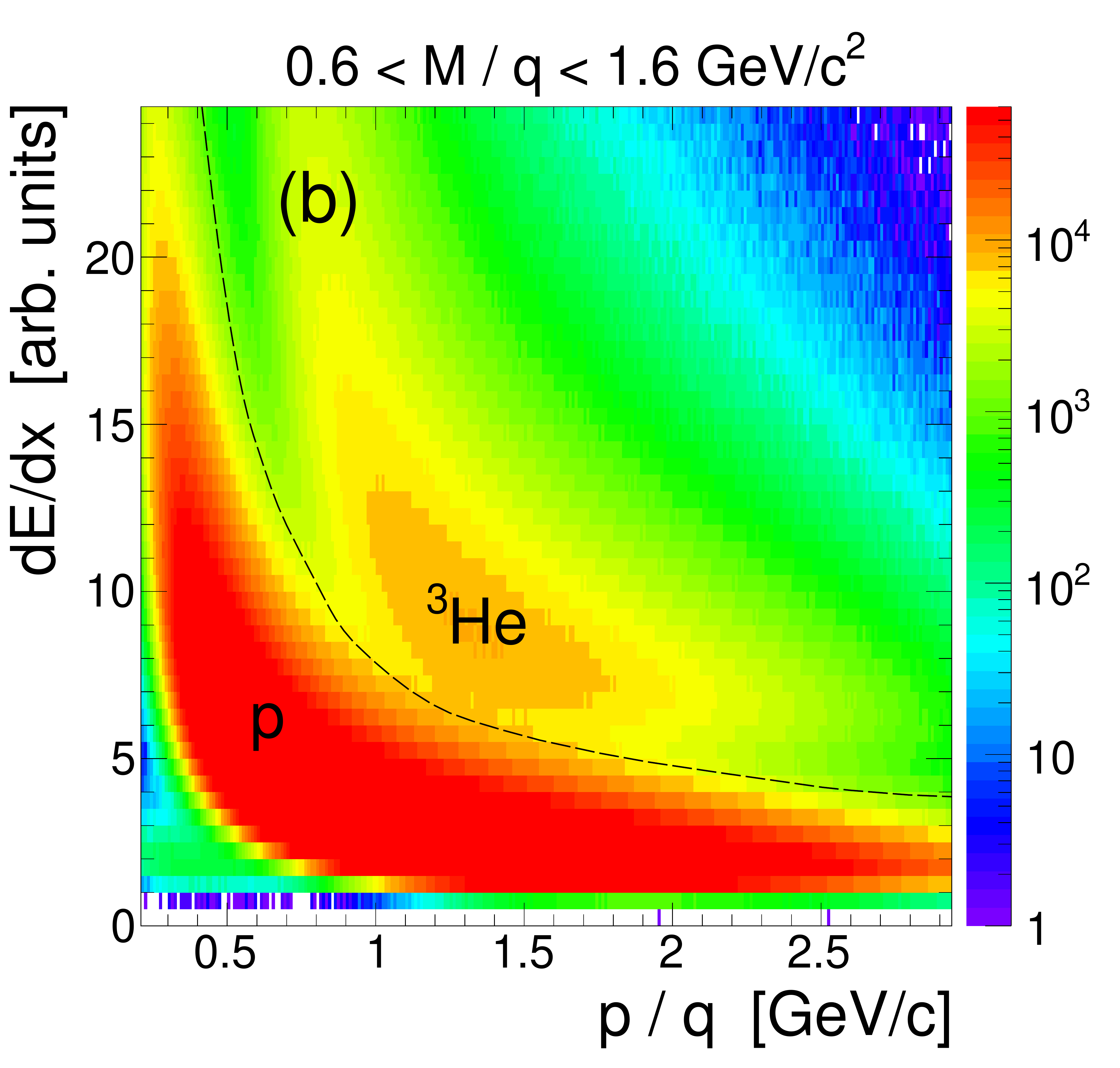}
     }
  \end{center}
  \vspace*{-0.2cm}
  \caption[] {(Color online) Particle identification (PID) based on particle velocity vs. momentum correlations
               as well as on energy loss vs. momentum correlations.  Shown in (a) is $\beta = v/c$ of the detected particle,
               with velocity $v$ obtained from the time of flight, as a function of its $p / q$, where $p$ is the particle
               momentum and $q$ is its charge.  Note that the weak branch lying between p and d/$\alpha$ are $^3$He nuclei.
               The dashed curves delineate the condition used to select protons.  \textcolor{black}{The weak $^3$He contamination remaining
               in this selection is suppressed by applying an additional condition on the energy-loss relation, $dE/dx$ vs. $p / q$,
               shown in (b) for the masses fulfilling $0.6 < M/q < 1.6$~\gevcc.}
             }
  \label{fig:pid}
\end{figure}

\begin{figure}[!htb]
  \begin{center}
     \resizebox{0.9\linewidth}{!} {
       \includegraphics{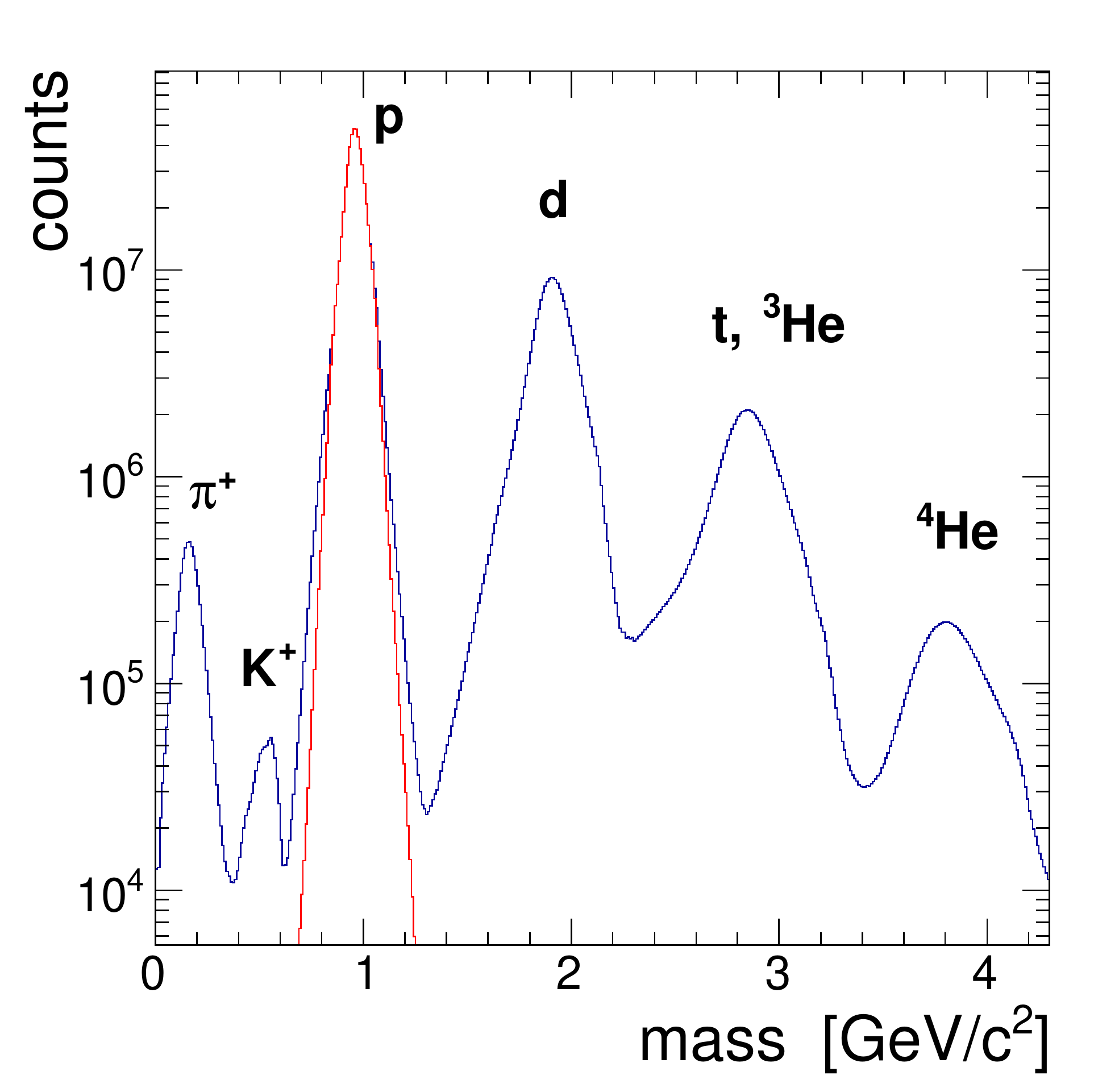} 
     }
  \end{center}
  \vspace*{-0.2cm}
  \caption[] {(Color online) Particle mass distribution of reconstructed tracks --
               as deduced from particle velocity, momentum, and energy loss -- for positive particles within
               the phase-space bin of interest ($y = y_{0} \pm 0.5$ and $0.4 < p_{t} < 1.6$~\gevc).
               The masses selected by the proton cut (see Fig.~\ref{fig:pid}) as well as the $Z=1$ condition are shown in red.
               Contaminations from lower masses ($\pi^+$, K$^+$) and higher masses (d, t, He)
               are below 0.1\%, resulting in an overall proton purity $>0.999$.
             }
  \label{fig:mass}
\end{figure}

Protons were identified by using their velocity vs. momentum correlation as well as their characteristic energy loss in the MDC. 
As seen in Fig.~\ref{fig:pid}(a), the proton branch is the most prominent one next to the charged pions and light nuclei
(deuterons, tritons, and He isotopes).  A $\pm 2\sigma$-wide cut on this branch\footnote{Applying a $\pm 2\sigma$ cut on
velocity per 40~\mevc\ momentum bin.} was used to select the protons.  An additional condition on the energy-loss signal
in the MDC, shown in (b), was applied to further suppress a potential residual contamination caused by the adjacent $^3$He branch,
resulting in a proton purity of $\geq0.999$ for tracks with $0.4 <p_t < 1.6$~\gevc\ and $y = y_0 \pm 0.5$, where $y_0 = 0.74$ is the
Au+Au center-of-mass rapidity.  This is plainly visible in the reconstructed particle mass spectrum, shown in Fig.~\ref{fig:mass}
with and without the proton selection cuts: the about thousandfold weaker K$^+$ signal on the left side of the proton peak
is evidently of no concern and the few-\% $^3$He signal, visible as a weak branch in Fig.~\ref{fig:pid}, is indeed efficiently
relocated to its correct position when assigning the charge $Z=2$ with help of the $dE/dx$ information.  Notice that, with the
charge assignment, the $^4$He hits are likewise moved to their proper position in the mass spectrum.

\begin{table}
\caption{Nuisance effects on the proton multiplicity measured in \sqrtsNN\ = 2.4~\gev\ Au+Au collisions with HADES.
         Listed are the estimated maximum relative contributions of background events (top rows)
         and of background to the proton yield arising within the events of interest (bottom rows).
         \textcolor{black}{For comparison, the expected antiproton yield, estimated from a thermal model fit to the various
         particle yields observed at freeze-out in the 10\% most central events, is listed as well.}
        }
 \label{tab:nuisance}
 \vspace*{3mm}
 \setlength{\extrarowheight}{2pt}  
 \begin{tabular}{l c}
  \hline \hline
  Nuisance effect~~~~~ & ~~Relative contribution \\
  \hline
  Event pileup & $\leq 3 \times 10^{-5}$ \\
  Au+C reactions & $\leq 2.5 \times 10^{-5}$ \\
  \hline
  PID impurities & $\leq 10^{-3}$ \\ 
  Knockout reactions &  $ \leq 3 \times 10^{-3}$ \\
  Hyperon decays & $\leq 6.5 \times 10^{-4}$ \\
  \hline
  Antiprotons (model fit) & $\simeq 2 \times 10^{-8}$/evt \\  
  \hline \hline
 \end{tabular}
\end{table}

We have investigated the production of secondary protons, i.e.\ protons knocked out from target and near-target material
by primary hadrons (mostly neutrons, protons, and pions), using our GEANT3 detector simulation with GCALOR as hadronic
interaction package \cite{Zeitnitz1994}.  We found that their relative contribution to the proton yield in the phase space bin
usable for the fluctuation analysis ($y \in y_0 \pm 0.5$ and $0.4 < p_t < 1.6$~\gevc, see below) is of order $\leq3 \times 10^{-3}$
(50\% from $p$, 45\% from $n$, and 5\% from $\pi$ reactions).  Furthermore, the relative contribution to the
total yield due to protons stemming from weak decays of $\Lambda$ and $\Sigma^0$ hyperons produced in the collision can be
estimated from data to be of order $6.5 \times 10^{-4}$ only \cite{Adamczewski2019lamb}.  These contributions to the proton multiplicity
are also listed in Table~\ref{tab:nuisance} and their implications on the fluctuation analysis are discussed in Sec.~\ref{Sec:Results}.
\textcolor{black}{Note finally that the production of antiprotons is far sub-threshold at our bombarding energy and does not contribute any observable
hits in the detector.  From a thermal model fit to the particle yields measured at freeze-out, we estimate the antiproton yield to be of order
$2 \times 10^{-8}$/event in the 10\% most central Au+Au collisions.}

\subsection{Proton acceptance}

\begin{figure}[!htb]
  \begin{center}
     \resizebox{0.9\linewidth}{!} {
       \includegraphics{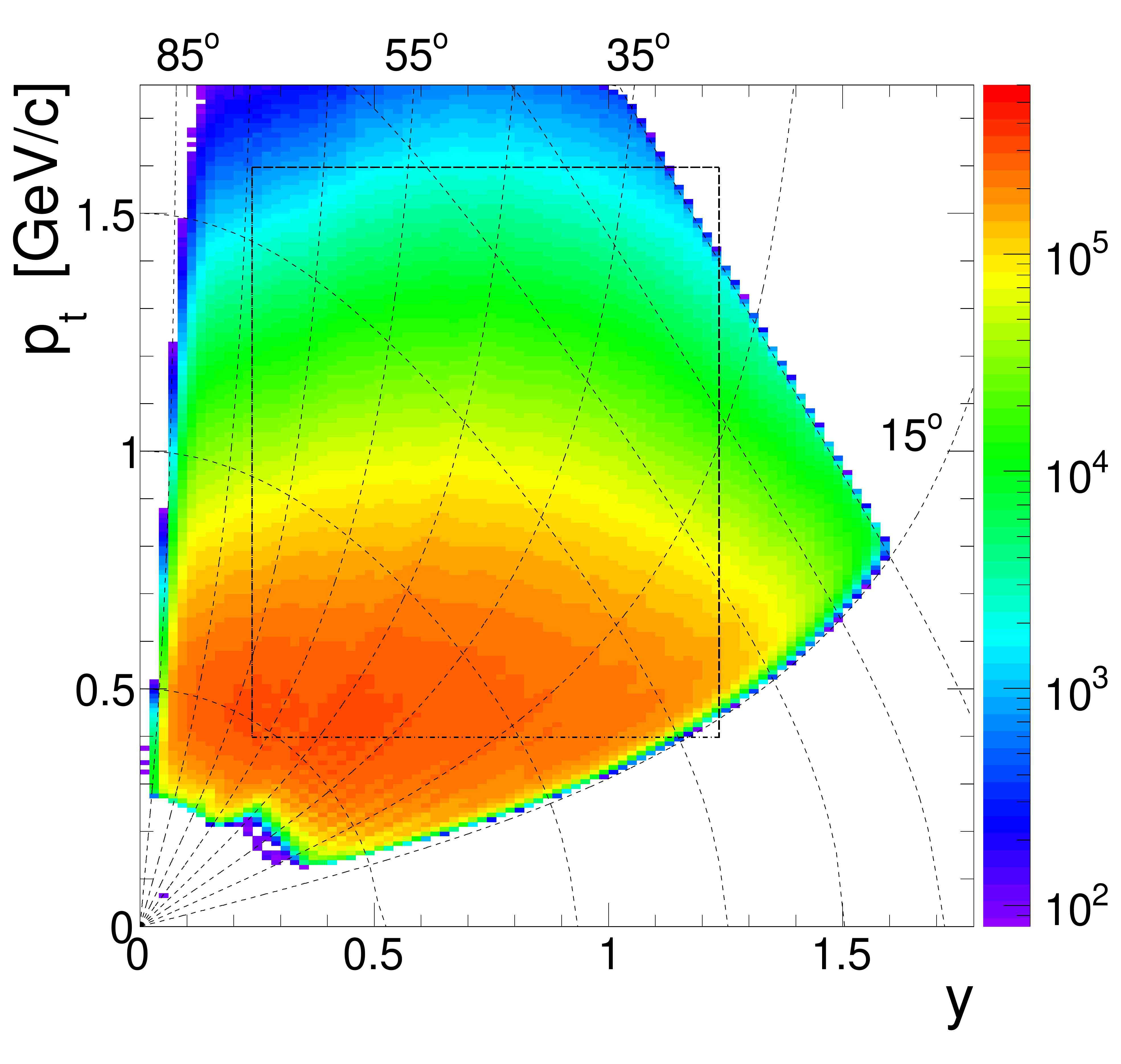} 
     }
  \end{center}
  \vspace*{-0.2cm}
  \caption[] {(Color online) Identified proton yield as a function of laboratory rapidity $y$ and
              transverse momentum $p_t$.  The acceptance is constrained by the geometry of
              HADES ($\theta = 15^o - 85^o$) and by momentum cuts ($0.3 < p <3$~\gevc).
              Thin dashed curves indicate the polar angle, in steps of $10^o$, and proton momentum,
              in steps of 0.5~\gevc.  The dashed rectangle corresponds to the phase space selected
              for the fluctuation analysis: $y = y_0 \pm 0.5$ and $0.4 < p_t < 1.6$~\gevc,
              where mid-rapidity $y_0 = 0.74$.
             }
  \label{fig:ypt}
\end{figure}

The yield of identified protons is shown in Fig.~\ref{fig:ypt} as a function of laboratory rapidity $y$ and transverse
momentum $p_t$.  The proton phase-space coverage is constrained by the polar angle acceptance of HADES
($\theta_{lab} = 15^{\circ} - 85^{\circ}$) as well as by a low-momentum cut ($\approx$~0.3~\gevc) due to energy loss in material
and deflection in the magnetic field, and an explicit high-momentum cut (3~\gevc) applied in the proton identification.
This results in a useful rapidity coverage of about 0.1 -- 1.5, which is quite well centered on the mid-rapidity $y_0 = 0.74$
of the 1.23\agev\ fixed-target reaction.  However, to guarantee a close to uniform and symmetric about mid-rapidity acceptance,
we have restricted the proton fluctuation analysis to the rapidity range $y = 0.24 - 1.24$ and transverse momentum range
$p_t = 0.4 - 1.6$~\gevc\, resulting in the dashed rectangle overlaid on Fig.~\ref{fig:ypt}.  \textcolor{black}{Notice that these selections
leave two small open corners in the acceptance.}  In addition, the azimuthal acceptance of HADES is also not complete because
of the six gaps occupied by the magnet cryostat \cite{Agakishiev2009tech}.  All of these are taken into account in the efficiency
corrections based on full detector simulations, as discussed in Sec.~\ref{Sec:Eff}.

\subsection{Characterizing the proton multiplicity distributions}

\begin{figure*}[!hbt]
  \begin{center}
     \resizebox{0.6\linewidth}{!} {
       \includegraphics{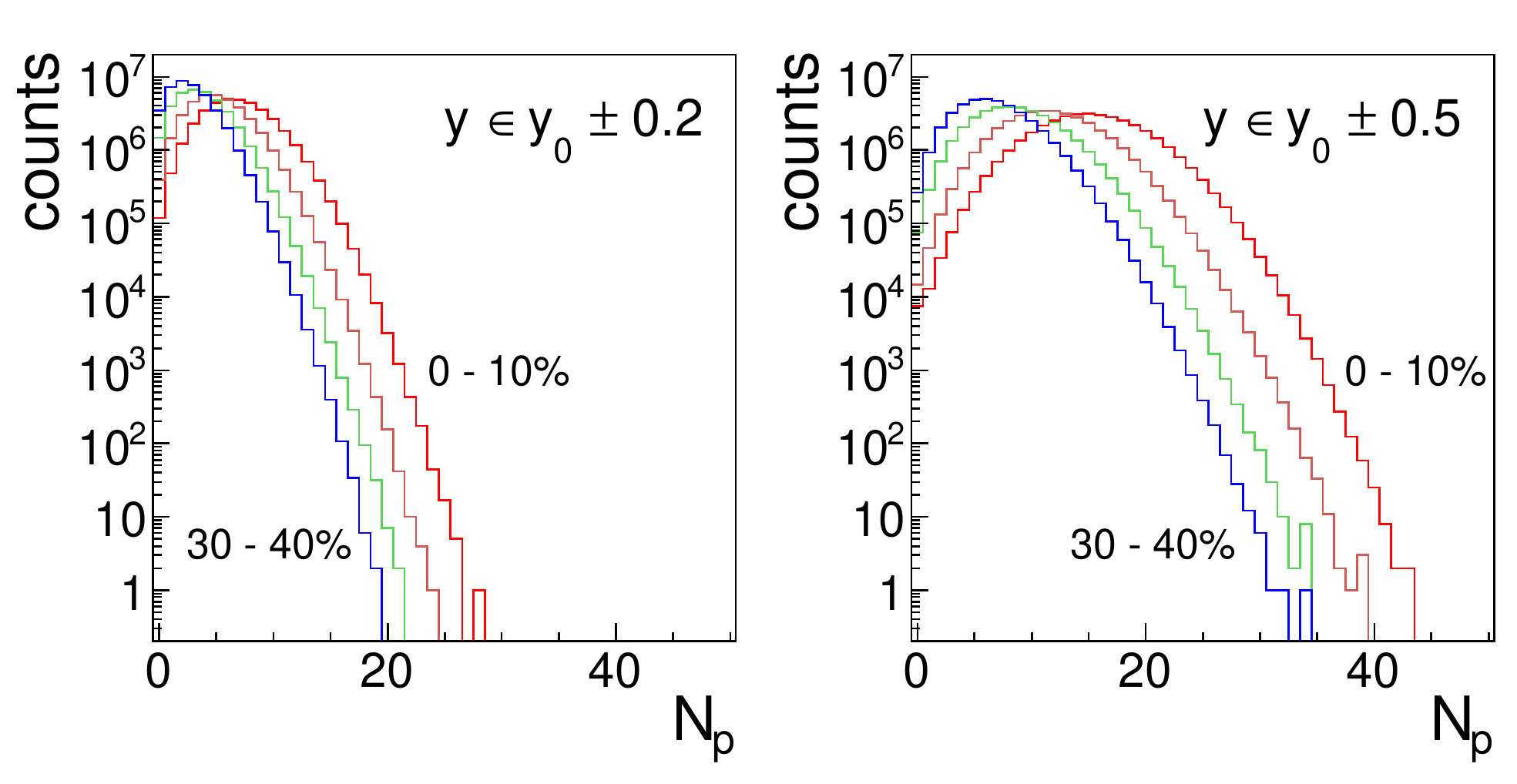}
     }
     \resizebox{0.6\linewidth}{!} {
       \includegraphics{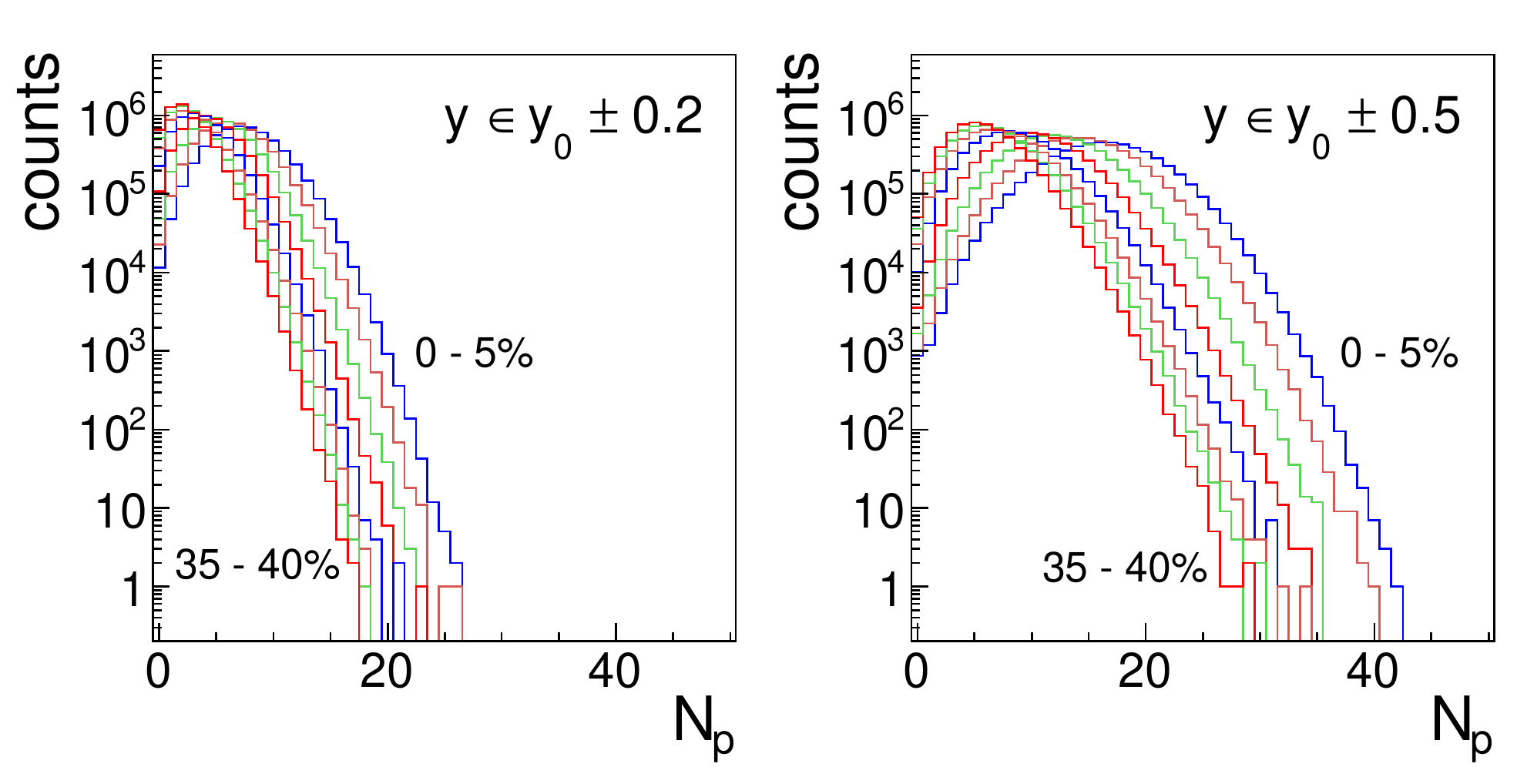} 
     }
  \end{center}
  \vspace*{-0.2cm}
  \caption[] {(Color online) Raw multiplicity distributions of fully reconstructed and identified protons in Au+Au collisions
              within a given phase-space bin ($y \in y_0 \pm \Delta y$ and $p_t = 0.4 - 1.6$~\gevc)
              and for different centrality selections based on the observable $\Sigma Q_{\textrm{FW}}$, using 10\%-wide bins (top)
              or 5\%-wide bins (bottom).
             }
  \label{fig:rawdata}
\end{figure*}

Before discussing corrections for detector inefficiency, we first take a look at the observed proton multiplicity
distribution, that is the distribution of the number $N_p$ of protons reconstructed and identified within the phase space
delineated in Fig.~\ref{fig:ypt}.  We have histogrammed this distribution for various centrality selections based
on the FWALL $\Sigma Q$ signal, as discussed before, and for various phase-space bins $y \in y_0 \pm \Delta y$
(with $\Delta y = 0.05 - 0.5$) and $p_t = 0.4 - 1.6$~\gevc.  Figure~\ref{fig:rawdata} shows the proton multiplicity 
distributions obtained in 5\% or 10\% centrality bins, and for a rapidity bin of $\Delta y = 0.2$ or $\Delta y = 0.5$,
respectively.  The basic idea of the fluctuation analysis is to remove from these ``raw'' distributions
any distorting nuisance effects, namely detector inefficiencies and reaction volume fluctuations, and then to systematically
characterize their shape in terms of higher-order moments and/or cumulants.  Procedures to achieve this are
discussed in the following two sections.  Because of the comfortably large size of our proton sample,
resulting in $\simeq 2 \times 10^7$ and $\simeq 4 \times 10^7$ events per centrality selection for the 5\% and 10\% bins
respectively, the shape of the proton distributions can be followed in Fig.~\ref{fig:rawdata} over more than six orders
of magnitude.  From this one may expect \cite{Luo2012,Luo2015a} that their cumulants can be extracted up to 4\textsuperscript{th}
order at least with sufficient statistical accuracy (i.e. $<5 - 10$\%) to quantify significant deviations from a simple Poisson
or binomial baseline.  In fact, in case of large deviations from such a baseline, much smaller statistics may be needed,
as has been argued in \cite{Bzdak2019}.

As already stated in the introduction, conserved quantities like baryon number, electric charge, or strangeness within a restricted
phase space and, in particular, their critical or pseudo-critical behavior are usually characterized by the higher-order
cumulants of the observed particle number distribution.  In an experiment, it is often convenient to first determine
the moments $\langle N^n \rangle$ or the central moments $\langle (N - \langle N \rangle)^n \rangle$ about the first moment.
Then, from the moments, all other quantities like factorial moments $\langle N (N-1) (N-2) \cdots \rangle$, cumulants, or 
factorial cumulants can be computed with ease (see e.g.\ \cite{Broniowski2017}).  In the literature \cite{Bzdak2012,Asakawa2016,LuoXu2017} various
notations\footnote{Asakawa and Kitazawa use e.g.\ $\langle N^n \rangle$ for moments, $\langle N^n \rangle_f$ for factorial moments,
$\langle N^n \rangle_c$ for cumulants, and $\langle N^n \rangle_{fc}$ for factorial cumulants \cite{Asakawa2016}.} for all
of those quantities are in use and we do not want to propose yet another one here.  We just follow Bzdak and Koch \cite{Bzdak2012}:
$M_n$ stands for moments of order $n$, $K_n$ for cumulants, $F_n$ for factorial moments, and $C_n$ for factorial cumulants.
For the acceptance and efficiency affected experimentally observed quantities, we use the corresponding lower-case letters,
$m_n$, $f_n$, and $k_n$.  Note also that deviations of a distribution from normality are usually characterized by non-zero values of
the dimension-less quantities named skewness, $\gamma_1 = K_{3}/K^{3/2}_{2}$, and excess kurtosis, $\gamma_2 = K_{4}/K^{2}_{2}$.
Yet other quantities referred to later in the text will be defined as needed.

The various moments and cumulants characterize a distribution in equivalent ways and a particular choice may just result from
convenience of use in a given situation.  Let us recall that moments $M_n$ and factorial moments $F_n$ transform into each other
via the following relationships \cite{Kendall2004,Broniowski2017}

\begin{equation}
  \begin{split}
 M_n = & \sum_{l=1}^{n} \: s_2(n,l) F_l \,,\\
 F_n = & \sum_{l=1}^{n} \: s_1(n,l) M_l \,,
  \end{split}
 \label{eq:momfacmom}
\end{equation}


\noindent
where $s_1(n,l)$ and $s_2(n,l)$ are the Stirling numbers of 1\textsuperscript{st} and 2\textsuperscript{nd} kind, respectively.  Note that these
relationships also hold between cumulants $K_n$ and factorial cumulants $C_n$.  Cumulants and moments, that is $K_n$ and $M_n$, or $C_n$ and $F_n$,
can be related via recursion \cite{Smith1995,Zheng2002}

\begin{equation}
  \begin{split}
  K_n = M_n - & \sum_{l=1}^{n-1} \: \binom{n-1}{l-1} \: K_l \: M_{n-l} \,,\\
  C_n = F_n - & \sum_{l=1}^{n-1} \: \binom{n-1}{l-1} \: C_l \: F_{n-l} \,,
  \end{split}
 \label{eq:momcum}
\end{equation}

\noindent
with the $\binom{n-1}{l-1}$ being binomial coefficients.

\section{Efficiency corrections}
\label{Sec:Eff}

The measured proton number distributions were obtained within the geometric acceptance of HADES, which is constrained
in polar and azimuthal angles, as well as in a limited momentum range only.  Furthermore, the proton yields are affected by
inefficiencies of the detector itself and of the hit finding, hit matching, track fitting, and particle identification algorithms
used in the reconstruction.  Geometric acceptance losses are minimized in our analysis by restricting the proton phase space to
the region indicated in Fig.~\ref{fig:ypt}.  Note also that, in the following, we do not explicitly distinguish between losses
due to finite acceptance and losses caused by hardware or analysis inefficiencies, we subsume all effects into one number which
we just call detection efficiency $\epsilon_{\textrm{det}}$.

\begin{figure}[!htb]
  \begin{center}
     \resizebox{0.9\linewidth}{!} {
       \includegraphics{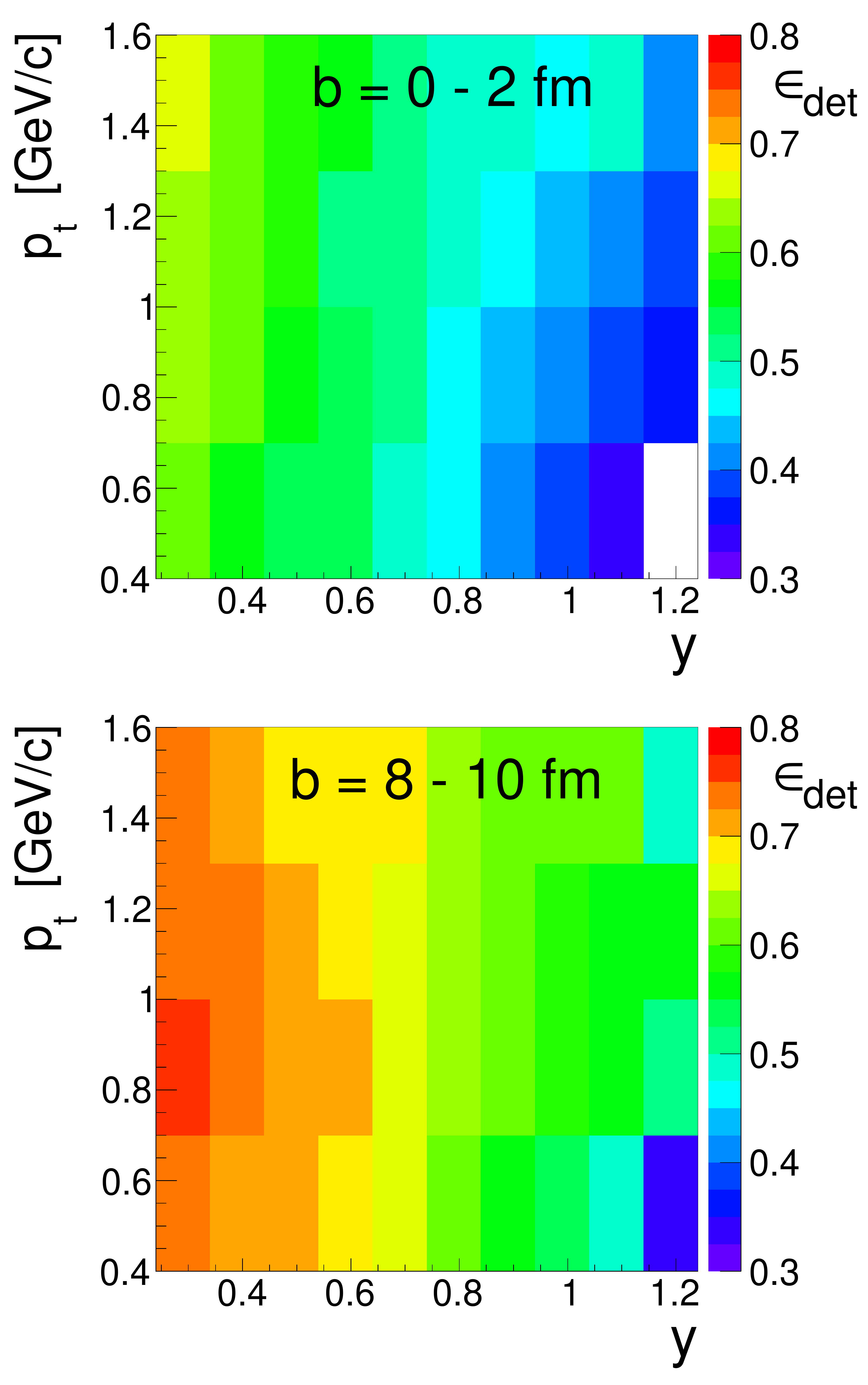} 
     }
  \end{center}
  \vspace*{-0.2cm}
  \caption[] {(Color online) Simulated proton detection efficiency $\epsilon_{\textrm{det}}$ in one sector of HADES as a function
              of rapidity $y$ and transverse momentum $p_t$, and for two impact parameter ranges
              (left, $b = 8 - 10$~fm and right, $b = 0 - 2$~fm).
             }
  \label{fig:iqmdeff1}
\end{figure}

\begin{figure}[!htb]
  \begin{center}
     \resizebox{0.9\linewidth}{!} {
       \includegraphics{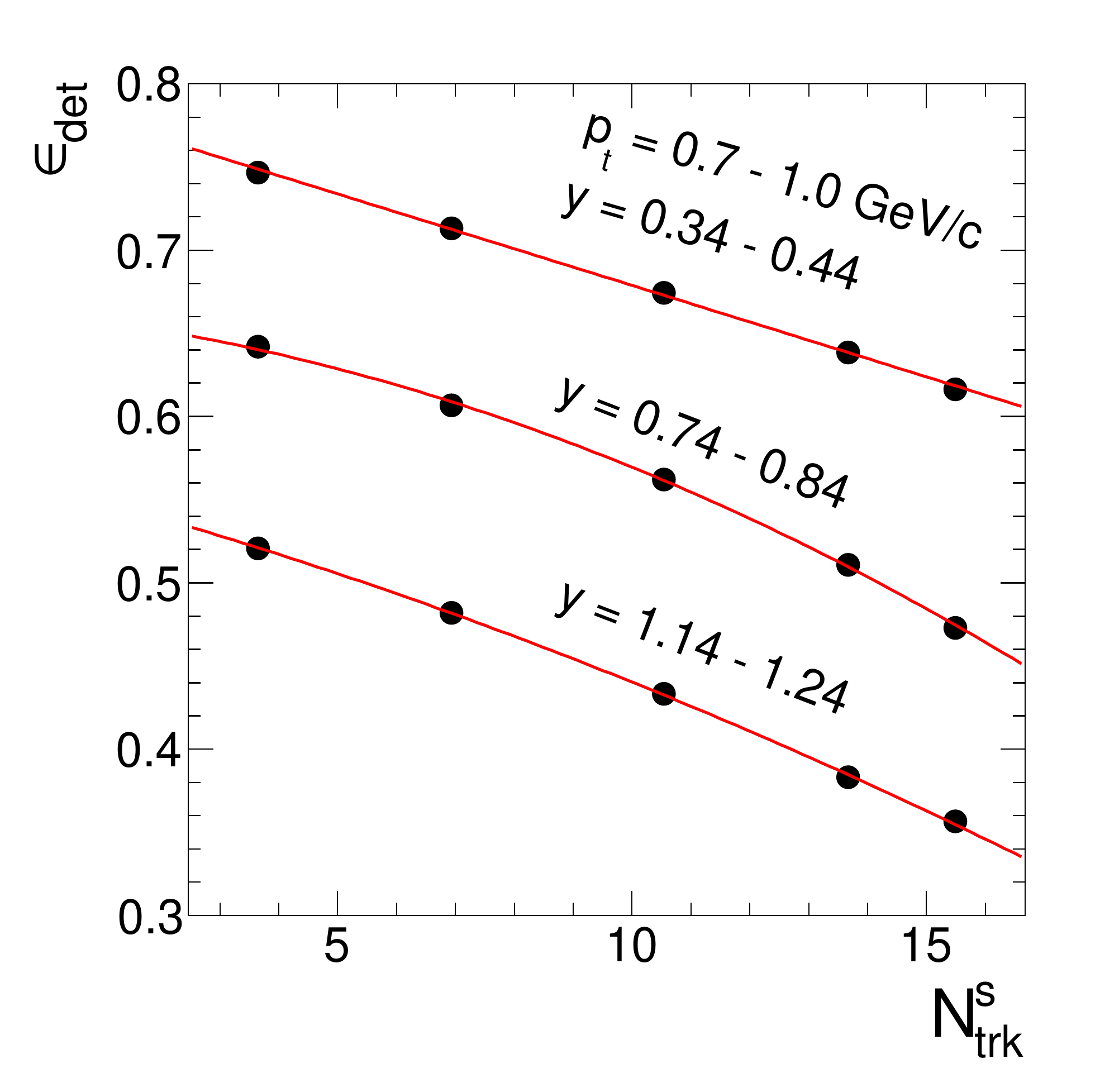} 
     }
  \end{center}
  \vspace*{-0.2cm}
  \caption[] {Simulated proton detection efficiency $\epsilon_{\textrm{det}}$ as a function of the
              number of reconstructed tracks per sector $N^s_{\textrm{trk}}$, shown here for three out
              of the 40 narrow phase-space bins used in our analysis.  Efficiency points correspond
              to 2~fm wide impact-parameter slices centered at $b = $ 9, 7, 5, 3, and 1~fm, respectively;
              curves are the polynomials adjusted to model the occupancy dependence of the efficiency.
             }
  \label{fig:iqmdeff2}
\end{figure}

With the HADES setup fully implemented in the detector modeling package GEANT3 \cite{GEANT3} and with appropriate digitizing
algorithms emulating the physical behavior of all detector components we have performed realistic simulations of its response
to Au+Au collisions.  In particular, using IQMD\footnote{In this paper, a reference to IQMD always implies that a MST algorithm
was used to add nuclear clusters to the event.} generated events, these simulations allowed to systematically investigate the proton
detection efficiency throughout the covered phase space.  Figure~\ref{fig:iqmdeff1} displays the calculated efficiency $\epsilon_{\textrm{det}}$ 
as a function of rapidity $y$ and transverse momentum $p_t$ for two centralities corresponding to narrow ranges of the collision impact
parameter $b$.  It appears clearly from these plots that the detection efficiency in HADES depends not only on phase-space bin (more strongly
on $y$ than on $p_t$) but also on the event centrality: overall the efficiency is considerably reduced in central events with
respect to the more peripheral ones.  This reduction can be connected to an increase of the hit and track densities in the detector
with increasing particle multiplicity.  In other words, larger occupancy in the detector lead to some deterioration of the reconstruction
procedures.  As both detection and reconstruction in HADES operate on one sector at a time, it is sufficient to investigate the
efficiency on a per-sector basis.  We illustrate this in Fig.~\ref{fig:iqmdeff2} by plotting for a few narrow phase-space bins
the simulated detection efficiency $\epsilon_{\textrm{det}}$ in one HADES sector as a function of the number of particle tracks $N^s_{\textrm{trk}}$
reconstructed in this sector, obtained by selecting events within a sequence of narrow $b$ bins.  Low-order polynomials have been fitted to the efficiency,
as also shown in the figure.  Such fits were done in all sectors and for 40 phase-space bins, using ten-fold segmentation in $y$ and four-fold
in $p_t$.  In fact, in most cases, a linear function turned out to be sufficient to model the observed behavior of $\epsilon = \epsilon(N^s_{\textrm{trk}})$.
Next, we discuss how these modeled efficiencies can be used to correct the measured proton number moments and cumulants.

Efficiency corrections of particle number cumulants have been extensively discussed
in the literature \cite{Bzdak2012,Bzdak2015,Luo2015a,Kitazawa2016,Nonaka2017}.
They usually rely on the premise of a binomial efficiency model, that is on the assumption that the detection processes of many
particles in a detector are independent of each other.   This implies that each particle has the same probability of being detected,
irrespective of the actual number of particles hitting the detector in a given event.  In such a case, the resulting multiplicity
distribution of detected particles is a binomial distribution.  A convenient property of the binomial efficiency model is that
it leads to a particularly simple relationship\footnote{This has also been extended to net particle numbers $N - \bar{N}$, where
$N$ and $\bar{N}$ refer to the particles and antiparticles, respectively.  However, at our beam energy we have to deal with
protons only.} for any order $n$ between the factorial moments $f_n$ of the distribution of detected particles and the factorial
moments $F_n$ of the true particle distribution \cite{Kirej2004,Bzdak2012,Kitazawa2016,HeLuo2018}:

\begin{equation}
  f_n = \epsilon^n F_n \,,
 \label{eq:facmom}
\end{equation}

\noindent
where $\epsilon$ is the detection efficiency of a particle.

If $\epsilon$ is known, the true $F_n$ can be calculated easily from the measured $f_n$, and all other true moments and cumulants
can be obtained with the help of Eqs.~\eqref{eq:momfacmom} and \eqref{eq:momcum}.  We have to handle, however, two more problems:
first, as the efficiency depends on phase-space bin (see Fig.~\ref{fig:iqmdeff1}), we have to do efficiency corrections bin by bin
and merge the corrected values into one global result.  Second, we also have to take into account that the efficiencies depend on
the number of particles actually hitting the detector (see Fig.~\ref{fig:iqmdeff2}) and therefore change from event to event.
However, ways to pass both hurdles have been found and are discussed next.

The need to handle more than one efficiency value, $\epsilon$ depending e.g.\ on particle species or on the $y - p_t$ bin,
had been recognized before and is discussed in \cite{Bzdak2015,Luo2015a,Kitazawa2016}.  In particular, in Ref.~\cite{Bzdak2015}
so-called local factorial moments were introduced, i.e.\ factorial moments of particles in a given phase-space bin and
obeying Eq.~\eqref{eq:facmom} individually, so that efficiency corrections can be done bin-wise.  In addition, based on a
multinomial expansion of the full factorial moments $F_n$ in terms of the corrected local moments, a prescription how to sum over
all phase-space bins was presented.  Although formally correct, the procedure is rather awkward to implement, and it quickly
turns prohibitively memory and CPU-time intensive if applied to big event samples and/or a large number $N_{b}$ of phase-space bins.

A much more efficient scheme based on factorial cumulants has been proposed in \cite{Kitazawa2016,Nonaka2017,Kitazawa2017}.
It omits the full multinomial expansion in terms of local factorial cumulants (or factorial moments), leading to a vast reduction
in memory needs and computing time\footnote{E.g., the number of terms to be evaluated and stored at 4\textsuperscript{th} order
for $N_b$ bins decreases from $[(N_{b}+3) (N_{b}+2) (N_{b}+1) N_{b}]/24$ to a mere 13 terms, independent of $N_b$ \cite{Nonaka2017}.}.
We have implemented this scheme in our analysis, using factorial moments however, and applied it directly to the efficiency correction
of the proton moments.  In doing so, we have partitioned the phase space covered by HADES (see Fig.~\ref{fig:ypt}) into 240 bins
in total, namely 6~sectors $\times$ 10~rapidity bins $\times$ 4~$p_t$ bins.

We have investigated two ways to overcome the second complication, that is the dependence of the efficiency on the number of tracks:
either by introducing an event-by-event recalculation of the efficiency correction or by using an unfolding procedure to directly
retrieve the true particle distribution from the measured one.

\subsection{Occupancy-dependent efficiency correction}

The dependence of the detection and reconstruction efficiencies on track number leads evidently to an event-by-event change of $\epsilon_{\textrm{det}}$.
On condition that the binomial efficiency model remains valid or at least a good approximation, one can consider grouping the events into
classes of identical efficiency and apply the efficiency correction of Eq.~\eqref{eq:facmom} individually to each one of these classes.
As has indeed been shown in \cite{HeLuo2018}, the efficiency-corrected factorial moments $F_n$ of a superposition of particle distributions,
stemming e.g.\ from different event classes, can be obtained from the weighted means of the observed class factorial moments $f_n^{(i)}$ as

\begin{equation}
  F_n = \sum_{i=0}^{k} a_i \frac{f_n^{(i)}}{\epsilon^n_i} \,, 
\label{eq:superposition}
\end{equation}

\noindent
where $\epsilon_i$ are the individual class efficiencies, $a_i$ are the class weights normalized such that $\sum a_i = 1$, and the index $i$
runs over all classes.  Note that this is a generalization of Eq.~\eqref{eq:facmom} and it is also applicable to factorial cumulants,
but not to moments and cumulants in general \cite{HeLuo2018}.  In particular, when the efficiency changes from event to event, each event
of the sample analyzed can be considered as a class of its own and the relation still remains valid.  This provides therefore a convenient
way to apply efficiency corrections within the event analysis loop by, first, recalculating the efficiencies on the fly for all phase-space
bins of interest as a function of the number of reconstructed tracks per sector as discussed in Sec.~\ref{Sec:Eff} and, second, computing
the average factorial moments (using e.g.\ Eq.~(17) of Ref.~\cite{Bzdak2015}) or factorial cumulants (using e.g.\ Eq.~(61)
of Ref.~\cite{Nonaka2017}).  We have investigated this procedure in GEANT3 detector simulations using Au+Au events calculated with
the IQMD transport model and we found good agreement of the corrected and true proton moments.  This also supports the
underlying assumption of event-wise binomial efficiencies in the HADES detector.

Note for completeness that non-binomial efficiencies based on the hypergeometric or beta-binomial distributions have been discussed
in \cite{Bzdak2016}.  Although the properties of these somewhat {\em ad hoc} models are well known, they lack an obvious connection to
physical phenomena playing a role in the actual detection process.  In Appendix~\ref{Sec:Occupancy} we propose yet another model,
known as the urn occupancy model \cite{Mahmoud2008}, that in fact possesses such an intuitive connection.  It is, however, specifically 
tailored for detectors with a well-defined hardware segmentation like tiled hodoscopes, pixel telescopes, modular calorimeters, etc.
The HADES setup, as a whole, does not fall into either category but we can still define for it a virtual subdivision and treat
the number of virtual segments $N_{seg}$ as a free parameter that, together with the single-hit efficiency $\epsilon_0$, can be adjusted
to simulated proton distributions.  Further below we show and discuss the result of such a fit.

\subsection{Response matrix and unfolding}

\begin{figure}[!htb]
  \begin{center}
     \resizebox{1.0\linewidth}{!} {
       \includegraphics{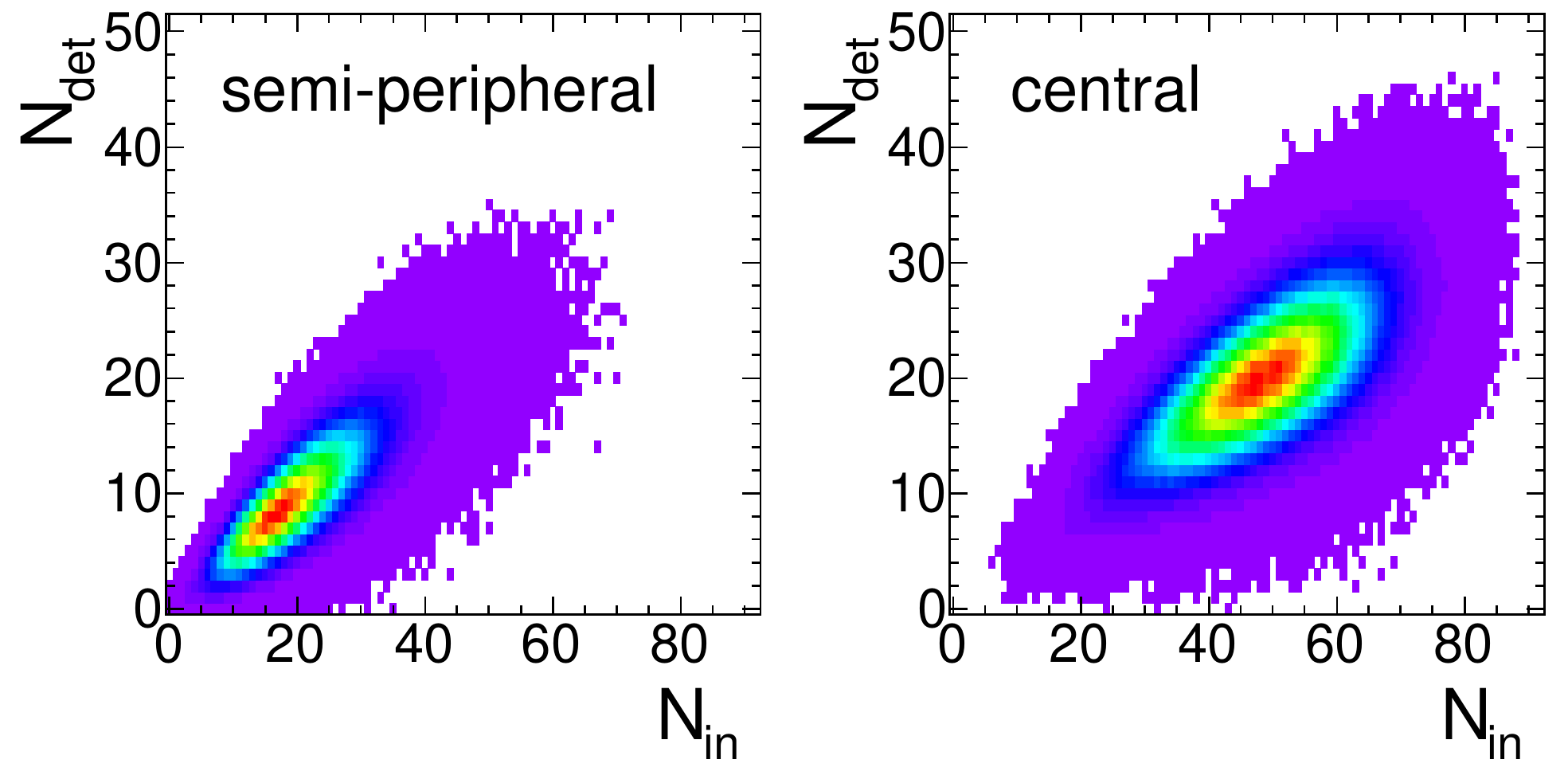} 
     }
  \end{center}
  \vspace*{-0.2cm}
  \caption[] {Proton response matrices of the HADES detector simulated with IQMD clustering
              for two centrality selections, semi-peripheral (30 -- 40\%) and central (0 -- 10\%).  These matrices
              encode the distribution of the number $N_{det}$ of reconstructed proton tracks -- within the phase-space bin
              of interest (here $y \in y_0 \pm 0.5$) -- for any given number $N_{in}$ of protons emitted into this bin.
             }
  \label{fig:protresp}
\end{figure}

Figure \ref{fig:protresp} shows simulated distributions of the number of detected protons $N_{det}$ in HADES as a function
of the number $N_{in}$ of protons emitted into the phase space of interest for IQMD events with 0 -- 10\% (30 -- 40\%) centrality.
These two-dimensional histograms represent the response of the detector for proton production in Au+Au collisions at a given centrality.
Their shape is not only determined by the multi-proton detection and reconstruction efficiency, but also by the shape of the true
proton multiplicity distribution presented as input.\footnote{Note that the finite momentum resolution of the detector also leads to small
cross-boundary effects, visible e.g. in Fig.~\ref{fig:protresp} for peripheral collisions: $N_{det}>0$ although $N_{in} = 0$.} 
These response matrices have been obtained from the same simulation used to
extract the efficiency per phase-space bin and centrality bin, discussed in the occupancy-dependent correction scheme, which basically assumes
binomial efficiencies.  On the other hand, any deviation from the binomial efficiency model would also be encoded in this response matrix.
It is therefore appropriate to investigate whether unfolding of the measured proton distributions
with the help of such response matrices is a useful approach to efficiency correction.  In the context of particle number fluctuation
analyses, a proof-of-principle simulation study of Bayesian unfolding has indeed been presented in \cite{Garg2013}.  On our side,
we have investigated this approach by making use of those unfolding procedures implemented in the ROOT analysis framework \cite{ROOT1993}.
It is well known that unfolding by straightforward inversion of the detector response matrix is mostly not successful, in particular
if the matrix is generated by Monte Carlo and is hence affected by limited event statistics.  The simulated proton response matrices
shown in Fig.~\ref{fig:protresp} clearly display the inevitable signs of resulting inaccuracies.  Such a matrix is generally ill-conditioned
and its quasi-singularity leads to unstable or even plainly unphysical results.  Various techniques to solve ill-behaved equation
sets have been proposed and are widely discussed in the literature: for example, Bayesian unfolding \cite{DAgostini1994}, singular value
decomposition (SVD) \cite{Hocker1995}, matrix regularization schemes \cite{Schmitt2012}, and Wiener filtering \cite{Tang2017}.

In our simulation study, we have investigated Tikhonov-Miller regularization and, for comparison, also unfolding with SVD.
In both cases, the aim is to solve an over-determined system of equations $\bf A \cdot x = y$ with a robust least-squares
procedure, where $\bf A$ is the response matrix, $\bf x$ is the unknown input vector, and $\bf y$ is the measured output vector.
In our context, $\bf x$ corresponds to the true particle number distribution and $\bf y$ to the actually measured distribution.
In an over-determined system, the dimension of $\bf y$ is larger than the dimension of $\bf x$ and the solution will only be
approximate.  Solving such a system in a least-squares sense is then equivalent to finding the minimum of the functional
\( \chi^2 = |\mathbf{A \cdot x - y}|^2 \).  To achieve a more robust solution, a Tikhonov regularization term \cite{Schmitt2012} can
be added to this functional:  \( \chi^2 = |\mathbf{A \cdot x - y}|^2  + \lambda \mathbf{x \cdot H \cdot x} \), where $\lambda$ is a
Lagrange multiplier controlling the strength of the regularization and $\bf H$ is a square matrix built from first-order or higher-order
finite differences of $\bf x$.  The basic idea of Tikhonov is that the additional quadratic term serves as a constraint that dampens
instabilities in the solution $\bf x$.  A typical choice for $\bf H$ is to use second-order finite differences in $x$ which favors minimum
overall curvature of the vector $\bf x$ and suppresses higher-order oscillations.  The optimal regularization strength is usually found
by a $\lambda$ scan and the ROOT implementation\footnote{ROOT class TUnfold.} provides two different methods to do so: the L-curve scan
and the minimization of global correlation coefficients.

We have furthermore explored an unfolding method based on singular value decomposition \cite{Hocker1995}, made available as well
in ROOT.\footnote{Implemented in ROOT class TSVDUnfold.}  The starting point of SVD unfolding is to write the response matrix $\bf A$
as a product $\bf A = U \cdot S \cdot V^T$, where $\bf U$ and $\bf V$ are orthonormal matrices, and $\bf S$
is a diagonal matrix the elements $s_i$ of which are the singular values of matrix $\bf A$.  All $s_i \geq 0$ and, no matter
how ill-conditioned $\bf A$ is, this decomposition can always be done.  For one, SVD gives us a clear diagnosis of
the degree of singularity of the response matrix and, arranging the singular values in decreasing order, it allows to
cure instabilities by truncation, that is by removing all terms with $s_i$ smaller than a given threshold value $s_{thr}$.
The solution $\bf x$ of the least-squares problem can be written as a linear combination of columns of matrix $\bf V$

\[ \mathbf{x} = \sum_i \left( \frac{\mathbf{U}_{(i)} \cdot \mathbf{y}}{s_i} \right) \mathbf{V}_{(i)} \,,\]

\noindent
with the summation going over all $i$ for which $s_i \geq s_{thr}$.  The threshold value $s_{thr}$ can be determined using
statistical significance arguments (details are given in \cite{Hocker1995}).  Note that SVD can also be combined with
a regularization, enforcing e.g.\ positiveness of the solution or minimum curvature.  Further below we show how well these
unfolding methods fare in our simulations.

\subsection{Moment expansion method}

\begin{figure}[!htb]
  \begin{center}
     \resizebox{1.0\linewidth}{!} {
       \includegraphics{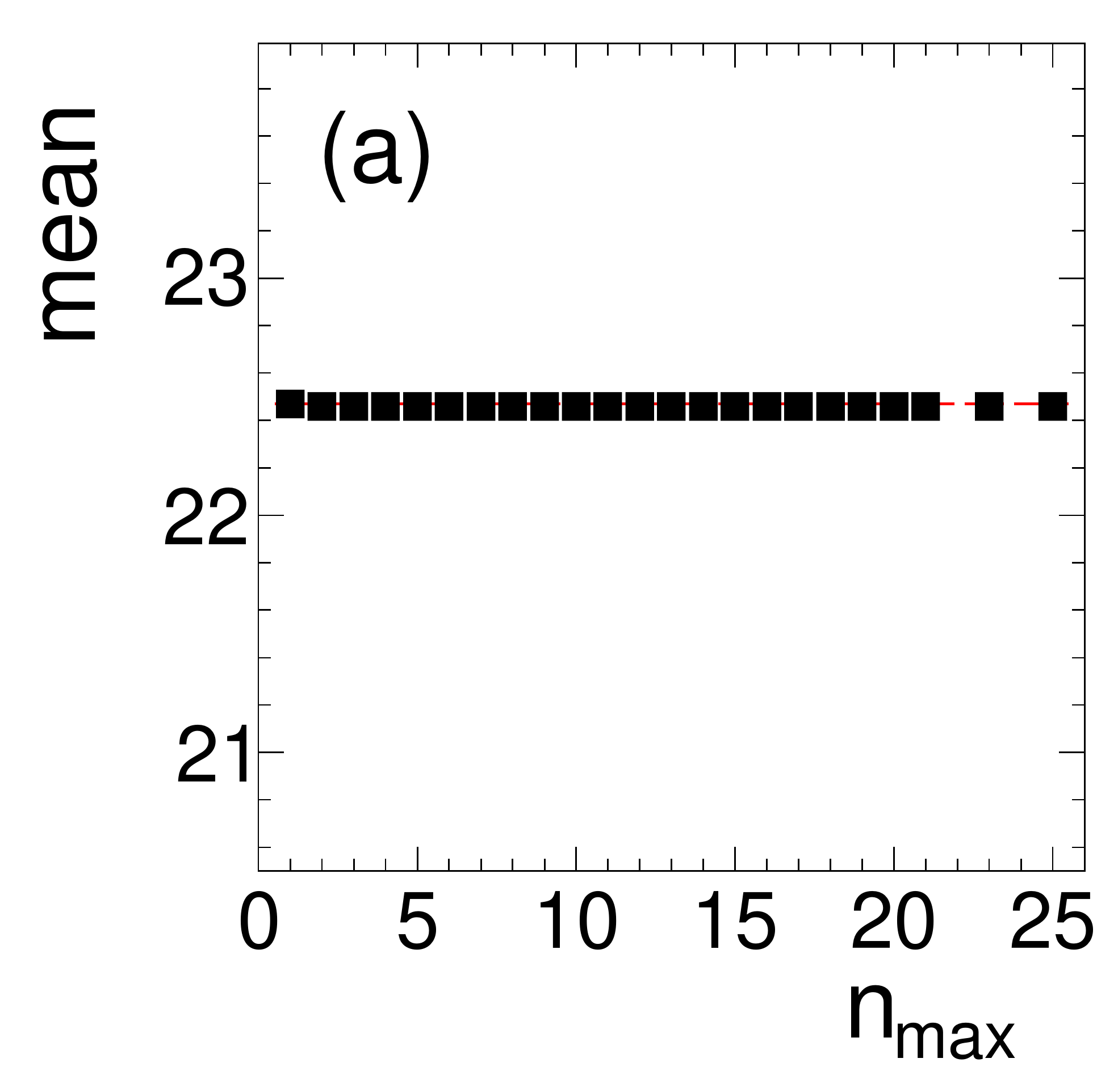} 
       \includegraphics{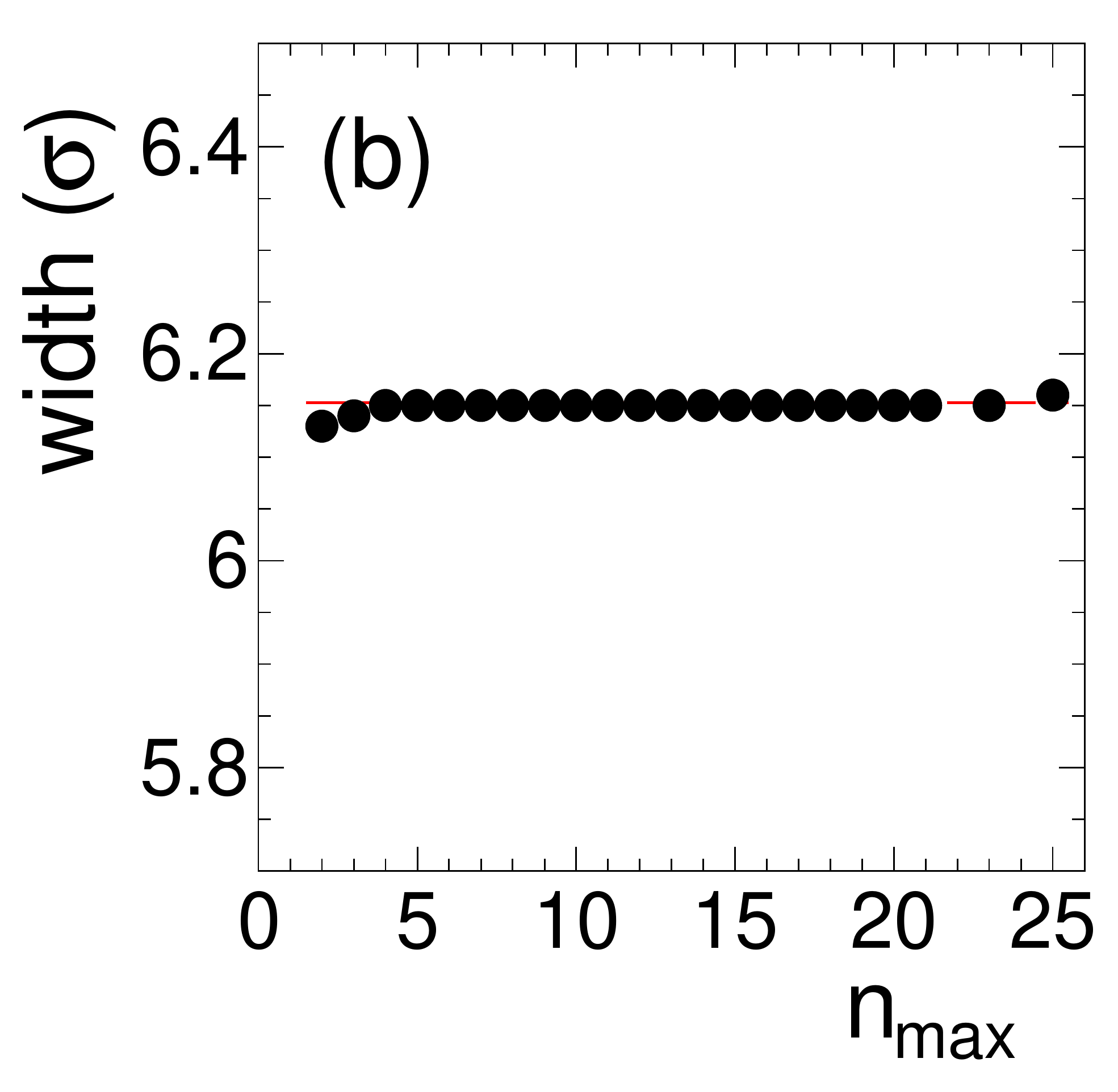} 
     }
     \resizebox{1.0\linewidth}{!} {
       \includegraphics{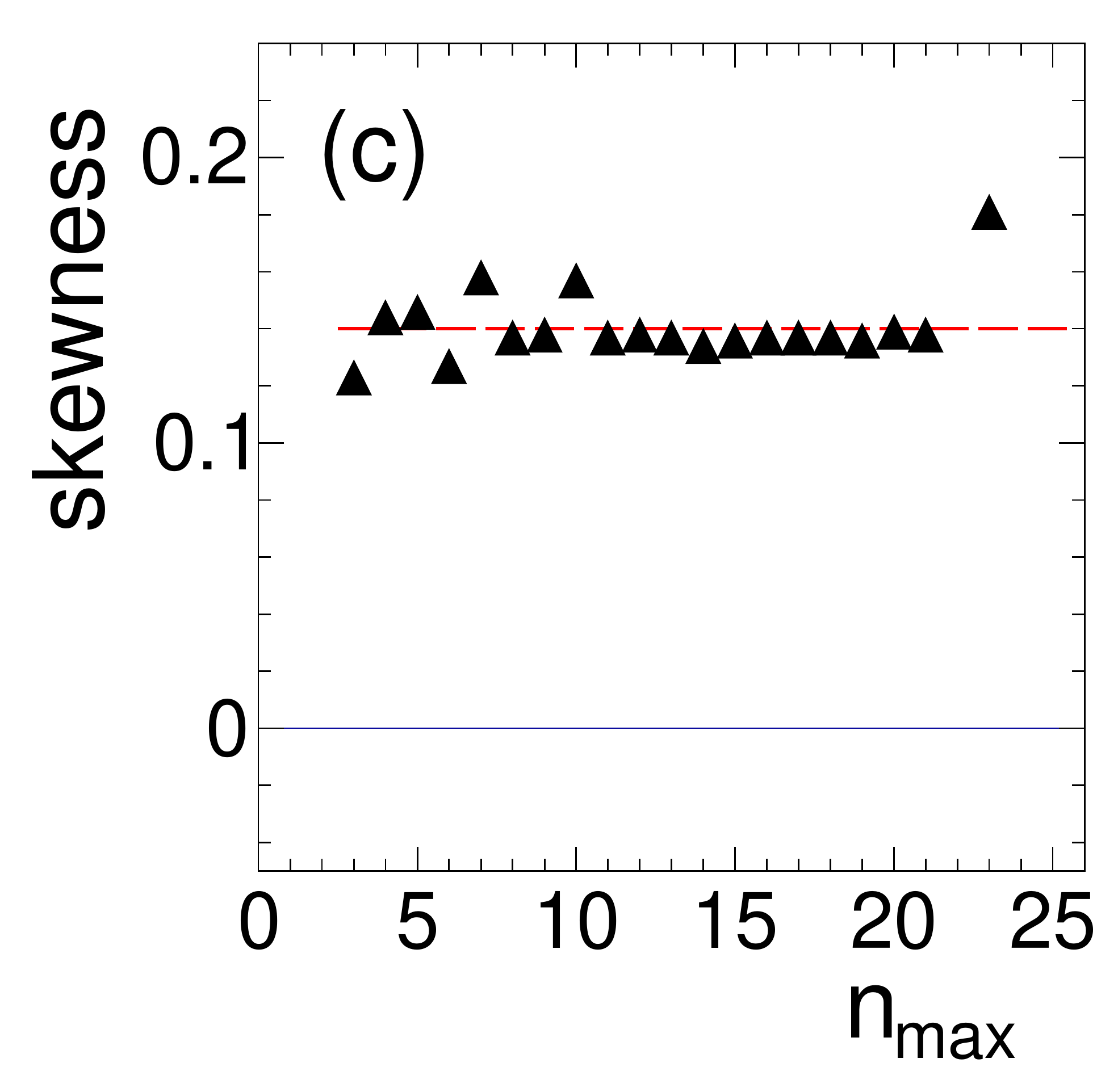} 
       \includegraphics{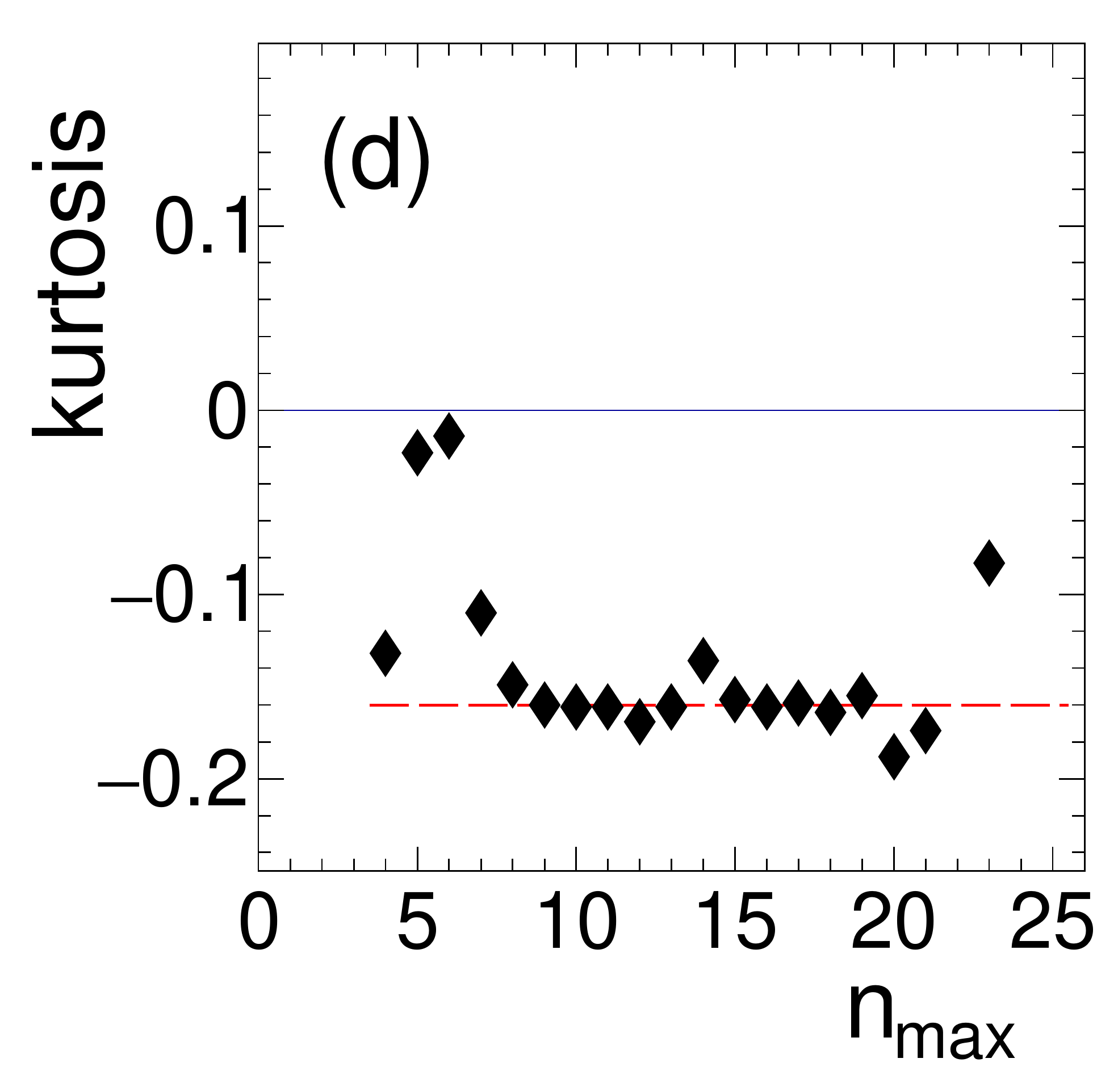} 
     }
  \end{center}
  \vspace*{-0.2cm}
  \caption[] {(Color online) Moment expansion technique applied to proton distributions from IQMD events (0 -- 10\% most central, $y \in y_0 \pm 0.2$).
              Shown is the dependence of the corrected moments on the truncation order $n_{max}$ of the moment matrix $\mathcal{A}$.
              The plateau region ($n_{max} \geq 10$), corresponding to stable behavior, is close to the true IQMD values,
              namely $\langle N_p \rangle=22.5$ (a), $\sigma=6.15$ (b), skewness $\equiv \gamma_1=0.14$ (c),
              and kurtosis $\equiv \gamma_2=-0.16$ (d), indicated by red dashed lines. 
             }
  \label{fig:momexp}
\end{figure}

For completeness, we would like to mention yet another approach to the efficiency correction of distribution moments, namely the
recently proposed method of moment expansion based on the detector response matrix \cite{Nonaka2018}.  Arguing that in most cases
the true particle number distributions are not really needed but typically only their cumulants, the authors of \cite{Nonaka2018}
proposed to bypass the unfolding altogether and instead establish a direct formal relation between the measured moments $m_n$ 
(or cumulants $k_n$) and the true moments $M_n$ (or cumulants $K_n$).  Indeed, the relevant information to do so is encoded
in the response matrix $\bf A$, more specifically in its column-wise moments which can be used to expand the observed $m_n$
in terms of the true $M_n$.  Depending on the efficiency model used (e.g.\ binomial, hypergeometric, beta-binomial) this expansion
is closed and, by solving the resulting system of equations \( \mathbf{m} = \mathcal{A} \cdot \mathbf{M} \), the $M_n$ are
expressed in terms of the $m_n$.  Note that the moment matrix $\mathcal{A}$ has a much lower dimension than the response
matrix $\bf A$ itself, typically $5^2 - 10^2$ versus $50^2 - 100^2$, which greatly eases its inversion.  For efficiency models
that do not lead to a closed form, for example models where the efficiency depends in a non-trivial way on particle multiplicity,
the expansion must be truncated at some order $n_{max}$ to be amenable to a solution.  In that case, one must study the inversion
as a function of $n_{max}$ to control the stability of the result obtained.  The effect of the truncation on the moments retrieved with
the expansion method is exemplified in Fig.~\ref{fig:momexp} for the first four moments of a simulated proton distribution.
The solution stabilizes on a plateau at $n_{max} \simeq 10$ meaning that the expansion can be safely truncated at this value of $n$.
However, note that in this analysis based on the simulated proton response matrix (the one used also in the unfolding investigations)
numerical instabilities start to set in for $n_{max} > 20$. 

Unfolding as well as expansion methods use as input a response matrix typically produced by Monte Carlo with the help of a
given event generator.  However, as often pointed out (see e.g.\ \cite{Blobel2013,Nonaka2018}), sufficiently large statistics
and a proper choice of the simulation input are of importance.  The input model can have an influence on the resulting
response and it is therefore mandatory to carefully check its validity, not only in the region of phase space covered
by the detector acceptance, but also beyond because of the inherent migration of yield from the latter to the former.

\subsection{Validation with IQMD transport events}

\begin{figure*}[!thb]
  \begin{center}
     \resizebox{0.8\linewidth}{!} {
       \includegraphics{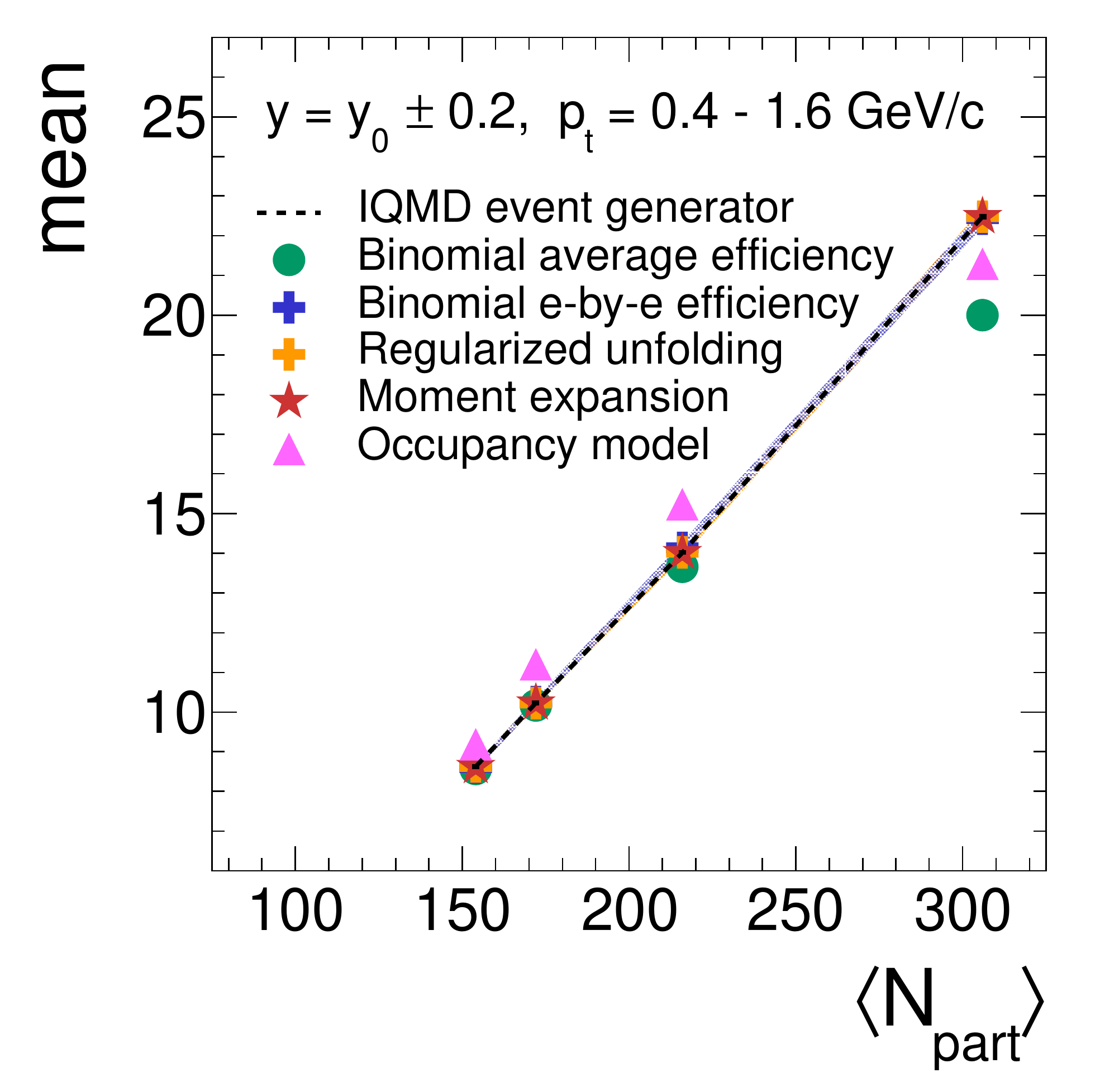} 
       \includegraphics{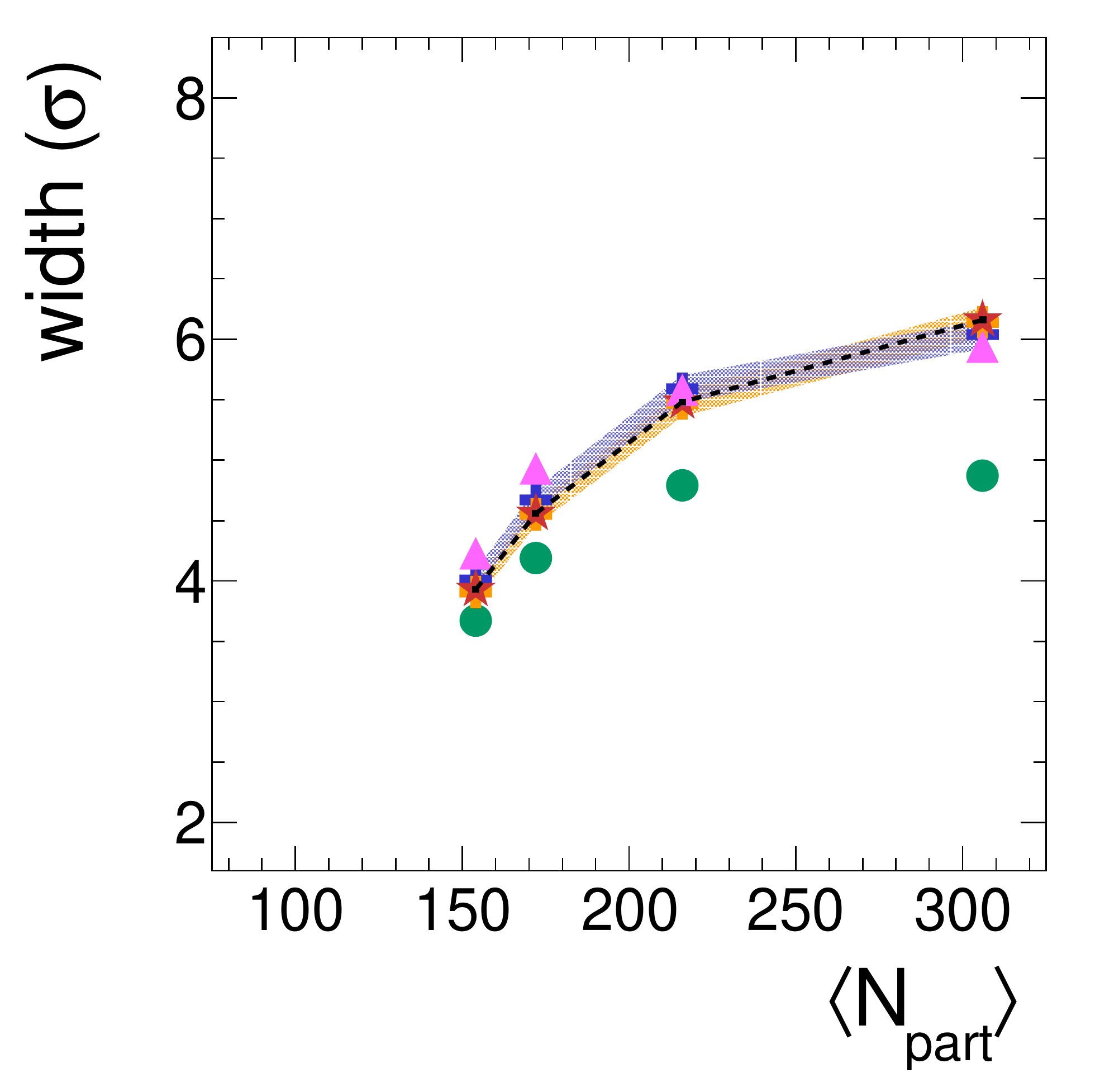} 
     }
     \resizebox{0.8\linewidth}{!} {
       \includegraphics{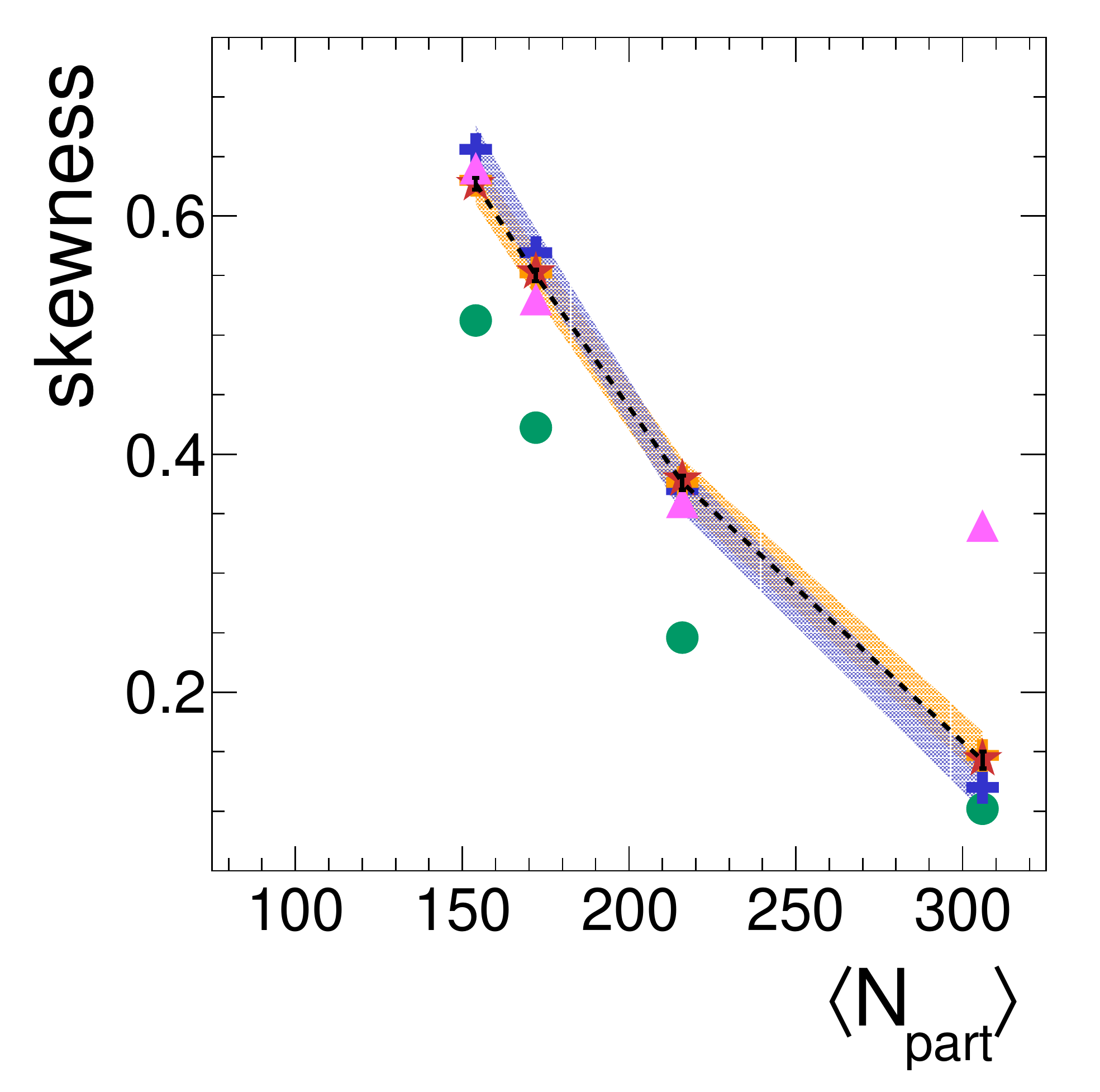} 
       \includegraphics{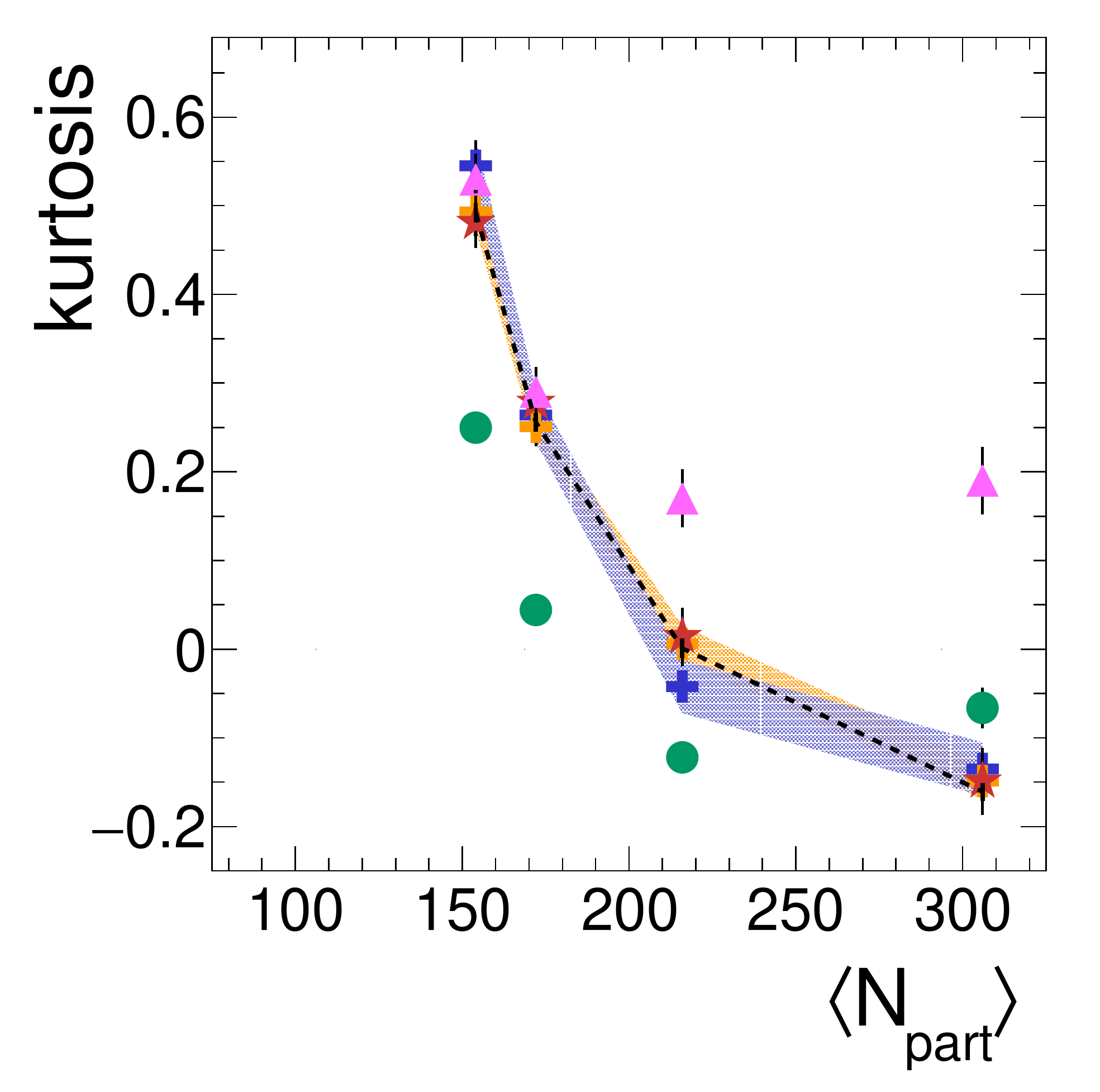} 
     }
  \end{center}
  \vspace*{-0.2cm}
  \caption[] {(Color online) Comparison of various efficiency correction methods applied
              to a simulation using IQMD events and selecting protons within $y \in y_0 \pm 0.2$.
              Plotted are the corrected proton number mean (a), width (b), skewness (c), and kurtosis (d)
              as a function of $\langle N_{\textrm{part}}\rangle$.  Error bars are statistical and
              colored shaded bands correspond to systematic errors of the correction technique:
              blue for the binomial event-by-event correction, ocher for unfolding, respectively.
              The IQMD truth is shown as black dashed lines.
             }
  \label{fig:effiqmddy02}
\end{figure*}

\begin{figure*}[!htb]
  \begin{center}
     \resizebox{0.8\linewidth}{!} {
       \includegraphics{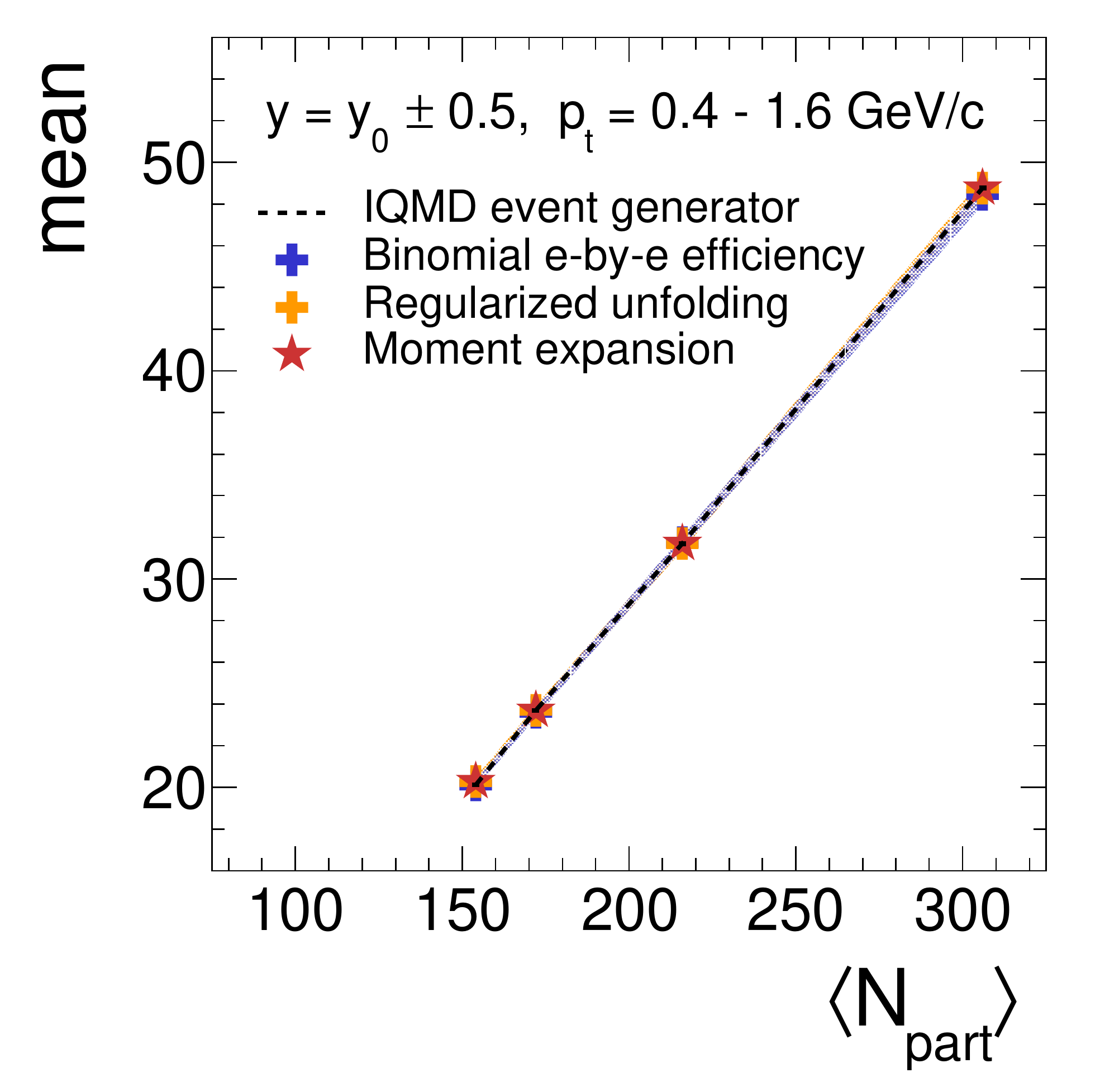} 
       \includegraphics{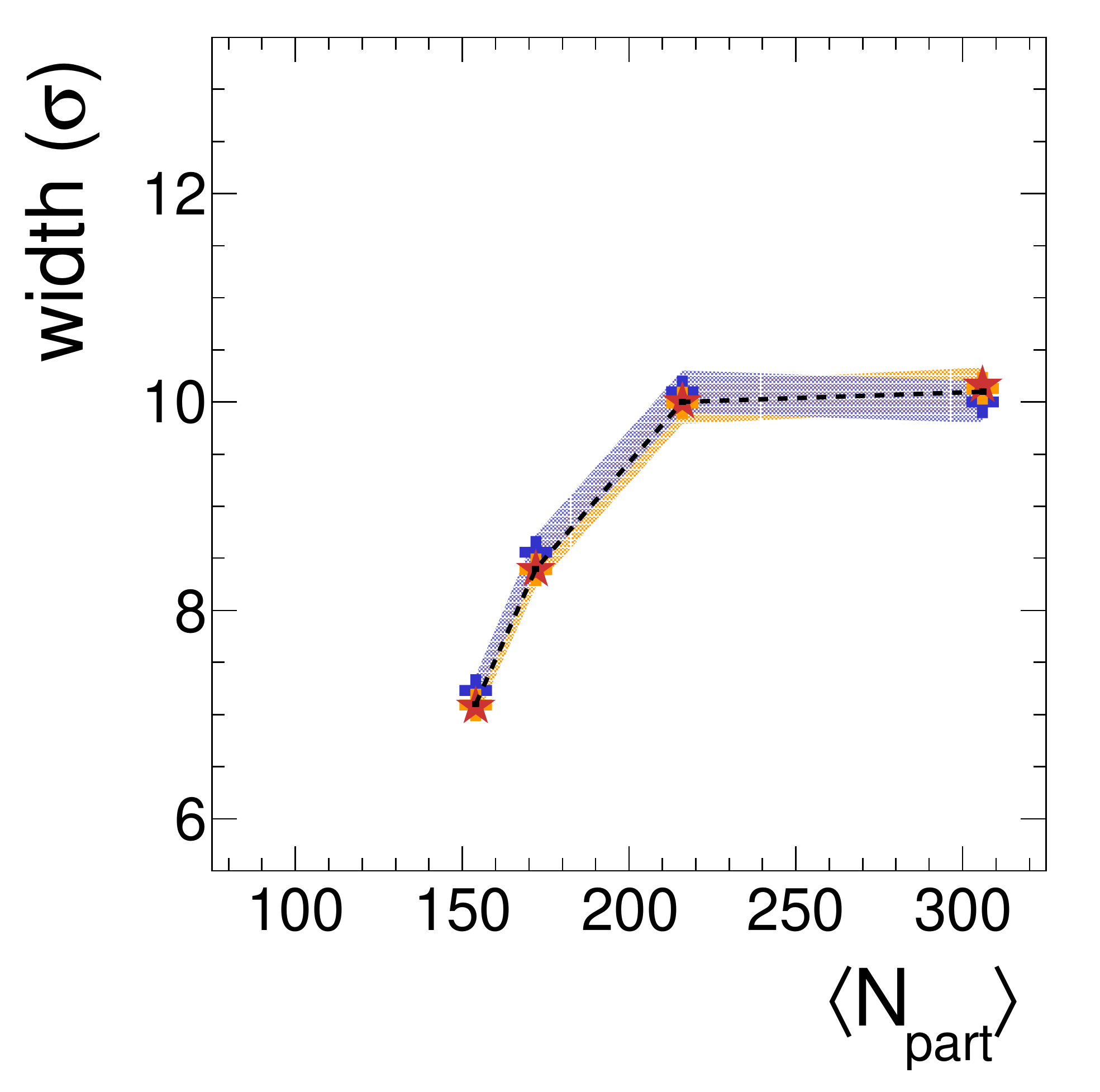} 
     }
     \resizebox{0.8\linewidth}{!} {
       \includegraphics{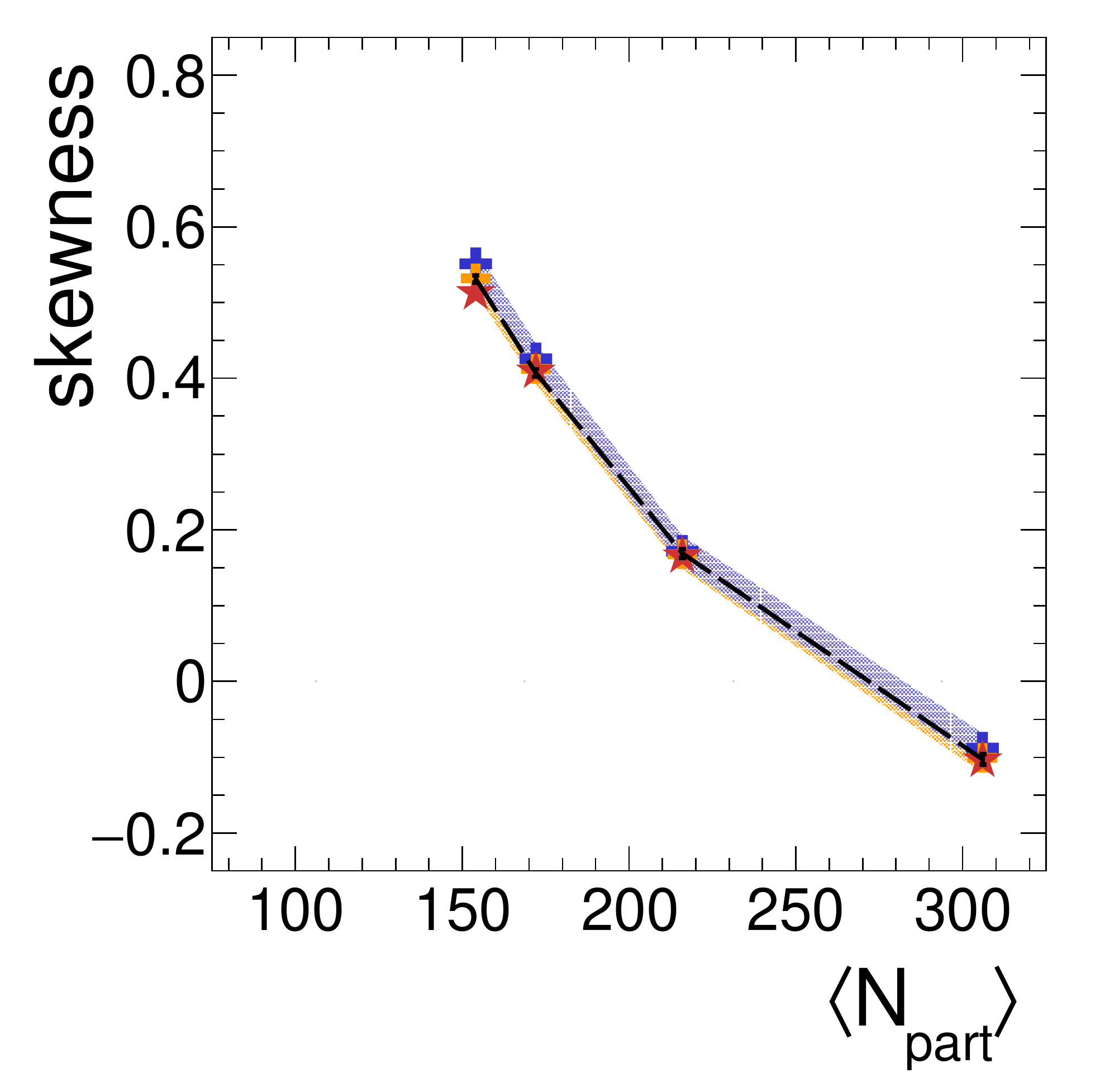} 
       \includegraphics{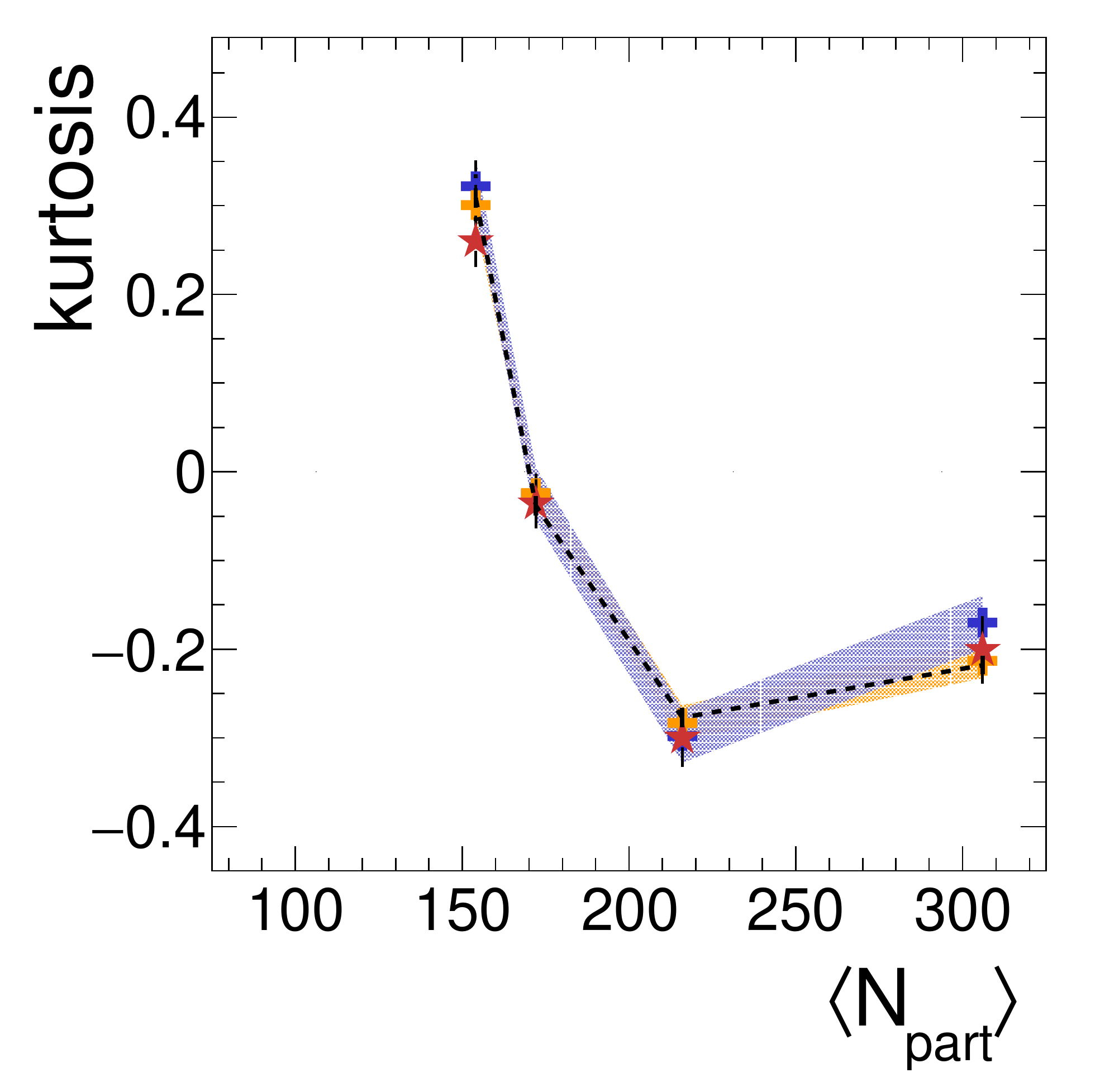} 
     }
  \end{center}
  \vspace*{-0.2cm}
  \caption[] {(Color online) Same as Fig.~\ref{fig:effiqmddy02} but for protons within a rapidity bin of $y \in y_0 \pm 0.5$.
              Note that only the most promising techniques are shown here (see text).
             }
  \label{fig:effiqmddy05}
\end{figure*}

In an extensive Monte Carlo investigation, we have validated and compared the various presented efficiency correction schemes.
As stated above, we have implemented the full HADES setup in GEANT3 and have run high-statistics simulations with as input
$10^8$ IQMD + MST clusterized Au+Au events in the impact parameter range $b = 0 - 10$~fm.  This range covered roughly the centralities
accepted by the PT3 trigger,\footnote{Already in the model calculation, a given impact parameter leads to a distribution
of participant nucleons and further smearing in the actual centrality observable.} namely the 0 -- 43\% most central events
(see \cite{Adamczewski2018} for details).  With reconstructed and identified proton tracks we histogrammed the multiplicity
of detected protons for various centrality and phase-space bins.  Applying efficiency corrections to these distributions,
we obtained the corresponding corrected proton multiplicity distributions as well as their corrected moments.  Finally, making
use of the event generator information, we also have the truth, i.e.\ the {\em a priori} proton distributions.  Results obtained
with the different correction procedures are compared in Fig.~\ref{fig:effiqmddy02} which displays the respective efficiency-corrected mean,
width ($\sigma$), skewness ($\gamma_1$), and kurtosis ($\gamma_2$) for protons emitted into the $y \in y_0 \pm 0.2$ and
$0.4 \leq p_t \leq 1.6$~\gevc\ phase space, as a function of the mean number of participant nucleons
$\langle N_{\textrm{part}} \rangle$.\footnote{$N_{\textrm{part}}$ is defined as the number of nucleons in the overlap volume of the
two colliding nuclei.}  Shown are the true IQMD moments (dashed lines) and results from various correction schemes:
constant efficiency, occupancy-dependent efficiency, unfolding, moment expansion, and, for comparison, also the ``occupancy'' model
(see Appendix~\ref{Sec:Occupancy}) directly adjusted to the IQMD truth.  It is clearly visible that by using a constant efficiency
(green full circles), i.e.\ an efficiency independent of track density, the correct moments are not retrieved.  The occupancy
model (pink triangles), while giving at least a fair description for the most peripheral centralities, is overall not satisfying.
On the other hand, the three correction methods discussed in detail above -- either applying an event-by-event correction or
unfolding\footnote{SVD and Tikhonov regularized unfolding give comparable results within statistical errors.} the measured proton
distribution or using a moment expansion -- all succeed in producing a result that agrees with the truth within given error bars.
Besides the statistical errors due to the finite event samples simulated, also systematic uncertainties occur caused, for example,
by different choices of the fit function used to model the effect of occupancy on the efficiency or of the optimal regularization applied
in the unfolding;  these systematic errors are indicated in Fig.~\ref{fig:effiqmddy02} as shaded bands ($\pm 1$\% for the mean,
$\pm2$\% for the width, $\pm 0.02$ for $\gamma_1$, and $\pm 0.03$ for $\gamma_2$).  Figure \ref{fig:effiqmddy05} shows the same
investigation done for protons emitted into a larger phase-space bin, namely $y \in y_0 \pm 0.5$.  Again, the agreement with IQMD truth
of the three favored methods is very satisfactory.  The constant efficiency correction and the occupancy model fail, however, and are
therefore omitted from the figure.  From these simulation studies we conclude that small point-to-point deviations between methods
do exist but no evident systematic trend is apparent and no clear preference for either of the three correction schemes emerges.
However, applying the efficiency corrections to our Au+Au data, the occupancy-dependent scheme turned out to become our favorite:
this was motivated, first, by its ease of implementation within the event analysis loop and, second, by the fact that both unfolding
and moment expansion are more sensitive to the particular choice of event generator used to produce the response matrices.
In the end, we treated differences between the correction methods as a contribution to our total systematic error
(see Sec.~\ref{Sec:ErrorBars}). 

\section{Volume corrections}
\label{Sec:Vol}

In heavy-ion collision experiments, the centrality determination can be based on various observable quantities, like the number
of hits or tracks in the detector, the total energy measured or the ratio of transverse to longitudinal energies, or the sum
of charges at forward angles.  The only requirement for any such observable to serve as proxy for centrality is that it
be a monotonic function of the impact parameter $b$, which itself is not directly measurable.  In addition, as observed
quantities are obtained with finite resolution only, any cut meant to restrict centrality to a particular value,
can do so only with a limited selectivity.  This means that in all cases a finite range of impact parameters will be selected
resulting in a distribution of the reaction volume and of the corresponding number of participant nucleons $N_{\textrm{part}}$.
As particle yields typically scale with some power of $N_{\textrm{part}}$, their number distributions will be affected as well 
and will therefore depend on the volume (or $N_{\textrm{part}}$) fluctuations of a given centrality cut.  This effect has been
recognized and discussed in \cite{Luo2013} where also centrality bin width corrections (CBWC) have been proposed as
a possible remedy.  The idea of CBWC is to compute yield-weighted averages of distribution moments over a number of
narrow subdivisions of the given wider centrality selection.  This way the larger statistics of a wide selection
could be benefitted while palliating the noxious effects of its increased volume fluctuations.  This procedure is
indeed useful, but only when the centrality resolution of the observable is narrower than the width of the selection cut.
At low beam energies, where the hit and particle multiplicities tend to be small, the achievable centrality resolution
is often quite limited such that the CBWC method fails.

A more formal study of the effect of volume fluctuations on the particle number cumulants has been done in \cite{Skokov2013}
and, more recently, in \cite{Rustamov2017a} where also a simulation study of the situation at the ALICE and STAR experiments
is presented.  In both publications, the authors start from the assumption that particle production scales with the
reaction volume $V$ \cite{Skokov2013}, that is the number of wounded nucleons $N_w$ \cite{Rustamov2017a}, such that all
particle number cumulants $K_n$ are proportional to $V$ (or $N_w$);  this behavior corresponds to independent particle production.
Introducing reduced particle number cumulants $\kappa_n = K_n/V$ and characterizing the volume fluctuations by
volume cumulants $V_l$, they arrived at a general expression for the volume affected reduced cumulants 

\begin{equation}
 \tilde{\kappa}_n = \sum_{l=1}^{n} \: v_l \: B_{n,l}(\kappa_1,\kappa_2,...,\kappa_{n-l+1}) \:,
 \label{eq:Skokov}
\end{equation}

\noindent
where $v_l = V_l/V$ are reduced volume cumulants\footnote{With $V_1 = V$ and $v_1 = 1$.} and $B_{n,l}$ are Bell
polynomials \cite{Arndt2011}.  Then, if all volume cumulants up to order $n$ are known, the $\kappa_n$ can be retrieved
from the $\tilde{\kappa}_n$ by solving the system of Eqs.~\eqref{eq:Skokov} recursively.  Up to fourth order this gives

\begin{equation}
  \begin{split}
 \kappa_1 = & \; \tilde{\kappa}_1 \,,\\
 \kappa_2 = & \; \tilde{\kappa}_2 - \kappa_1^2 v_2\,, \\
 \kappa_3 = & \; \tilde{\kappa}_3 - 3 \kappa_1 \kappa_2 v_2 - \kappa_1^3 v_3\,, \\
 \kappa_4 = & \; \tilde{\kappa}_4 - (4 \kappa_1 \kappa_3 + 3 \kappa_2^2) v_2 - 6 \kappa_1^2 \kappa_2 v_3 - \kappa_1^4 v_4\,.
  \end{split}
 \label{eq:SkokovLO}
\end{equation}

\begin{figure*}[!hbt] 
  \begin{center}
     \resizebox{0.41\linewidth}{!} {
       \includegraphics{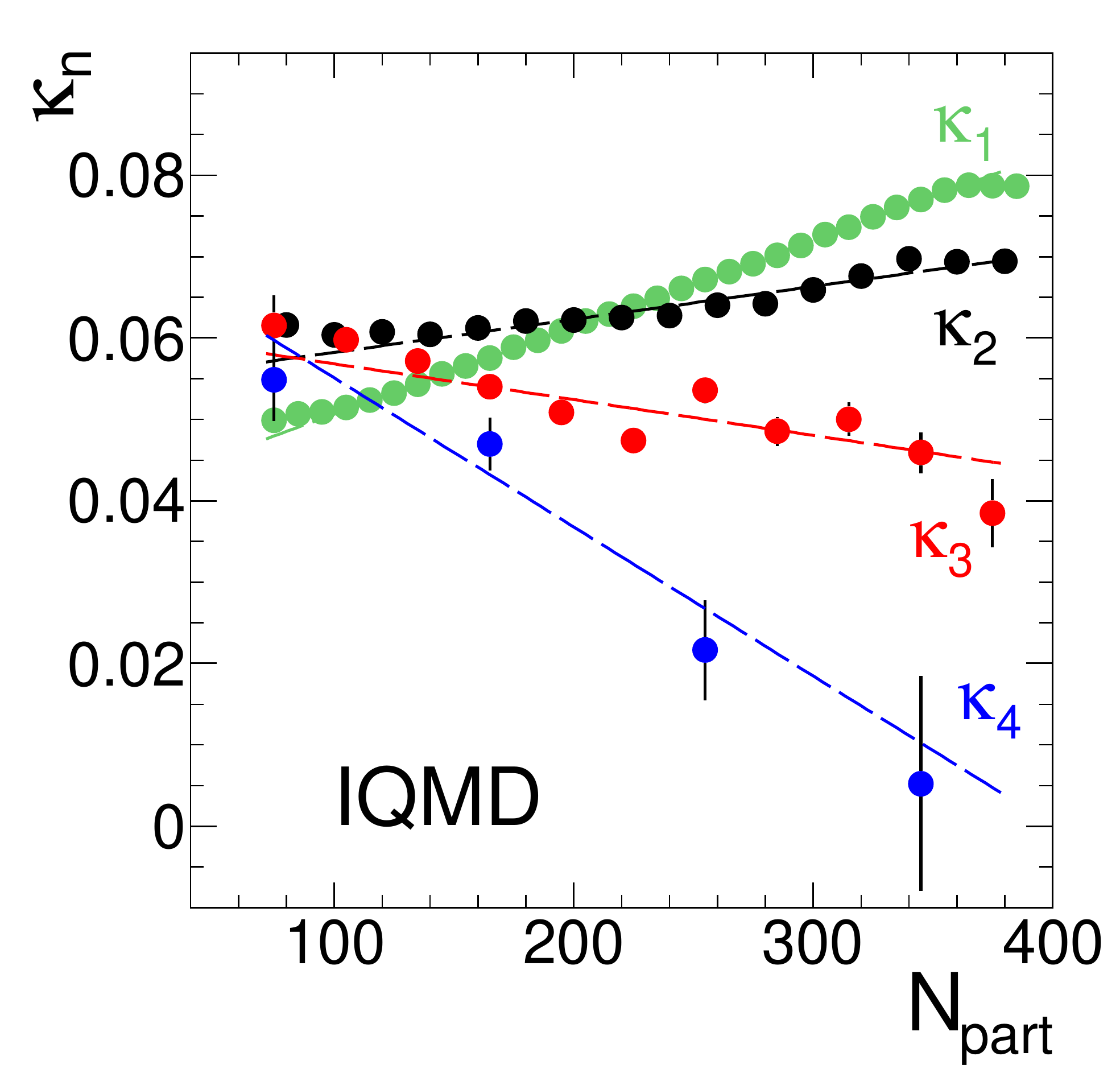}
     }
     \resizebox{0.41\linewidth}{!} {
       \includegraphics{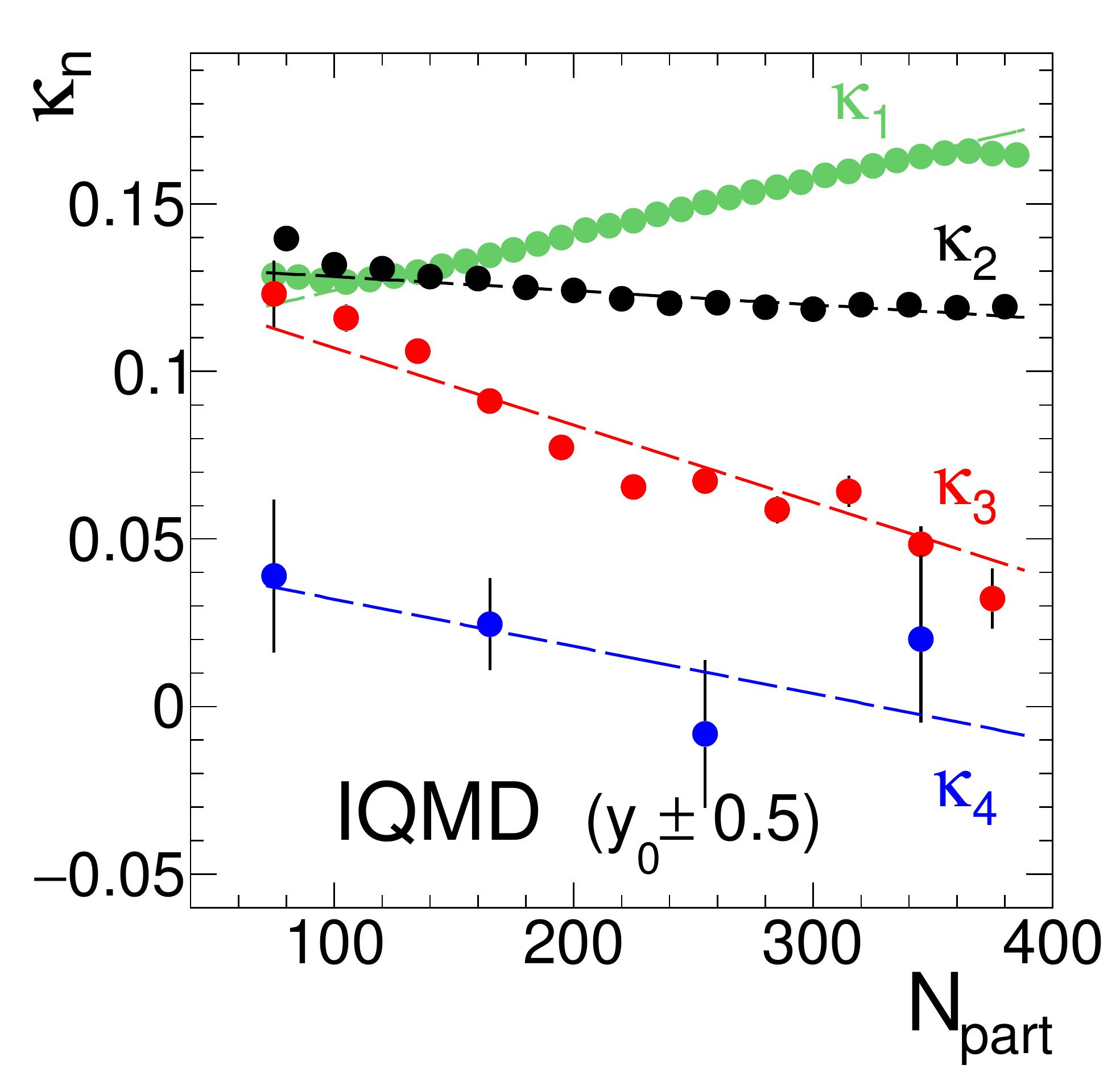}
     }
     \resizebox{0.41\linewidth}{!} {
       \includegraphics{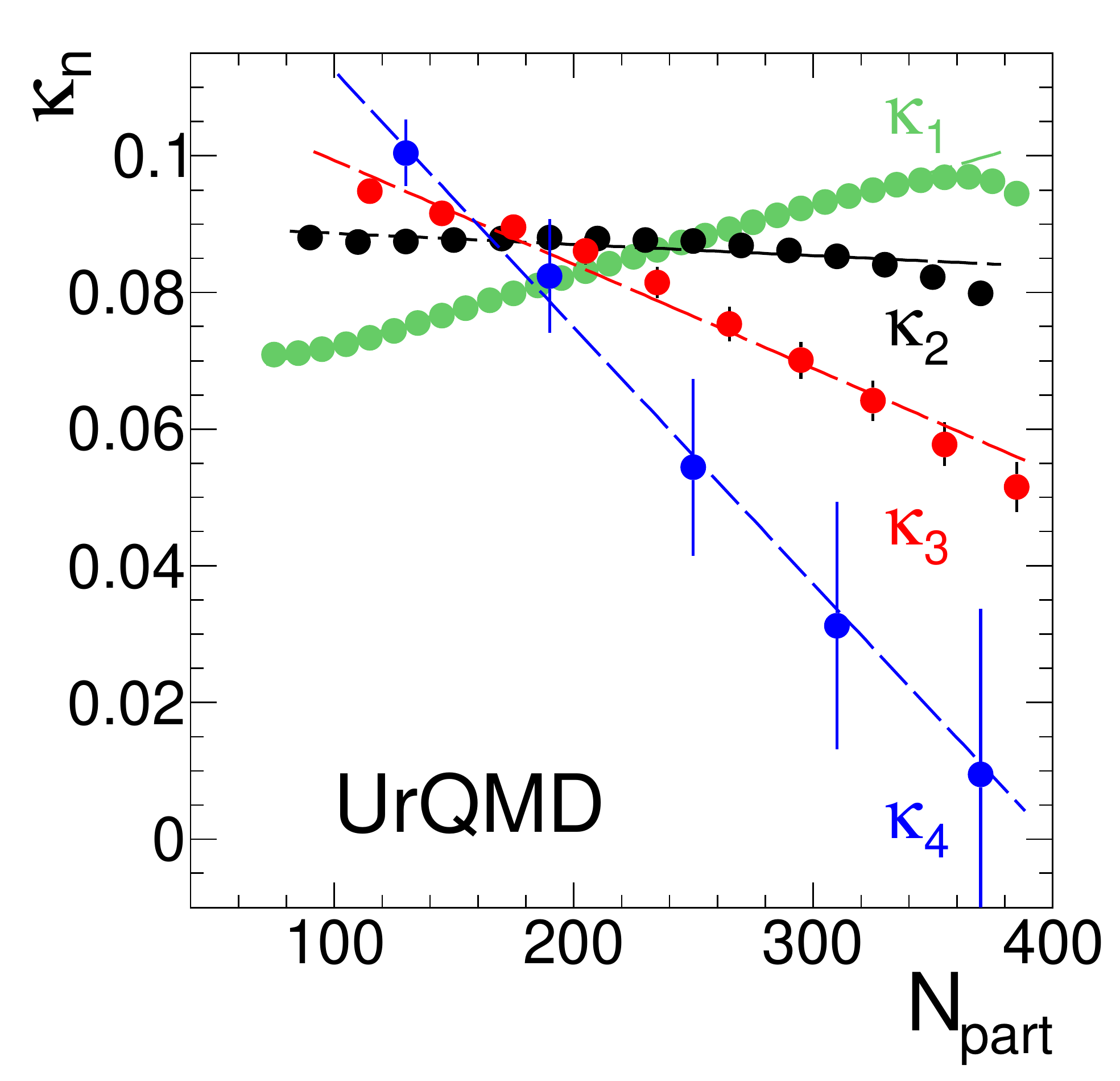}
     }
     \resizebox{0.41\linewidth}{!} {
       \includegraphics{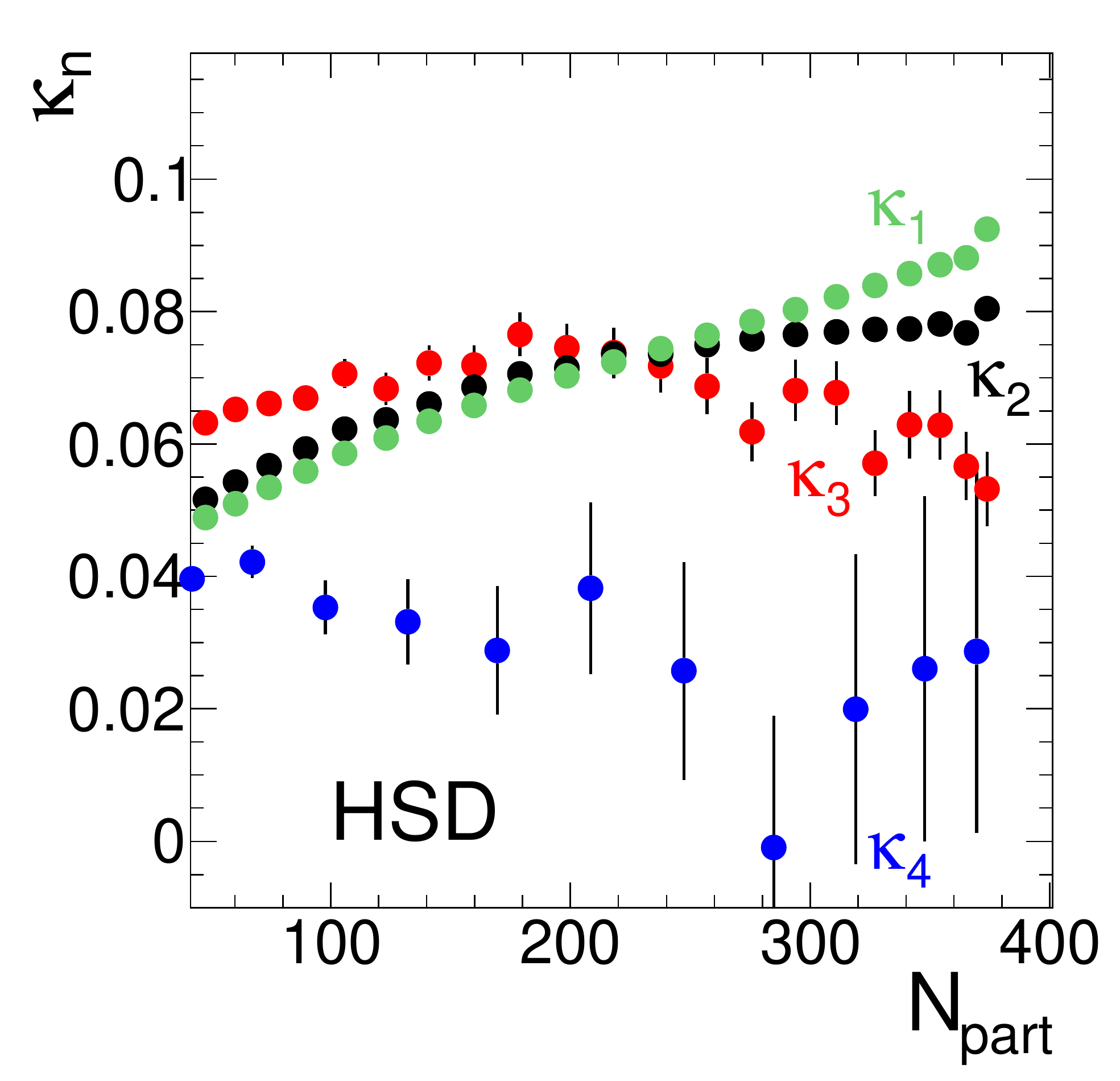}
     }
  \end{center}
  \vspace*{-0.2cm}
  \caption[] {(Color online) Reduced proton cumulants $\kappa_n$ as a function of $N_{\textrm{part}}$ in four
              different transport calculations done with the IQMD, UrQMD, and HSD models respectively.  The phase space
              chosen here was $y \in y_0 \pm 0.2$ (or $y \in y_0 \pm 0.5$) and $0.4 \leq p_t \leq 1.6$~\gevc\ for IQMD,
              respectively $y \in y_0 \pm 0.2$ and $0.4 \leq p_t \leq 1.6$~\gevc\ for UrQMD and HSD.
              Dashed lines are linear fits of function $\kappa(N_{\textrm{part}}) = \kappa_{0} + \kappa' \, N_{\textrm{part}}$ to 
              some of the presented reduced cumulants.
             }
  \label{fig:transport}
\end{figure*}

With these equations the observed cumulants can be corrected for the volume contributions resulting from the spread of the
applied centrality selection.  The corresponding volume distribution must be known of course, either from a model
calculation or, with the help of a procedure to be defined, from the data itself.  As proposed in \cite{Rustamov2017a},
for a more practical measure of volume one could use the number of wounded nucleons $N_w$ or, at low bombarding energy, rather
the number of participating nucleons $N_{\textrm{part}}$.  The open question, however, is to what extent the assumed scaling with volume
of the cumulants, which is at the basis of Eq~\eqref{eq:Skokov}, can be considered valid.  The authors of Ref.~\cite{Skokov2013}
argued that, while this is indeed a reasonable approximation in ultrarelativistic heavy-ion collisions probing the low-$\mu_B$,
high-$T$ region of the phase diagram, caution should be applied in the high-$\mu_B$ regime relevant for CEP searches.  To find some guidance,
we have investigated the respective behavior of transport codes, namely IQMD, UrQMD, and also HSD (version 711n) \cite{Cassing1999} run for the
\sqrtsNN\ = 2.4~\gev\ Au+Au reaction, i.e.\ we have analyzed their true proton number cumulants as a function of $N_{\textrm{part}}$ within
various phase-space bins.  Figure~\ref{fig:transport} shows examples of such calculations together with fits of the linear function
$\kappa(N_{part}) = \kappa_0 + \kappa' \, N_{part}$ to some of the simulation points.\footnote{The $N_{\textrm{part}}$ bin width used in these calculations
(10 for IQMD and UrQMD, 20 for HSD) was fine enough to keep volume fluctuation effects small.}
It is obvious from these plots that all three transport models
strongly violate the assumption of constancy of $\kappa_n$ versus $N_{\textrm{part}}$ by revealing a linear and, in some cases (e.g.\ for HSD),
even a quadratic dependence on $N_{\textrm{part}}$.  A systematic study also reveals a high degree of variability with respect to the
particular bin chosen in $y - p_t$ phase space.  The models differ, however, in the details of their rendering of the complex
dependency of $\kappa_n$ on centrality.

Evidently, the assumption of constancy of $\kappa_n$ has to be abandoned and at least a linear term or, better, linear
plus quadratic terms have to be taken into account when calculating the contributions of volume fluctuations.
Along the lines presented in \cite{Skokov2013}, we have extended the derivation of Eq.~\eqref{eq:Skokov} by replacing
the constant ansatz $\kappa_n(V) = \kappa_n$ with a 2\textsuperscript{nd}-order Taylor expansion of $\kappa_n(V)$ around
the mean $\left<V\right>$ of the volume distribution

\begin{equation}
 \kappa_n(V) = \kappa_n + \kappa_n' (V-\left<V\right>) + \kappa_n'' (V-\left<V\right>)^2 \,,
\label{eq:Taylor}
\end{equation}

\noindent
where the $\kappa_n$ are the leading constant terms, $\kappa_n'$ are slopes, and $\kappa_n''$ are curvatures,
all of which can depend on rapidity and transverse momentum.  With this new ansatz, a more complete set of volume
terms contributing to the reduced cumulants $\tilde{\kappa}_n$ has been derived.  Using slopes only (i.e.\ $\kappa_n''=0$),
the following relations were found for $n$ = 1, 2, 3, and 4:

\begin{equation} \label{eq:NL}
\begin{split}
  \tilde{\kappa}_1 &= \kappa_1 + v_2 \kappa_1' \,, \\
  \tilde{\kappa}_2 &= \kappa_2 + \kappa_1^2 v_2 + \kappa_2' v_2 + 2 \kappa_1 \kappa_1' V_2 + 2 \kappa_1 \kappa_1' v_3 \\ 
                   &+ 2 \kappa_1'^2 v_2 V_2 + \kappa_1'^2 V_1 V_2 + 2 \kappa_1'^2 V_3 + \kappa_1'^2 v_4 \,, \\
  \tilde{\kappa}_3 &= \kappa_3 + \kappa_1^3 v_3 + 3 \kappa_1 \kappa_2 v_2
                    + 3 (\kappa_1 \kappa_2' + \kappa_1'  \kappa_2) v_3  \\ 
                   &+ 6 \kappa_1' (\kappa_1^2 + \kappa_2') v_2 V_2 + 3 \kappa_1' (\kappa_1^2 + 2 \kappa_2') V_3 \\ 
                   &+ 3 \kappa_1' (\kappa_1^2 + \kappa_2') v_4 + 12 \kappa_1 \kappa_1'^2 V_2^2 + 3 \kappa_1 \kappa_1'^2 V_1 V_3 \\
                   &+ 24 \kappa_1 \kappa_1'^2 v_2 V_3 + 6 \kappa_1 \kappa_1'^2 V_4 + 3 \kappa_1 \kappa_1'^2 v_5 + \kappa_3' v_2\\
                   &+ 3 (\kappa_1 \kappa_2' + \kappa_1' \kappa_2) V_2 + 8 \kappa_1'^3 v_2 V_2^2 + 6 \kappa_1'^3 V_1 V_2^2 \\
                   &+ 10 \kappa_1'^3 v_3 V_3 + \kappa_1'^3 V_1^2 V_3 + 24 V_2 V_3 \kappa_1'^3 + 3 \kappa_1'^3 V_1 V_4 \\
                   &+ 12 \kappa_1'^3 v_2 V_4 + 3 \kappa_1'^3 V_5 + \kappa_1'^3 v_6 + 3 \kappa_1' \kappa_2' V_1 V_2\,, \\
  \tilde{\kappa}_4 &= \kappa_4 + \kappa_1^4 v_4 + 6 \kappa_1^2 \kappa_2 v_3 + (4 \kappa_1 \kappa_3 + 3 \kappa_2^2) v_2 \ldots \,.
\end{split}
\end{equation}

\noindent
Because of its length, the cumulant of order $n = 4$ is fully listed in Appendix~\ref{Sec:NL}, Eq.~(B4). 
Compared to Eq.~\eqref{eq:Skokov}, many additional terms that all depend on the slopes appear, including terms
involving volume cumulants up to order $2n$, that is for $\tilde{\kappa}_4$ up to order eight.  In a somewhat colloquial manner,
we designate the slope-related corrections by NLO, i.e.\ next to leading order, and the curvature affected terms (see below) by N2LO,
i.e.\ next to next to leading order.  For the higher orders, the number of terms quickly rises and the formulas become
cumbersome to derive by hand; we have instead used a symbolic computation program\footnote{Wolfram Mathematica.}
to generate them as well as the corresponding C code needed for their numerical evaluation.  Evidently, the relations derived with
all slopes and curvatures included are even lengthier (see Table~\ref{tab:terms}) and they require volume cumulants up to order $3n$.
The C code can be provided on request and, for illustration only, we list here the first two N2LO cumulants:

\begin{equation} \label{eq:N2L}
\begin{split}
  \tilde{\kappa}_1 &= \kappa_1 + v_2 \kappa_1' + (V_2 + v_3) \kappa_1'' \,, \\
  \tilde{\kappa}_2 &= \kappa_2 + \kappa_1^{2} v_2 + \kappa_2' v_2 + 2 \kappa_1 \kappa_1' V_2 + 2 \kappa_1 \kappa_1' v_3 \\
                   &+ 2 \kappa_1'^2 v_2 V_2 + \kappa_1'^2 V_1 V_2 + 2 \kappa_1'^2 V_3 + \kappa_1'^2 v_4 \\
                   &+ 6 \kappa_1 \kappa_1'' v_2 V_2 + 2 \kappa_1 \kappa_1'' (V_3 + v_4) + \kappa_2'' (V_2 + v_3) \\
                   &+ 10 \kappa_1' \kappa_1'' V_2^2 + 18 \kappa_1' \kappa_1'' v_3 V_2 + 2 \kappa_1' \kappa_1'' V_1 V_3 \\
                   &+ 4 \kappa_1' \kappa_1'' V_4 + 2 \kappa_1' \kappa_1'' v_5 + 15 \kappa_1''^2 v_2 V_2^2 \\
                   &+ 2 \kappa_1''^2 V_1 V_2^2 + 18 \kappa_1''^2 V_2 V_3 + 15 \kappa_1''^2 v_4 V_2 \\
                   &+ 9 \kappa_1''^2 v_3 V_3 + \kappa_1''^2 V_1 V_4 + 2 \kappa_1''^2 V_5 + \kappa_1''^2 v_6 \,.
\end{split}
\end{equation}

\begin{figure}[!hbt]
  \begin{center}
     \resizebox{1.0\linewidth}{!} {
       \includegraphics{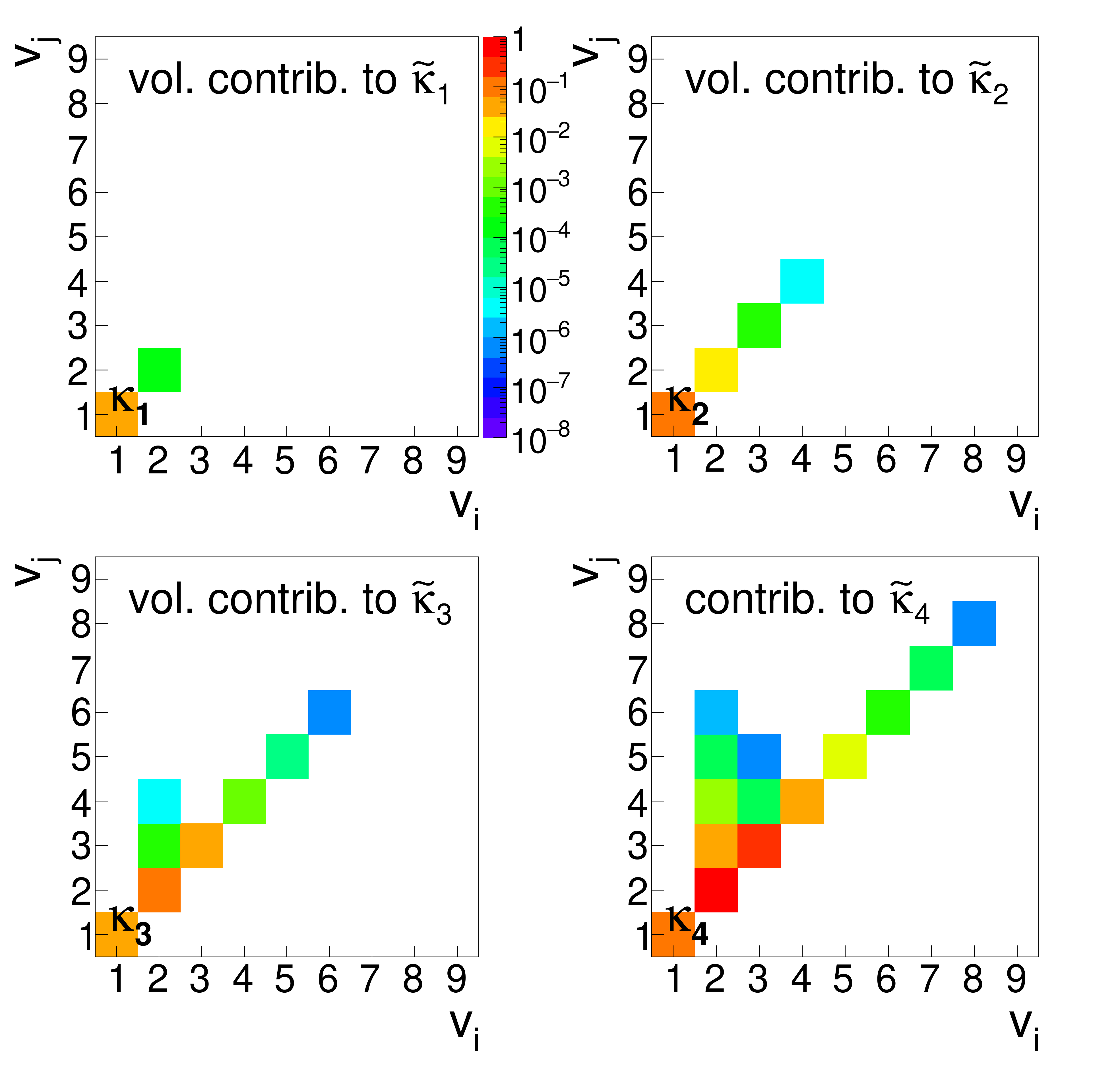} 
     }
  \end{center}
  \vspace*{-0.2cm}
  \caption[] {(Color online) Illustration of the magnitude of NLO + N2LO volume fluctuation terms contributing to
               the observed reduced proton number cumulants $\tilde{\kappa}_n$ with $n = 1, 2, 3,$ and 4.
               The magnitude of the terms decreases quickly with increasing order of $v_i$; low-order diagonal terms
               are dominant and off-diagonal terms, depending on a product $v_i \cdot v_j$, fall off even faster.
               The corrections shown were determined for a selection of the 0 -- 10\% most central
               Au+Au collisions measured in HADES (see Sec.~\ref{Sec:Npart} below).\\  
             }
  \label{fig:N2Lcontrib}
\end{figure}

\noindent
Despite the large number of contributing terms, one has to keep in mind that these expressions are just polynomials
which can be easily evaluated for given values of $\kappa_n$, $\kappa_n'$, $\kappa_n''$, and $v_l$.  And they can
be adjusted to simulated or real data to extract the cumulants of interest.  Doing such fits to proton cumulants obtained
with transport models, we find that typically $\kappa_n \gg \kappa_n' \gg \kappa_n''$ and consequently most of the higher-order
volume terms turn out to be very small.  This is also confirmed by fits of Eq.~\eqref{eq:N2L} to our data, as exemplified
in Fig.~\ref{fig:N2Lcontrib} which shows the rapid drop over nearly seven orders of magnitude of the contributing volume terms with
increasing order of $v_l$.  We conclude that in practice it is sufficient to consider terms with $v_l$ up to order~$l=5$ or~6 at most,
i.e.\ the full gamut of the $v_l$ up to $l = 8$ (NLO) or even $l = 12$ (N2LO) will most likely never be required.

\begin{table}
 \caption{Number of volume fluctuation terms contributing to the observed reduced cumulants $\tilde{\kappa}_n$.
          Column L (leading terms) corresponds to Eq.~\eqref{eq:Skokov}, L+NL (including slopes of
          the $N_{\textrm{part}}$ dependence) corresponds to Eq.~\eqref{eq:NL}, and L+NL+N2L (with slopes and curvatures)
          corresponds to Eq.~\eqref{eq:N2L}.
         } 
 \label{tab:terms}
 \vspace*{3mm}
 \setlength{\extrarowheight}{2pt}  
 \begin{tabular}{c c c c}
  \hline \hline
  $\tilde{\kappa}$~~ & ~~L~~ & ~L+NL~ & ~L+NL+N2L \\
  \hline
  $\tilde{\kappa}_1$ & 0 & 1  &   3 \\
  $\tilde{\kappa}_2$ & 1 & 8  &  26 \\
  $\tilde{\kappa}_3$ & 2 & 28 & 128 \\
  $\tilde{\kappa}_4$ & 4 & 84 & 527 \\
  \hline \hline
 \end{tabular}
\end{table}

\section{Centrality selection and $\mathbf{N_{\textrm{part}}}$ distributions}
\label{Sec:Npart}

\begin{figure}[!hbt]
  \begin{center}
     \resizebox{0.8\linewidth}{!} {
       \includegraphics{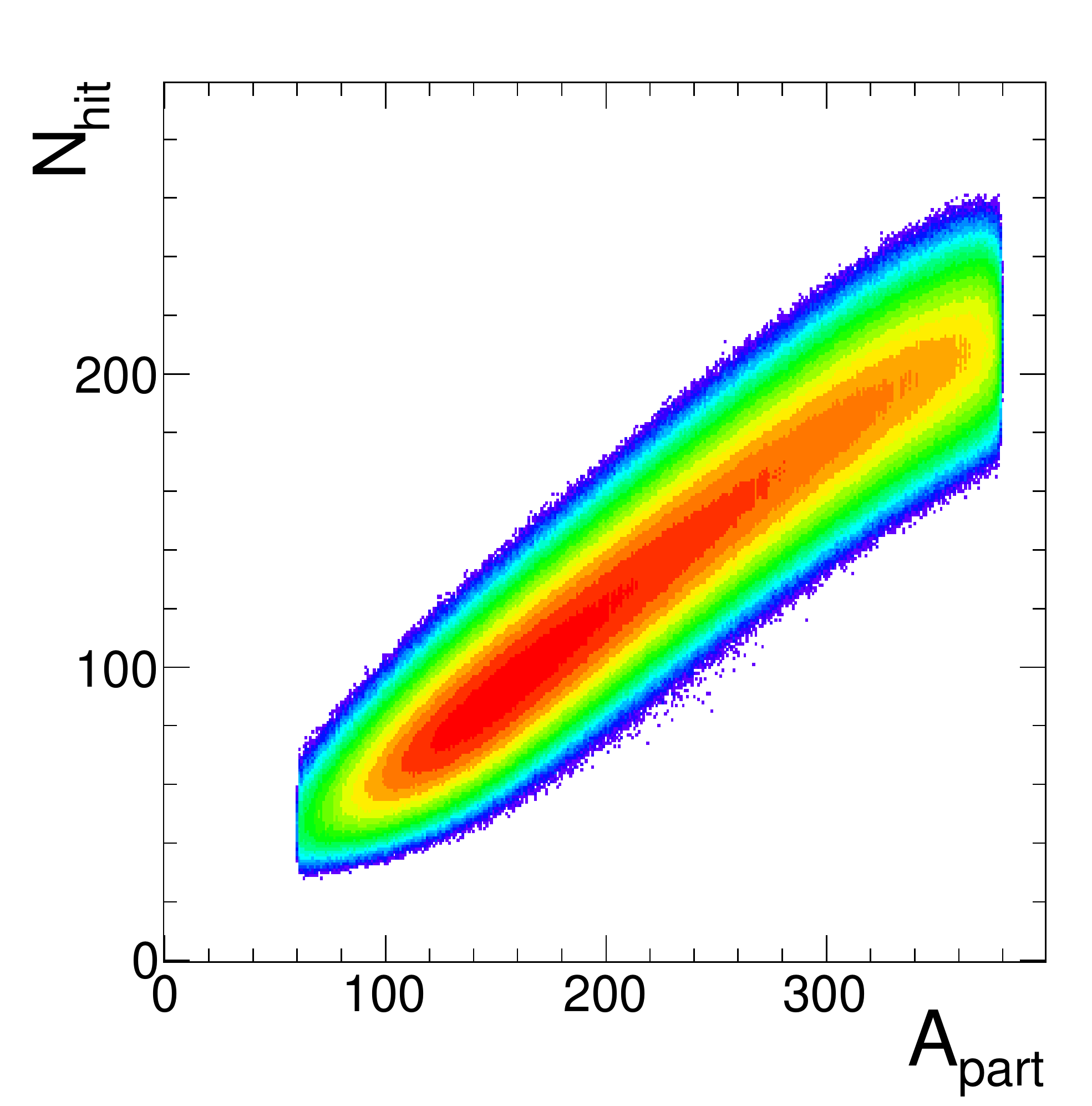} 
     }
  \end{center}
  \vspace*{-0.2cm}
  \caption[] {(Color online) Correlation between the total number of hits $N_{\textrm{hit}}$ in
              the HADES time-of-flight detectors and the number of participants $N_{\textrm{part}}$
              in 1.23\agev\ Au+Au events simulated with the IQMD transport code for impact parameters in the range of 0 -- 10~fm.} 
  \label{fig:NhitNpart}
\end{figure}

In order to apply volume corrections to particle-number cumulants measured within a given centrality selection, it is mandatory
to know the corresponding $N_{\textrm{part}}$ distribution or at least its cumulants up to sufficiently high order.
In simulations, the impact parameter $b$ is known event by event and the corresponding number of participants can be determined
from the geometric overlap or, better, with the help of a more involved Glauber model.  In the data, however, $b$ and $N_{\textrm{part}}$
are not directly observable; we must find a proxy for $N_{\textrm{part}}$, e.g. the number of observed hits $N_{\textrm{hit}}$ or of
reconstructed tracks $N_{\textrm{trk}}$, to quantify the volume effects.  The very strong and nearly linear correlation between $N_{\textrm{hit}}$
and the underlying $N_{\textrm{part}}$ is illustrated in Fig.~\ref{fig:NhitNpart} which shows a simulation done with the IQMD transport model.
Hence, a first approach to arrive at the volume cumulants $V_l$, required by Eqs.~\eqref{eq:Skokov}, \eqref{eq:NL}, and \eqref{eq:N2L},
could be to just use the cumulants of the observed $N_{\textrm{hit}}$ distribution.  Yet, the two quantities do not trivially relate
to each other: first, particle production per participant nucleon is a random process;  second, finite detector acceptance and efficiency make
the observation process random too.  Consequently, any given $N_{\textrm{part}}$ results in a spread of the observed number of hits, where the relation
$N_{\textrm{hit}}(N_{\textrm{part}})$ can be approximated by a negative binomial distribution \cite{Adamczewski2018}.  Because of this spread, simply
using $N_{\textrm{hit}}$ as a direct proxy for $N_{\textrm{part}}$ will lead to an overestimation of $V_2$ and will generally result in wrong higher-order
cumulants.  Of course, the same arguments also speak against using $N_{\textrm{trk}}$ as proxy for $N_{\textrm{part}}$.

\begin{figure}[!hbt]
  \begin{center}
     \resizebox{0.8\linewidth}{!} {
       \includegraphics{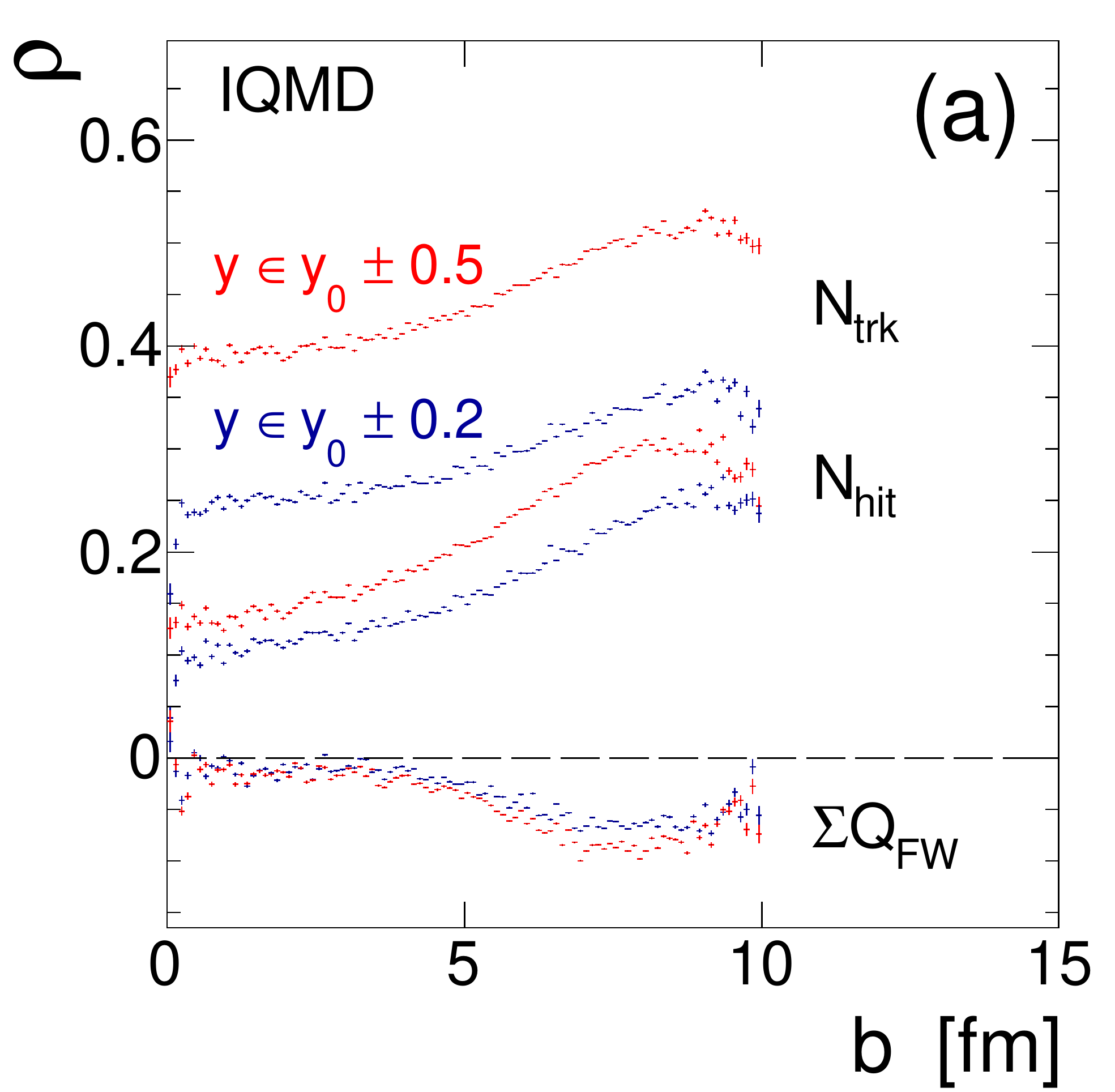}
       }
     \resizebox{0.8\linewidth}{!} {
       \includegraphics{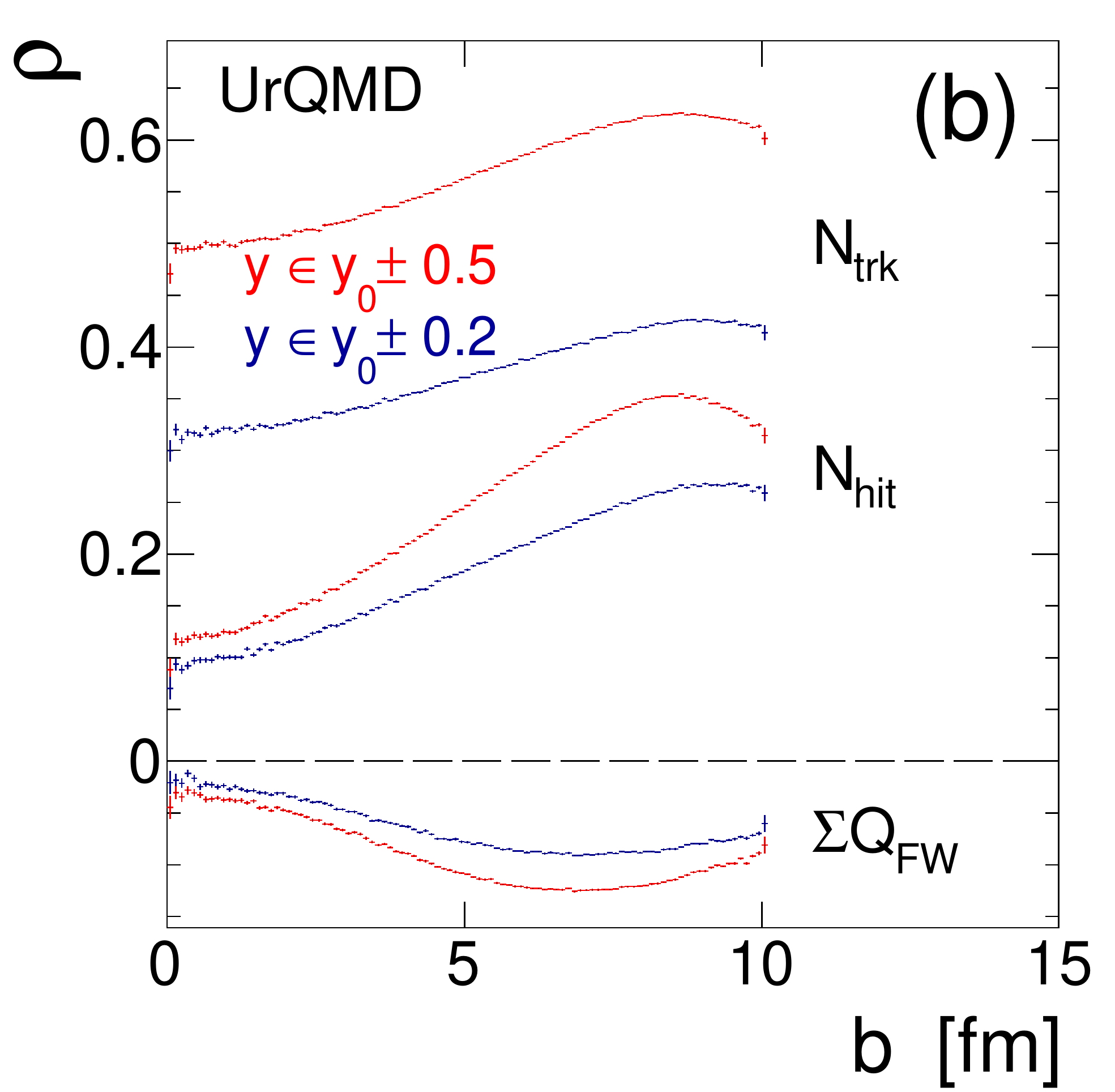}
       }
  \end{center}
  \vspace*{-0.2cm}
  \caption[] {(Color online) Transport simulations of Au+Au events with the IQMD (a) and UrQMD (b) models
              showing correlations between the number of identified protons $N_{\textrm{prot}}$ and the centrality defining
              observable $N_{\textrm{cen}} = N_{\textrm{hit}}$, $N_{\textrm{trk}}$, and $\Sigma Q_{\textrm{FW}}$, respectively.
              Pearson's linear correlation coefficient $\rho(N_{\textrm{prot}}, N_{\textrm{cen}})$ is displayed for two rapidity ranges,
              $y \in y_o \pm 0.2$ (dark blue) and $y \in y_o \pm 0.5$ (red), as a function of the impact parameter $b$. }
  \label{fig:rho}
\end{figure}

There is in fact yet another, more subtle effect that needs to be taken into account in the determination of the $V_l$,
namely the correlations between $N_{\text{prot}}$ and the centrality measure used, $N_{\textrm{hit}}$, $N_{\textrm{trk}}$, or $\Sigma Q_{\textrm{FW}}$.
A recent simulation study \cite{Sugiura2019} done with \sqrtsNN\ = 200~\gev\ UrQMD events concluded that the correlations between the proton number
and the centrality defining multiplicities do affect the volume fluctuation correction.  A qualitative argument for this
phenomenon is the following:  a centrality selection realized by applying cuts on the number of observed hits,
$N_{\textrm{min}} \leq N_{\textrm{hit}} \leq N_{\textrm{max}}$, constrains not only the mean number of hits $\langle N_{\textrm{hit}} \rangle$ but,
because $N_{\textrm{hit}}$ and $N_{\textrm{prot}}$ are strongly correlated, also the corresponding mean number of protons.  This cut will therefore
tend to curtail large excursions of $N_{\textrm{prot}}$ from its mean value $\langle N_{\textrm{prot}} \rangle$, leading to a reduction of its variance
and generally affecting the higher-order proton number cumulants in a non-trivial way.  At the low bombarding energies where HADES operates,
protons are the dominant particle species and they contribute most of the hits and tracks, causing correlations to be particularly pronounced.
This appears also from the transport model simulations (IQMD and UrQMD) displayed in Fig.~\ref{fig:rho}, where the linear correlation
coefficients $\rho$ between $N_{\textrm{prot}}$ and, respectively, $N_{\textrm{hit}}$, $N_{\textrm{trk}}$, and $\Sigma Q_{\textrm{FW}}$ are plotted
against impact parameter $b$.  Both models show strong positive correlations for $N_{\textrm{hit}}$ and $N_{\textrm{trk}}$, and negative,
but much weaker correlations for $\Sigma Q_{\textrm{FW}}$.  Ideally, one would want to incorporate correlations between the proton number
and the centrality-defining observable into a more comprehensive volume-fluctuation formalism expressed, if possible, as a function
of the experimentally accessible relevant correlation coefficient $\rho(N_{\textrm{prot}}, N_{\textrm{hit}})$, $\rho(N_{\text{prot}}, N_{\textrm{trk}})$,
or $\rho(N_{\textrm{prot}}, \Sigma Q_{\textrm{FW}})$.  Unfortunately,  such a complete model is not yet at hand, and  we have hence taken the
pragmatic approach to (1) use the centrality selector with lowest correlations and (2) modify the volume cumulants based on
the resulting $N_{\textrm{hit}}$ distributions such as to express the correlation-affected $N_{\textrm{part}}$ distributions. 

In the Au+Au data presented here, the observed correlation coefficient $\rho(N_{\textrm{prot}}, \Sigma Q_{\textrm{FW}})$ was found to be negative
as in the transport simulations but of somewhat larger magnitude, 0.15 -- 0.25.  However, these values also include a part caused by the
global volume fluctuations within the finite experimental centrality bins and the intrinsic correlations of the two observables might in fact be weaker.
All matters considered, the observable $\Sigma Q_{\textrm{FW}}$ displays the smallest correlations with $N_{\textrm{prot}}$ and we have hence used
it as centrality selector for the proton fluctuation analysis.  We developed an {\em ad hoc} scheme to handle in one swipe both, the blurring
of the $N_{\textrm{part}} \rightarrow N_{\textrm{hit}}$ mapping and the correlation-induced modifications of the volume cumulants $v_l$.  The core idea
is to introduce at every order ($l = 2, 3, \ldots, l_{\textrm{max}}$) a pair of modifiers, $f_l$ and $d_l$, and substitute $v_l$ in the cumulant
expressions, i.e. Eqs.~\eqref{eq:NL} and \eqref{eq:N2L}, with $v^{\textrm{mod}}_l = f_l \times v_l + d_l$; each volume cumulant is eventually
adjusted by applying an appropriate scaling factor $f_l$ as well as a modifying cumulant $d_l$.  This transformation yields modified
volume cumulants once the $f_l$ and $d_l$ are properly fixed.  For our analysis, we have determined these parameters in a multi-order fit
of Eq.~\eqref{eq:N2L} to the reconstructed proton cumulants $\tilde{\kappa}_n$ of a high-statistics sample of IQMD events run through the
HADES detector simulation and analysis pipeline, i.e.\ in a situation where the true $\kappa_n$, as well as their slopes $\kappa_n'$ and
curvatures $\kappa_n''$ were all fully known.  In this procedure, the reduced volume cumulants $v_l$ were taken from the $N_{\textrm{hit}}$ distribution,
with its abscissas rescaled by the factor $\langle N_{\textrm{part}} \rangle / \langle N_{\textrm{hit}} \rangle$, and using $N_{\textrm{part}}$
as a proxy for the source volume, i.e.\ $V \equiv N_{\textrm{part}}$.  Note that the $N_{\textrm{hit}} - N_{\textrm{prot}}$ correlations do not
preclude us from using the scaled $N_{\textrm{hit}}$ distribution as a proxy for the volume distribution because, in the determination
of the $v_l$, only event-averaged quantities enter.

\begin{figure}[!hbt]
  \begin{center}
     \resizebox{0.8\linewidth}{!} {
       \includegraphics{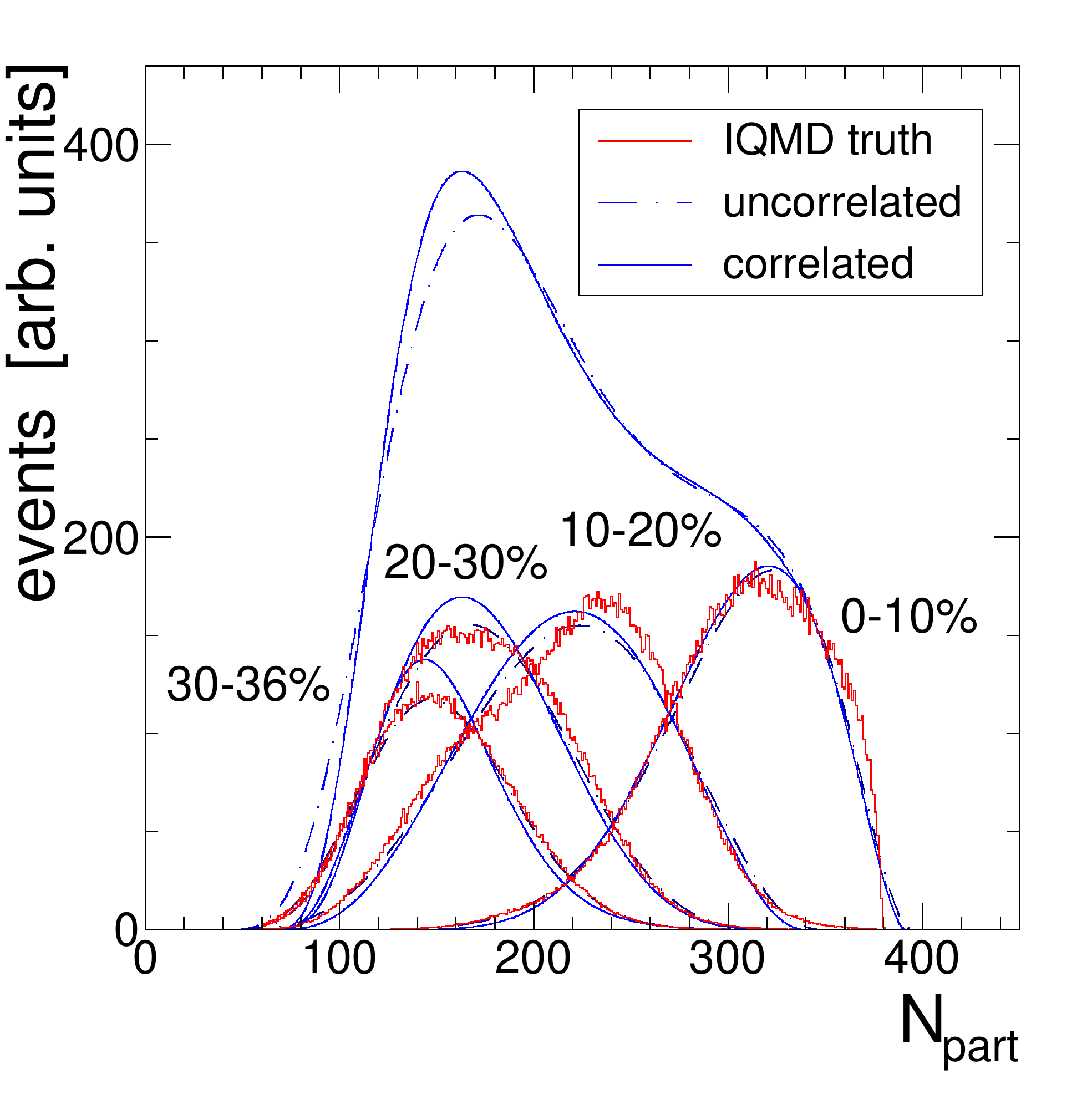} 
     }
  \end{center}
  \vspace*{-0.2cm}
  \caption[] {(Color online) $N_{\textrm{part}}$ distributions of simulated IQMD events
              shown for four centrality selections based on the $\Sigma Q_{FW}$ signal in the FWALL.
              IQMD truth (red histograms) is compared to reconstructions based on their first four volume cumulants, 
              displayed as 4-parameter beta functions, without correlation effects (dot-dashed curves) and
              with correlations (solid curves).  The sum curves correspond to the 0 -- 36\% most central events. 
             }
  \label{fig:iqmdNpart}
\end{figure}

With the adjusted set of parameters $f_l$ and $d_l$, the modified volume cumulants can be obtained and used to generate an approximation
(up to some order $l_{\textrm{max}}$) of the effective, i.e.\ correlation-affected $N_{\textrm{part}}$ distributions (see Appendix~\ref{Sec:Beta} for details).
Figure~\ref{fig:iqmdNpart} illustrates the procedure with simulated IQMD $N_{\textrm{part}}$ distributions corresponding to four different
centrality bins selected with cuts on the $\Sigma Q_{\textrm{FW}}$ observable.  The true distributions from the model (red histograms)
are compared with a reconstruction (dot-dashed curves) based on their first four reduced volume cumulants $v_l$ ($l = 1, 2, 3, 4$)
and plotted as 4-parameter beta distributions;  likewise, the approximation based on the modified reduced volume cumulants $v^{\textrm{mod}}_l$
is plotted (solid curves).  Although all volume cumulants up to 6\textsuperscript{th} order were included in the parameter fit,
the reconstructed $N_{\textrm{part}}$ is plotted as a 4-parameter distribution.  This approximation -- used here for display only --
evidently misses some of the more wobbly features of the true distribution, which are best visible in the 0 -- 10\% and 10 -- 20\%
centrality bins.  Most notable, however, are the changes caused by correlations, leading to a consistent reduction in width of the
effective $N_{\textrm{part}}$ distributions as compared to the model truth.

\begin{figure}[!hbt]
  \begin{center}
     \resizebox{0.8\linewidth}{!} {
       \includegraphics{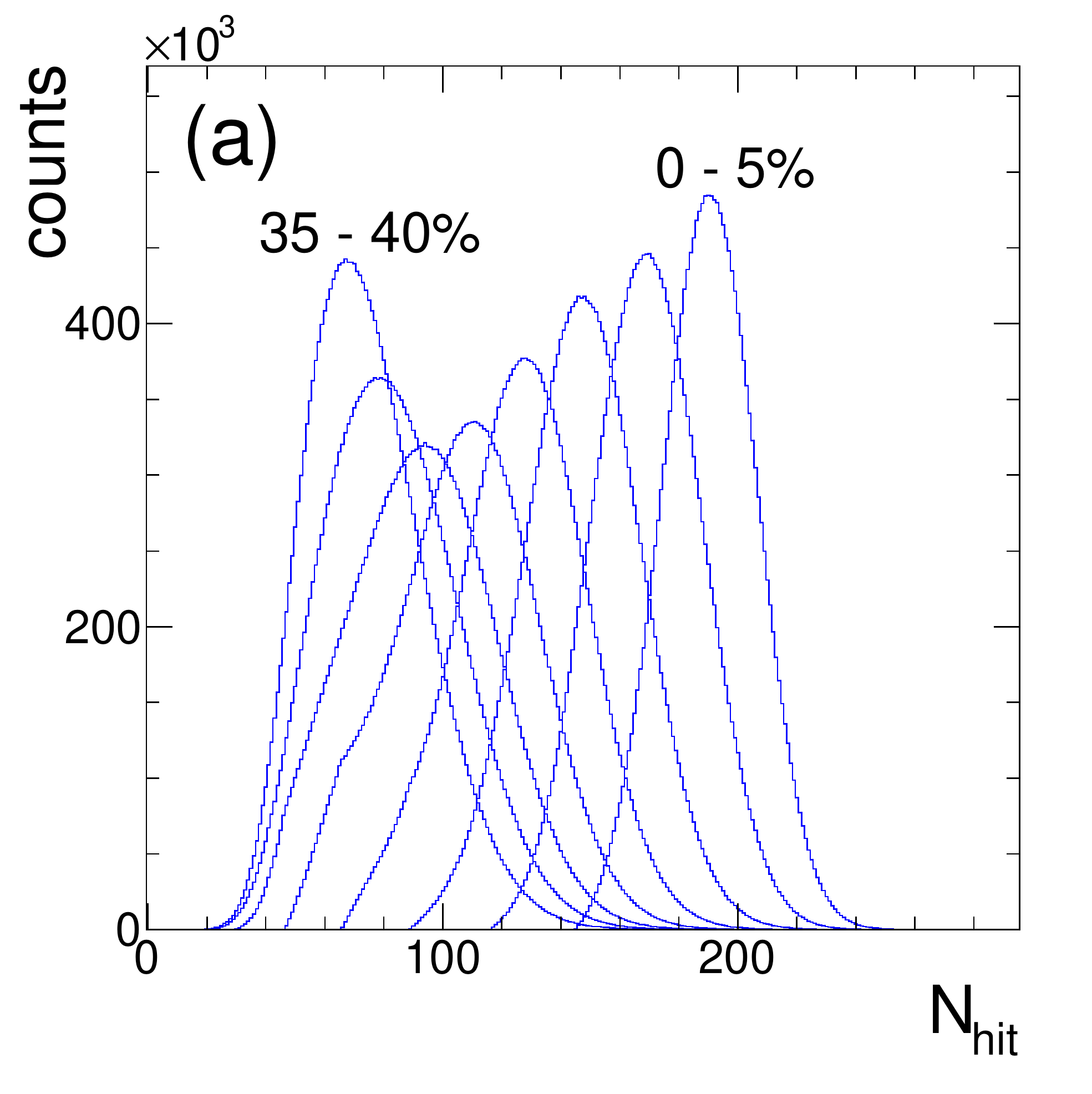}
     }
     \resizebox{0.8\linewidth}{!} {
       \includegraphics{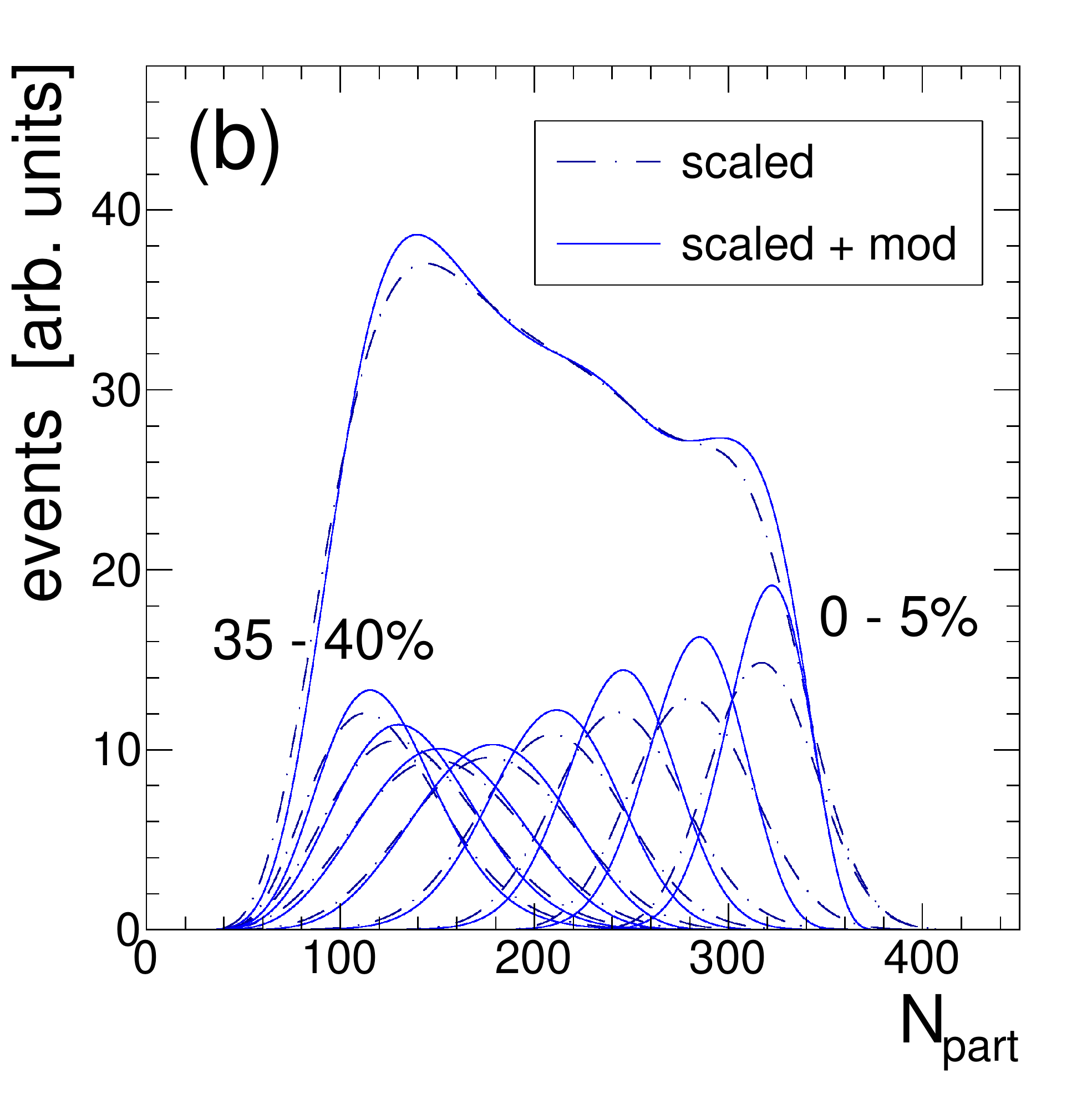}
     }
  \end{center}
  \vspace*{-0.2cm}
  \caption[] {(a) Measured Au+Au $N_{\textrm{hit}}$ distributions of eight 5\% centrality selections
              based on the $\Sigma Q_{FW}$ signal.  The latter is constrained by the 2D cut shown in Fig.~\ref{fig:sumfwnhit},
              causing also the barely visible ``knee'' in the 15 -- 20\% distribution.
              (b) Corresponding reconstructed $N_{\textrm{part}}$ distributions -- plotted as 4-parameter beta functions --
              when using the rescaled and modified volume cumulants $v_l^{\textrm{mod}}$ (solid curve), or the just rescaled ones $v_l$ (dot-dashed curve).
              The sum curves correspond to the 40\% most central events.
             }
  \label{fig:dataNpart8bins}
\end{figure}

In the data, the true event-by-event number of participating nucleons is not known; however, as explained above, we can reconstruct
the event-averaged $N_{\textrm{part}}$ distribution from its cumulants by using a volume-cumulant transformation of the observed
$N_{\textrm{hit}}$ distribution. This transformation is done by applying the modifiers $f^{\textrm{sim}}_l$ and $d^{\textrm{sim}}_l$,
determined previously in a simulation, to the cumulants $v_l$ of the rescaled $N_{\textrm{hit}}$ distribution,
namely $v^{\textrm{mod}}_l = f^{\textrm{sim}}_l \times v_l + d^{\textrm{sim}}_l$.  To do the $N_{\textrm{hit}} \rightarrow N_{\textrm{part}}$ scaling,
the mean number of participants $\langle N_{\textrm{part}} \rangle$ underlying the mean number of hits $\langle N_{\textrm{hit}} \rangle$ in a
given centrality bin was taken from a Glauber fit to the experimental hit distribution \cite{Adamczewski2018}.  As discussed above,
with the modified reduced cumulants $v^{\textrm{mod}}_l$, the effective $N_{\textrm{part}}$ distribution can be reconstructed up to a given order.
The result of this procedure is displayed in Fig.~\ref{fig:dataNpart8bins} for eight 5\% centrality selections based on the $\Sigma Q_{\textrm{FW}}$
observable; the measured $N_{\textrm{hit}}$ distributions are shown in (a) and the corresponding reconstructed $N_{\textrm{part}}$ distributions in (b),
the latter plotted as 4-parameter beta functions based on either the $v^{\textrm{mod}}_l$ (solid curves) or, for comparison,
the plain $v_l$ (dot-dashed curves).  Note again that, in all centrality selections, using modified cumulants leads to a substantial
narrowing of the reconstructed $N_{\textrm{part}}$ distributions.

After having determined the reduced volume cumulants $v^{\textrm{mod}}_l$ by scaling and transforming the measured $N_{\textrm{hit}}$ distributions,
we are finally in a position to fully remove volume fluctuation effects from the efficiency-corrected reduced proton number
cumulants $\tilde{\kappa}_n$.  This is done in a combined multi-order fit of Eq.~\eqref{eq:N2L} to the set of measured $\tilde{\kappa}_n$
values, thereby adjusting all $\kappa_n$, $\kappa_n'$, and $\kappa_n''$ at once while keeping the $v^{\textrm{mod}}_l$ fixed to their simulated values.
Errors arising from the correction procedure are discussed in Sec.~\ref{Sec:ErrorBars} below, while final results obtained with the full
analysis chain are presented in Sec.~\ref{Sec:Results}.

\section{Error treatment}
\label{Sec:ErrorBars}

\subsection{Statistical errors}

As discussed in the previous sections, fluctuation observables must be subjected to sophisticated analysis procedures
like efficiency correction and volume effect removal.  Unfortunately, the explicit propagation of the corresponding
statistical errors throughout this complex reconstruction pipeline is awkward at best \cite{Luo2015a} and, in our case,
not practical at all.  Instead, we have used the resampling method known as the bootstrap \cite{Efron1979a,Efron1987,Kendall2004}
as well as event subsampling \cite{Pandav2019} to determine statistical error bars.

The main idea of the bootstrap method is to repeatedly resample events with replacement from the total set of $N_{ev}$ measured
events which, by definition, are considered independent and identically distributed (iid).  For large $N_{ev}$, this procedure
reuses on average a fraction $1 - e^{-1} \simeq 0.632$ of all events in the set.  Each of our 5\% centrality selections comprises
$N_{ev}~\approx$~20~million events and a huge number\footnote{In fact, $\binom{2N_{\textrm{ev}}-1}{N_{\textrm{ev}}}$,
i.e.\ approximately an astounding $10^{12000000}$.} of non-identical event
sets can be resampled;  as our main goal is to get an error estimate, a few hundred resamplings are considered sufficient
in practice \cite{Kendall2004}.  Every one of the resampled event sets is processed through the full analysis pipeline, allowing
to construct the distribution of each observable of interest.  The histogrammed observables provide in turn estimates of their
respective mean and standard deviation, that is the statistical error bar aimed for.

An alternative method to determine statistical errors is provided by subsampling.  In that approach, the full set of observed
events is divided into a number of equal-sized subsets which are analyzed one by one, again producing distributions of all
observables with corresponding estimates of mean and standard deviation.  Subsampling is much faster than resampling as it
operates on smaller event sets with, however, a correspondingly larger error.  In order to use the subsampling standard deviation
as an estimator of the error on an observable obtained from the full data set, it has hence to be scaled by $1/\sqrt{N_{sub}}$,
where $N_{sub}$ is the number of subsamples \cite{Pandav2019}.  
Another important difference to resampling is that subsampling can be sensitive to long-term changes in the data properties,
resulting e.g.\ from instabilities of the experimental conditions affecting the detector and/or beam during the data taking.
Indeed, if the subsamples correspond to consecutive time periods, the resulting error bars will not only represent fluctuations
due to counting statistics but also incorporate a measure of mid- and long-term experimental changes.  It is then a matter of
discussion how to label these additional contributions: while random, mostly short-term instabilities may be presented as part
of the statistical fluctuations, long-term drifts may be considered more akin to systematic errors.  In our case, we have
corrected the measured proton yields for long-term, i.e.\ day-by-day changes by rescaling the average number of reconstructed 
tracks per event to a reference value. 
Comparing next the standard deviations from subsampling, based on a splitting into 2-hour long data-taking periods, with the ones
from resampling of the full set of events, we observe an overall increase resulting in about a doubling of the error
on 4\textsuperscript{th}-order moments and cumulants.  As systematic effects ultimately dominate the total error on our measurements
(see Sec.~\ref{Sec:Results}), we decided to accommodate the remaining short-term random variations in our statistical errors by
using the subsampling standard deviations instead of the resampling ones.

\subsection{Systematic uncertainties}

As already argued in Sec.~\ref{Sec:Exp}, various nuisance effects can potentially influence the measured proton multiplicity
distribution.  They result either from a contamination by other event classes, namely pileup events or Au+C reactions, or from
background processes within valid Au+Au events, like misidentified particles, decay protons, or knockout protons.  We determined
upper limits on this background (listed in Table~\ref{tab:nuisance}) and simulated how the proton cumulants are affected.
As within-event contributions just add particles to the event, their effect on the cumulants is of similar magnitude as the
background itself, i.e.\ well below the $1\%$ level.  Assuming purely poissonian processes \cite{Bzdak2019} in our simulation, the
estimated contributions of event classes with either larger average multiplicity (pileup) or lower (Au+C reactions) were found
to induce changes of maximally 5\%.  
And their influence would become even smaller if the relevant physics signal turned out to be of non-poissonian nature.  
We conclude that in the present analysis both nuisance effect classes are inconsequential. \\

\noindent
Systematic errors also arise at various stages of the analysis.  We classify those into three types:
\begin{enumerate}
\item Type A errors are caused by a global uncertainty on the proton efficiency, arising in the track reconstruction
and particle identification procedures.  The estimated efficiency error of 4--5\% in the phase-space bin of interest \cite{Szala2019}
results typically in about $n \times$(4--5)\% errors on cumulants and reduced cumulants of order $n$.  
\item Type B errors arise from imperfections of the cumulant correction schemes, i.e.\ event-by-event correction, unfolding,
or moment expansion.  From a systematic comparison of these methods in both, simulation (see discussion of Figs.~\ref{fig:effiqmddy02}
and \ref{fig:effiqmddy05}) and data,
we find a typical error of 1.5\% on $K_1$, 3\% on $K_2$, 7.5\% on $K_3$, and 15\% on $K_4$. 
\item Type C errors are due to an overall inaccuracy of about 8--9\% on the $N_{\textrm{part}}$ calibration of our centrality determination
(caused by model dependencies and the limited experimental resolution \cite{Adamczewski2018}).  This impacts the cumulants indirectly
through the applied volume correction, resulting in uncertainties of order $\leq$2\% for $K_2$, 3 -- 6\% for $K_3$, and 10 -- 30\% for $K_4$.
Reduced cumulants are however affected more directly through their normalization to $\langle N_{part} \rangle$.
\end{enumerate}
The factorial cumulants $C_n$ and, to some extend, also the cumulant ratios turn out to be more robust than the $K_n$,
showing generally a factor 2 -- 3 smaller relative systematic error.   \textcolor{black}{In the result section below,
we present the total systematic error, obtained as a combination of the three types A, B, and C.  This was achieved by applying
the efficiency and volume corrections to the respective observable of interest, $K_n$ or $C_n$, while varying the detection efficiency
and the volume proxy, i.e.\ $\langle N_{part} \rangle$, within the ranges specified above.  The resulting total spread of the corrected
observable was then assigned as a systematic error, thus complementing the statistical one.}

\section{Results}
\label{Sec:Results}

\subsection{Cumulants and moments}

\begin{figure}[hbt]
  \begin{center}
     \resizebox{0.80\linewidth}{!} {
       \includegraphics{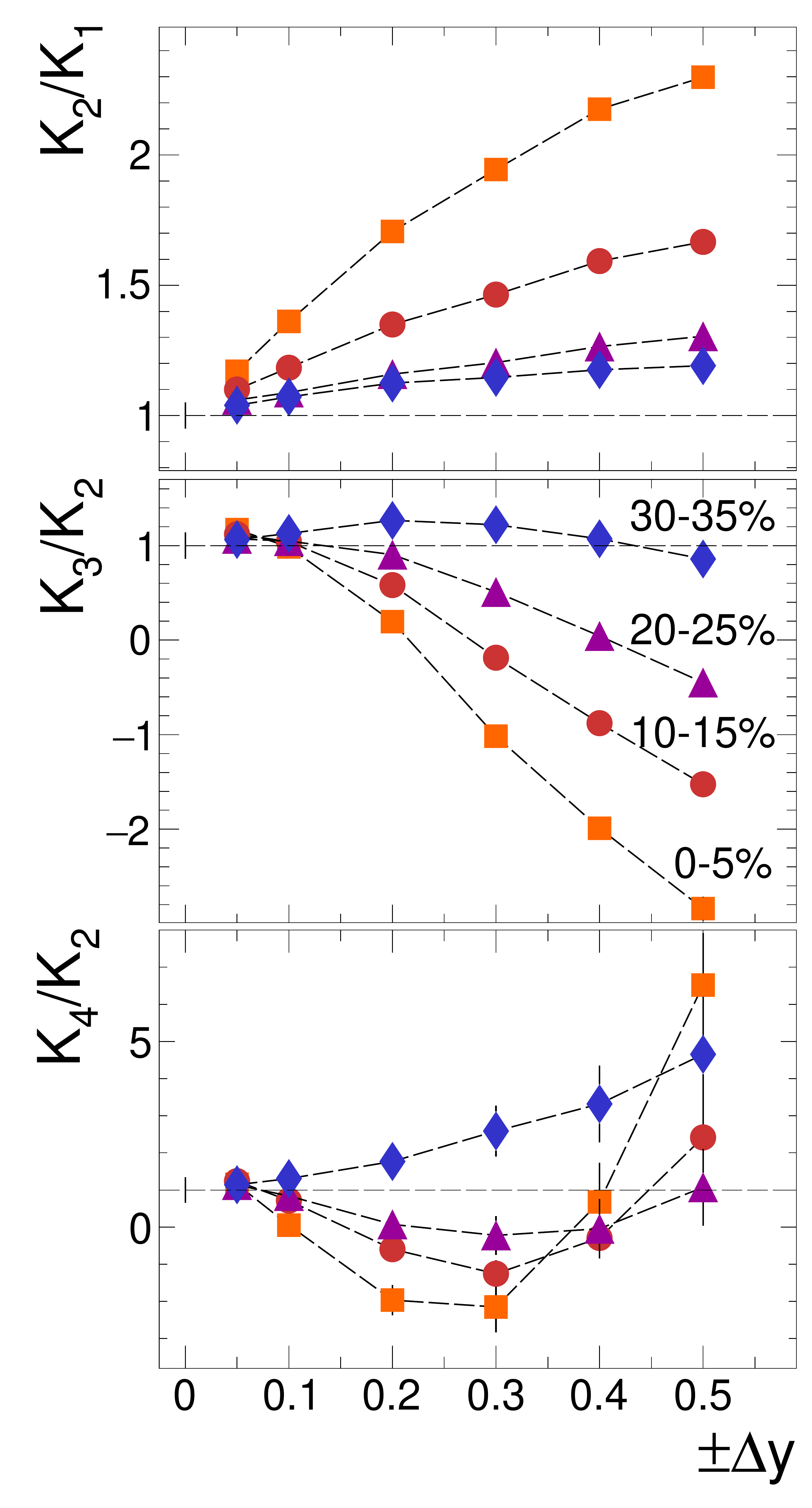}
     }
  \end{center}
  \vspace*{-0.2cm}
  \caption[] {(Color online) Au+Au data: Efficiency and N2LO volume corrected proton cumulant ratios plotted as a function
               of the width of the rapidity bin defined by $y \in y_0 \pm \Delta y$ and $0.4 \leq p_t \leq 1.6$~\gevc.
               Shown are $\omega = K_2/K_1$ (top), $\gamma_1 \times \sigma = K_3/K_2$ (middle),
               and $\gamma_2 \times \sigma^2 = K_4/K_2$ (bottom) for various 5\% centrality selections.  Error bars are statistical
               only, and dashed lines connect the data points belonging to a given centrality.  With decreasing $\Delta y$,
               all ratios tend towards unity (indicated also by a horizontal line), i.e.\ they approach the Poisson limit
               where $K_1 = K_2 = K_3 = K_4$. 
             }
  \label{fig:Poissonizer}
\end{figure}

Here we present the efficiency and volume corrected proton multiplicity moments and cumulants obtained in 1.23~\agev\ Au+Au collisions
(\sqrtsNN = 2.4~\gev).  To start, we show in Fig.~\ref{fig:Poissonizer} for a few centrality selections the ratios of fully corrected cumulants
($\omega = K_2/K_1$, $\gamma_1 \times \sigma = K_3/K_2$, $\gamma_2 \times \sigma^2 = K_4/K_2$, where $K_n$ are cumulants)
as a function of the width of the rapidity bin, namely $y \in y_0 \pm \Delta y$, centered at mid-rapidity $y_0$ = 0.74 and with
$0.4 \leq p_t \leq 1.6$~\gevc. These ratios were derived from the reduced cumulant expansions obtained by fitting one of Eqs.~\eqref{eq:NL}
or \eqref{eq:N2L} to the efficiency-corrected and centrality-selected data points.\footnote{For very narrow phase space, the NLO and N2LO
fits give very similar results.}  In this procedure, the modified volume cumulants $V_n$ obtained from the experimental $N_{hit}$ distributions,
as laid out in Sec.~\ref{Sec:Npart}, were inserted while the values of the $\kappa_n$, $\kappa_n'$, and $\kappa_n''$ were adjusted.
Error bars shown in Fig.~\ref{fig:Poissonizer} are statistical; they were obtained with the sampling techniques discussed
in Sec.~\ref{Sec:ErrorBars}.  As phase space closes more and more, ever fewer correlated particles contribute and one expects
their distribution to approach the Poisson limit \cite{Begun2004} where the $K_n$ converge, i.e.\ $K_n = \langle N_{\textrm{prot}} \rangle$
for all $n$.  From the figure it is apparent that the data follow indeed in all centrality selections such a behavior,
with the cumulant ratios approaching unity within their statistical errors.


\begin{figure*}[!htb]
  \begin{center}
     \resizebox{0.7\linewidth}{!} {
       \includegraphics{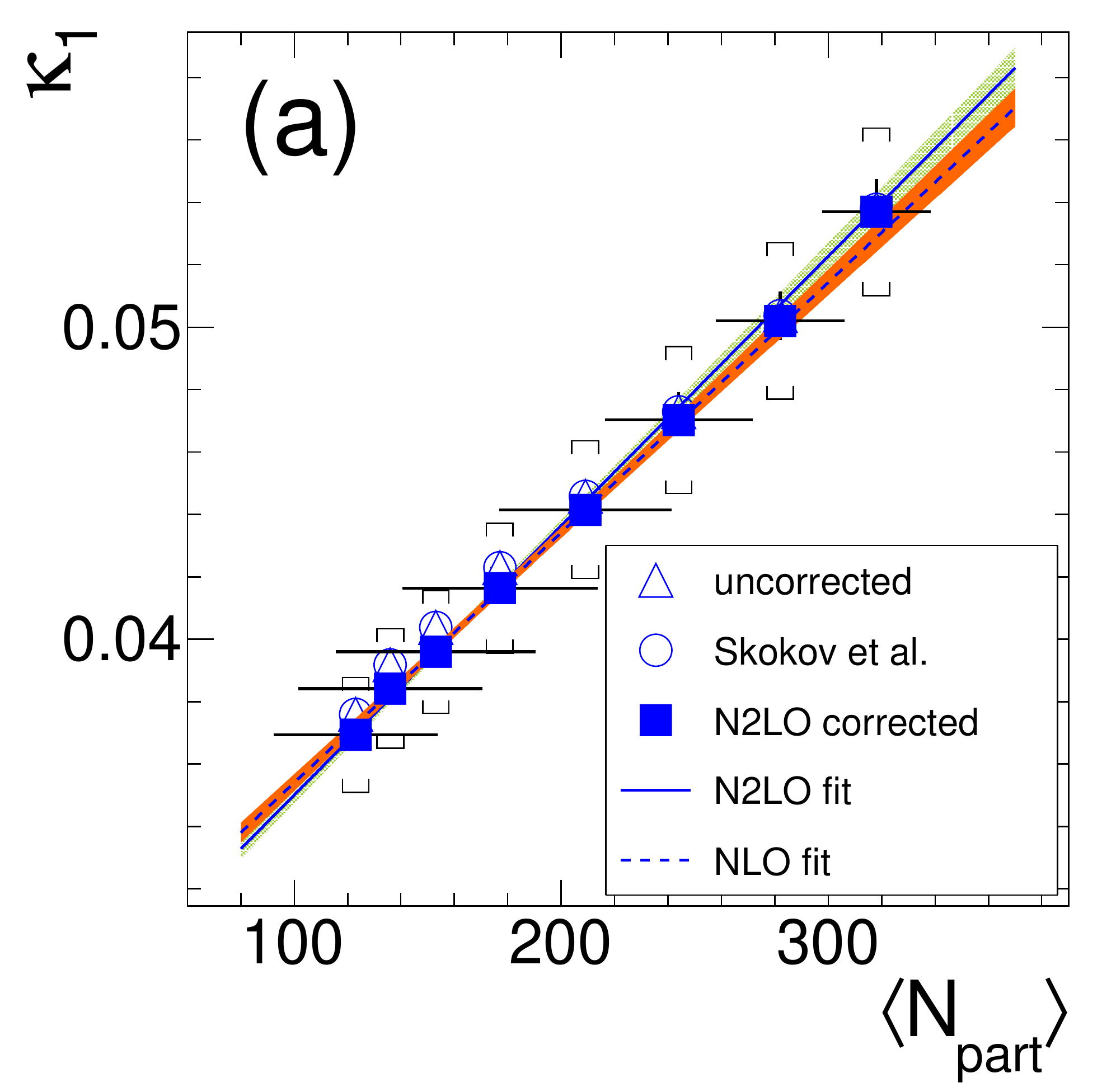}
       \includegraphics{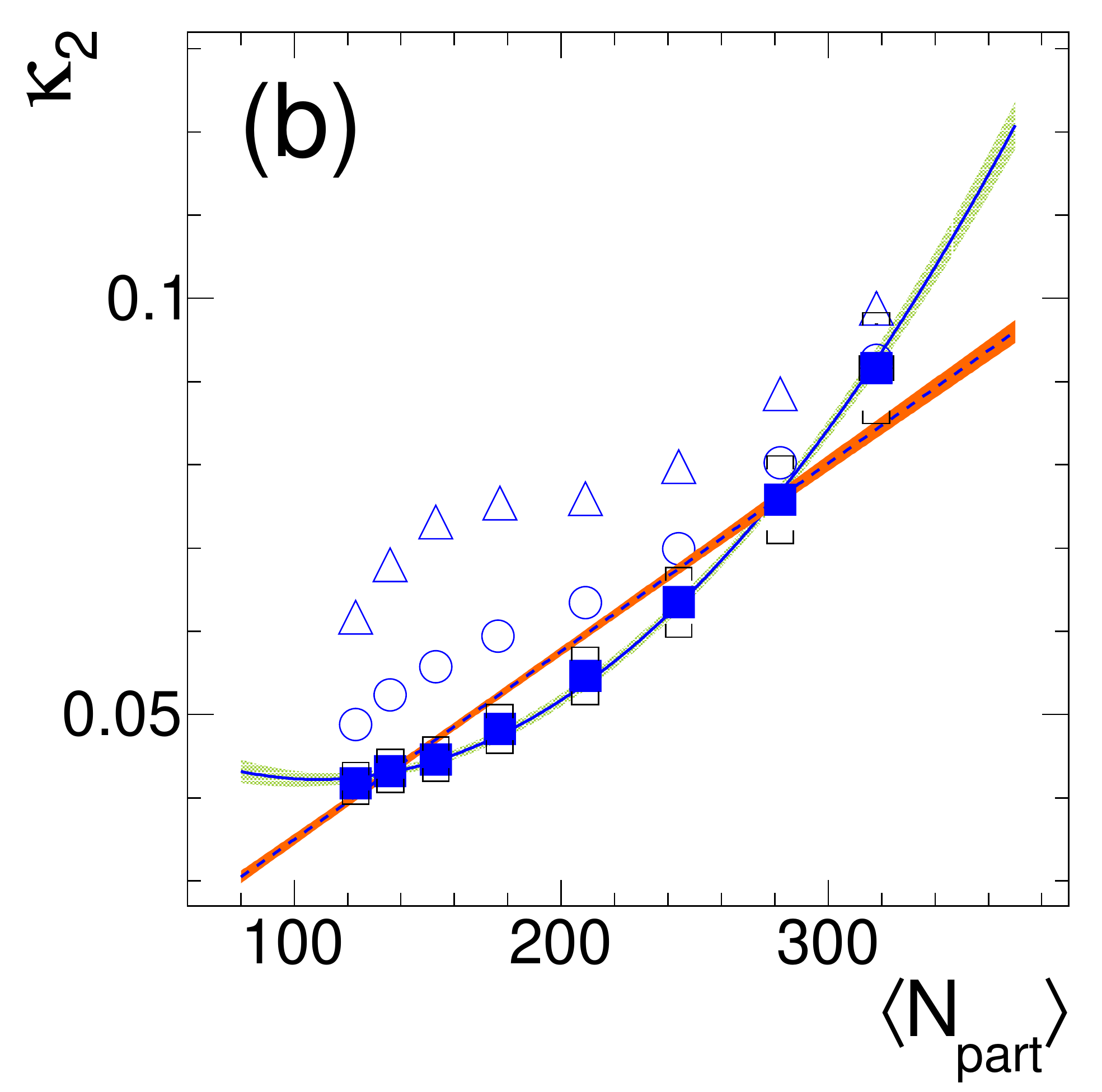}
     }
     \resizebox{0.7\linewidth}{!} {
       \includegraphics{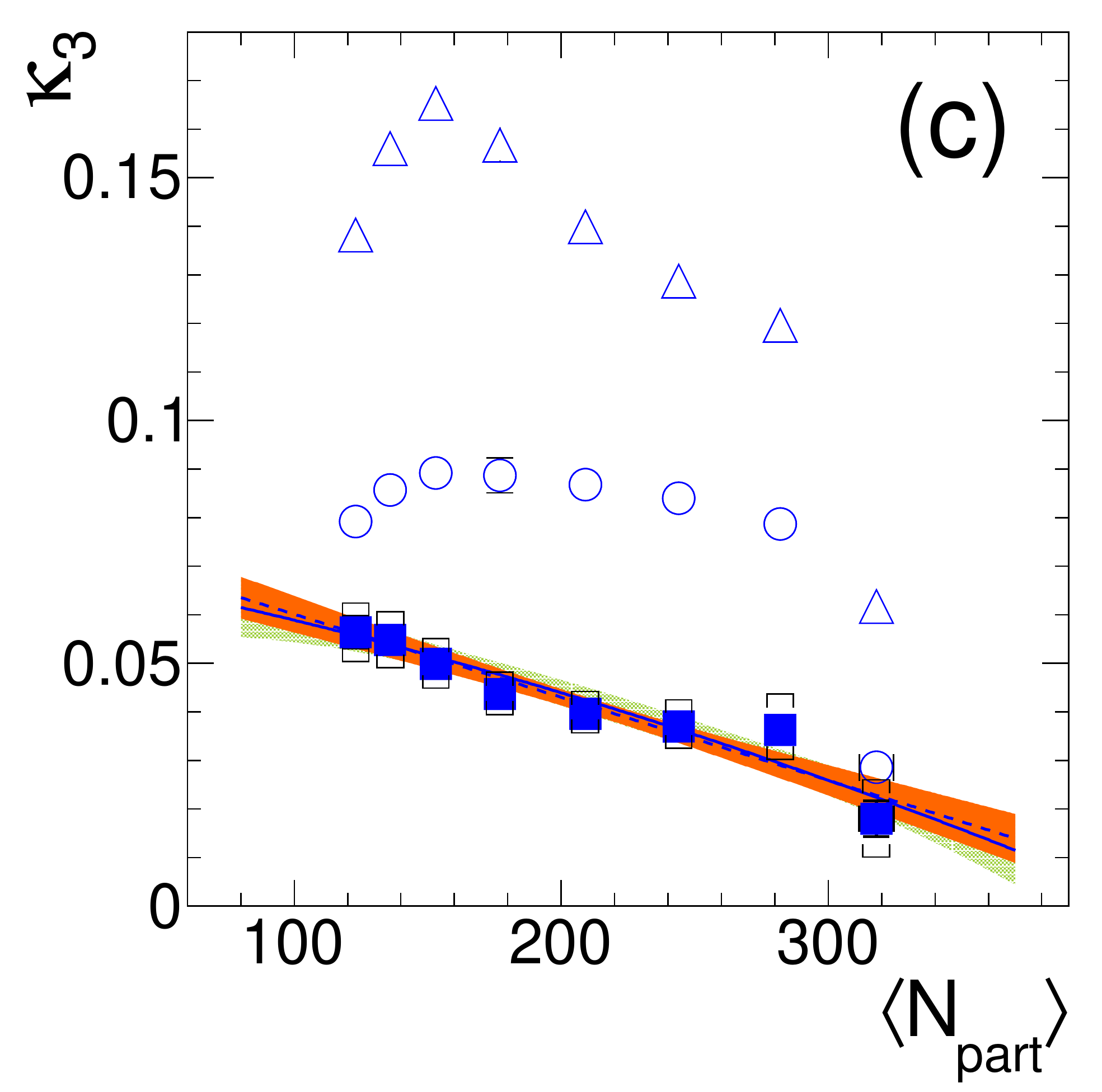}
       \includegraphics{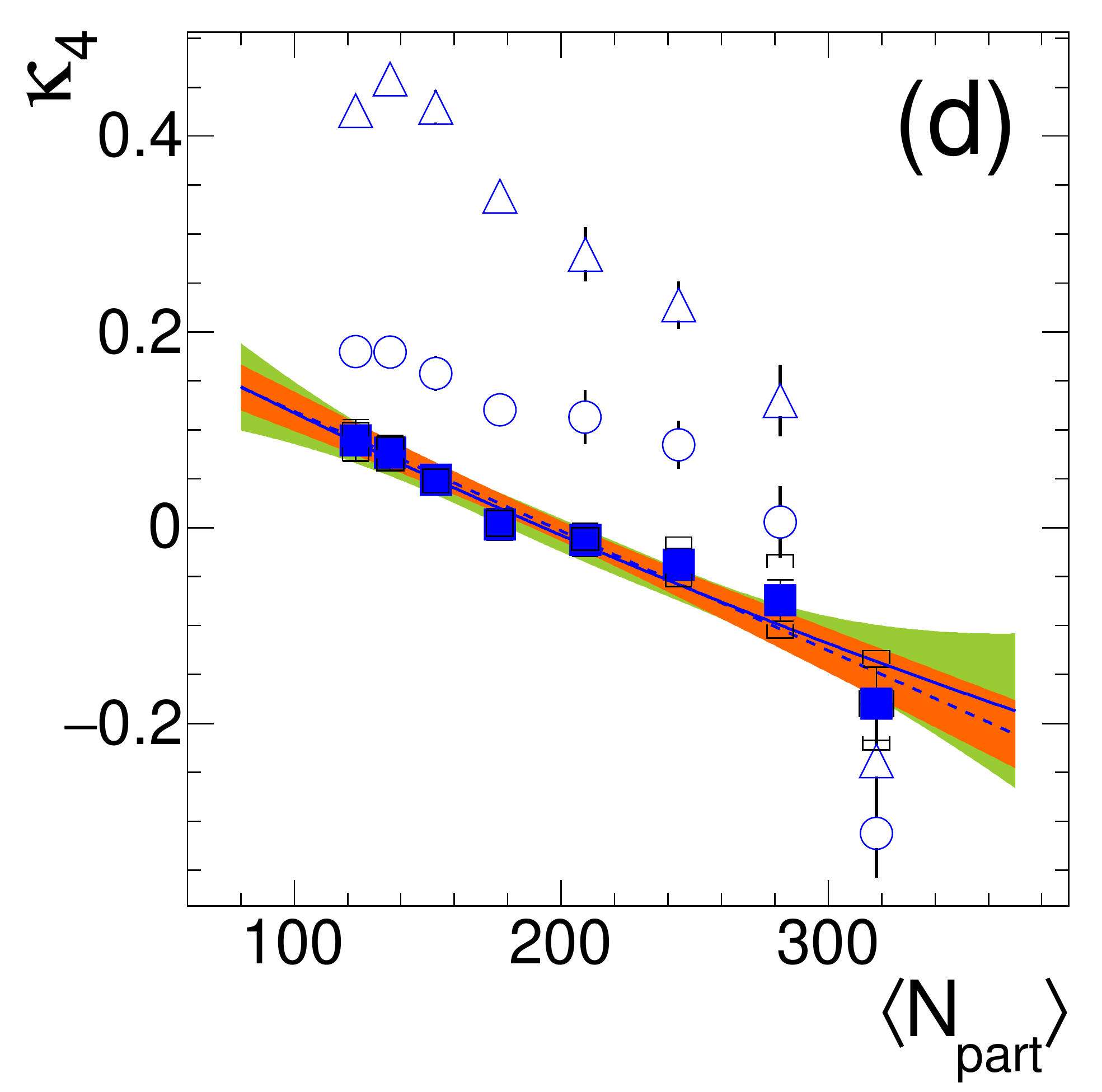}
     }
  \end{center}
  \vspace*{-0.2cm}
  \caption[] {(Color online) Au+Au data: Efficiency and volume corrected reduced proton cumulants $\kappa_{n}$ for the phase-space bin
               $y \in y_0 \pm 0.2$ and $0.4 \leq p_t \leq 1.6$~\gevc\ as a function of mean $N_{\textrm{part}}$, using 5\% centrality bins.
               Shown are the data without volume correction (open triangles), with Skokov {\em et al.} \cite{Skokov2013} correction
               (open circles), and with N2LO correction (solid squares).  Vertical bars are statistical errors, cups delimit
               full systematic errors (shown on the N2LO corrected points only), and horizontal bars shown in (a) correspond to the width
               ($\pm 1$~s.d.) of the $N_{\textrm{part}}$ distribution in the given centrality bin.  Solid curves are N2LO fits,
               dashed curves are NLO fits (for comparison), and shaded bands are the $\pm 1$~s.d. statistical errors of the fits
               (orange for NLO, olive for N2LO).
             }
  \label{fig:cumul02}
\end{figure*}

\begin{figure*}[!hbt]
  \begin{center}
     \resizebox{0.7\linewidth}{!} {
       \includegraphics{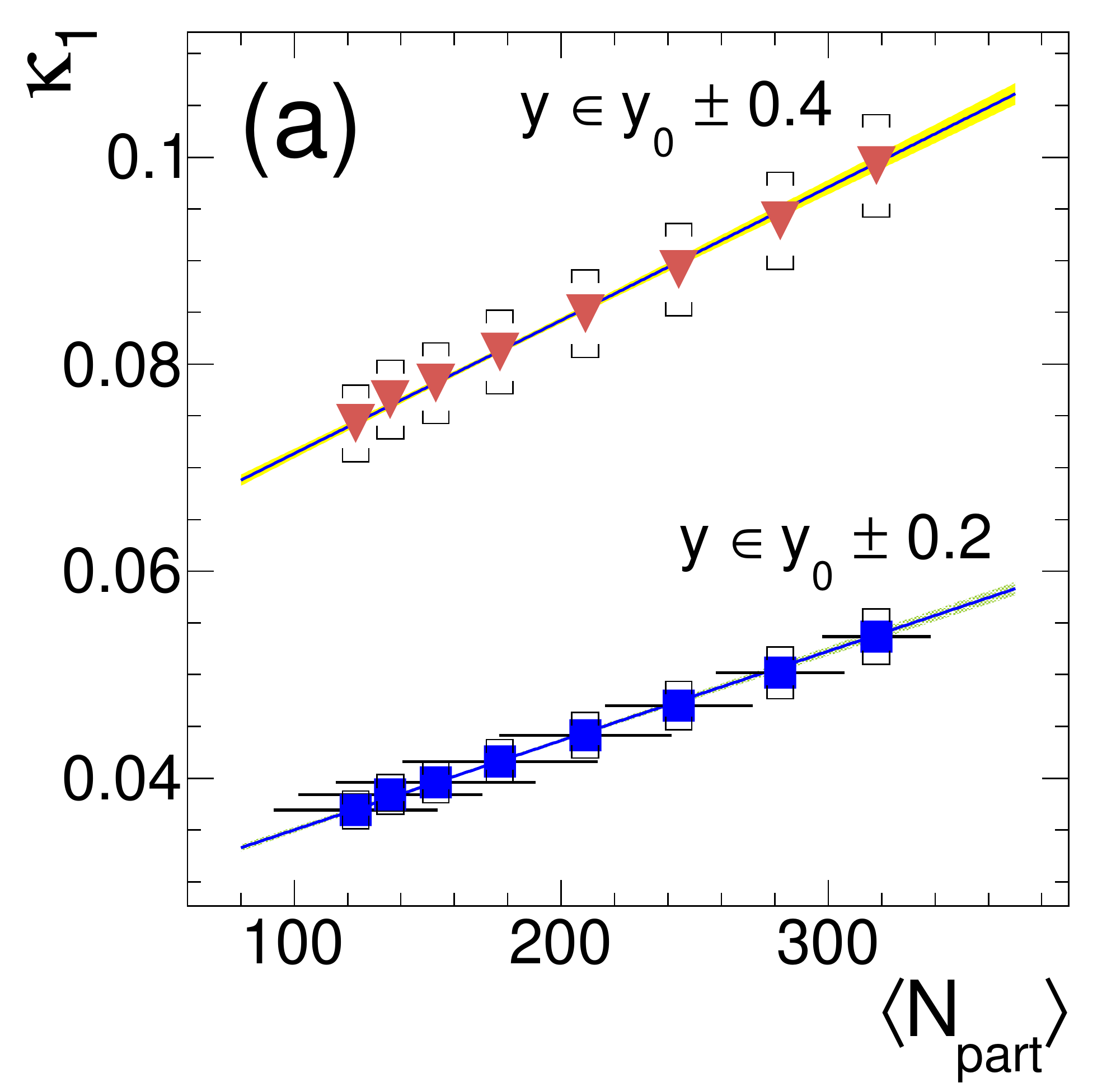}
       \includegraphics{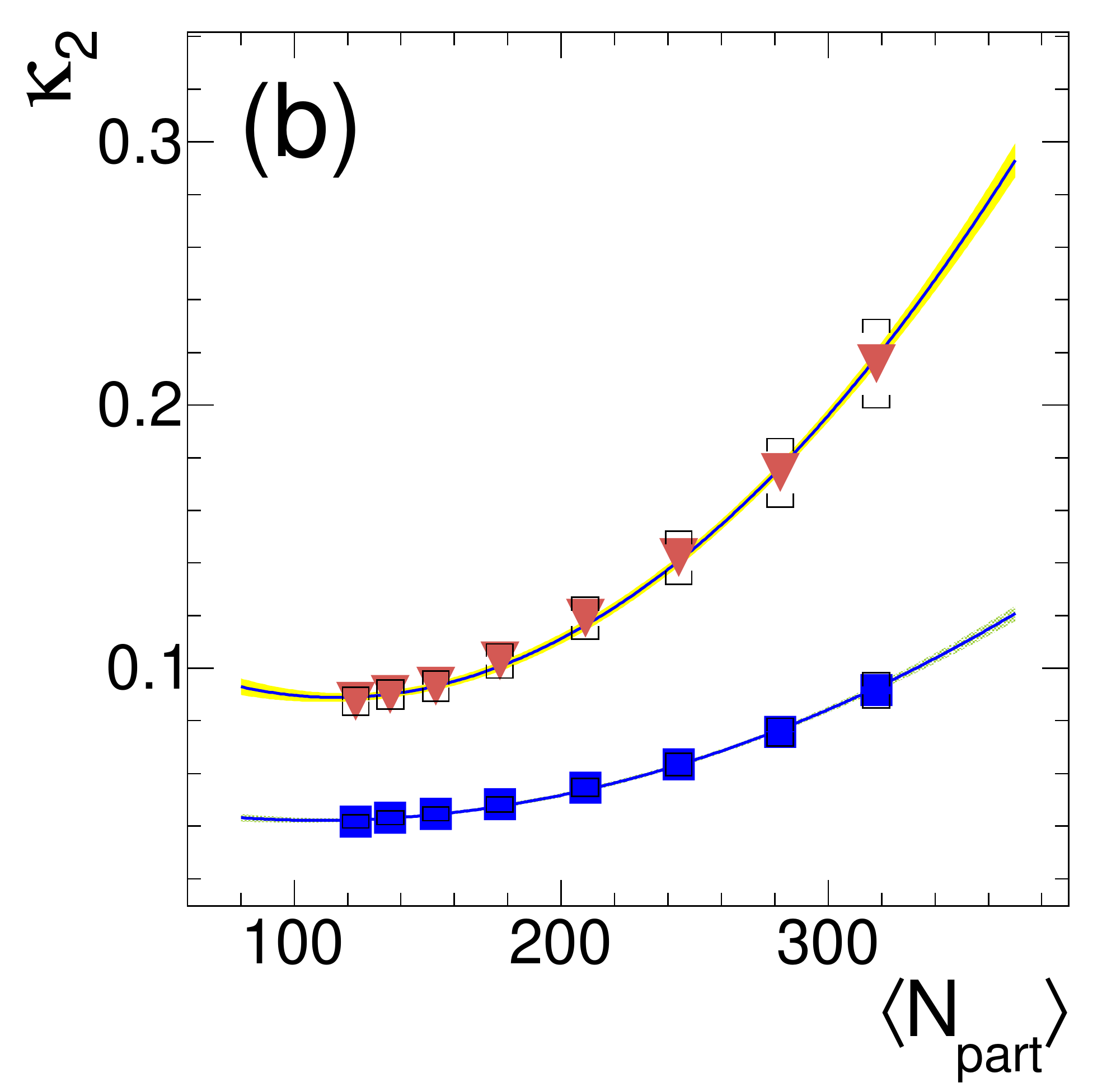}
     }
     \resizebox{0.7\linewidth}{!} {
       \includegraphics{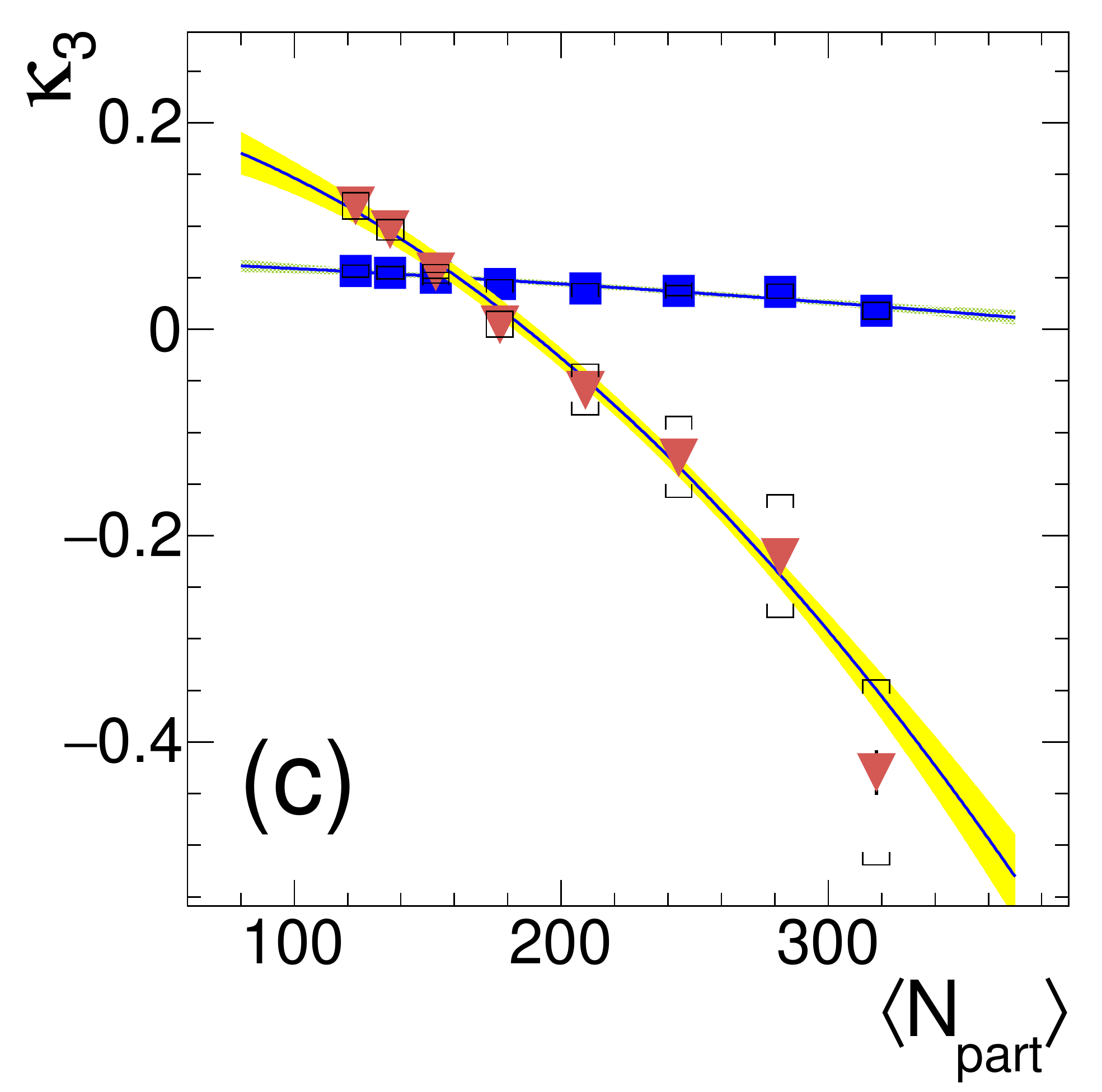}
       \includegraphics{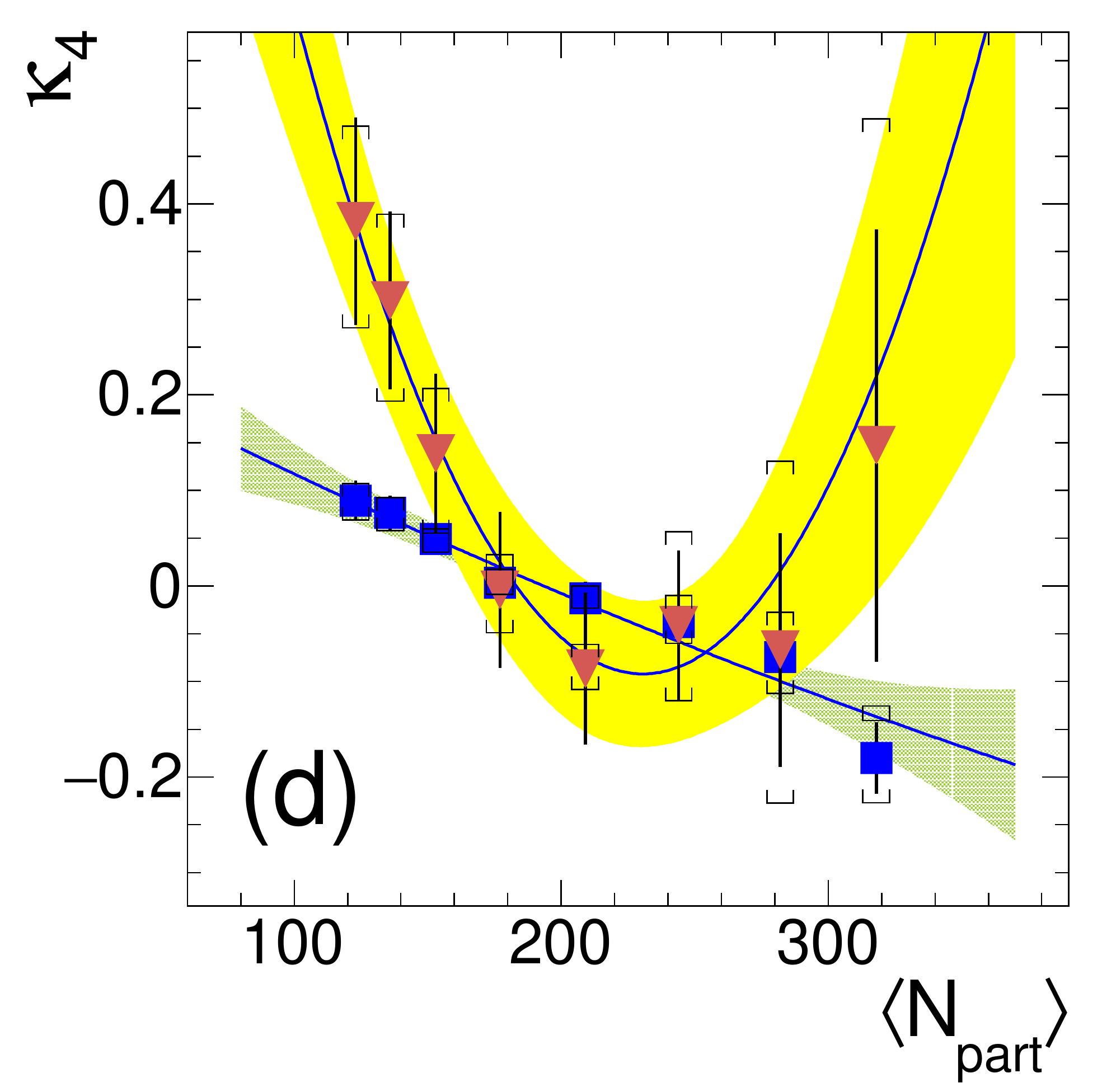}
     }
  \end{center}
  \vspace*{-0.2cm}
  \caption[] {(Color online) Au+Au data: Efficiency and N2LO volume corrected reduced proton cumulants as a function of mean $N_{\textrm{part}}$,
               using 5\% centrality bins.  Data from two phase-space selections are shown: $y \in y_0 \pm 0.2$ (blue squares) and
               $y \in y_0 \pm 0.4$ (red inverted triangles), both with $0.4 \leq p_t \leq 1.6$~\gevc.
               Vertical bars are statistical errors on data, cups delimit full systematic errors;  horizontal bars, shown in (a) only,
               correspond to the width ($\pm 1$~s.d.) of the $N_{\textrm{part}}$ distribution in the given centrality bin.
               Solid curves are N2LO fits to the data and shaded bands are the $\pm 1$~s.d. statistical fit errors.
             }
  \label{fig:cumul04}
\end{figure*}

Turning to rapidity bites substantially larger than $\pm 0.1$, we found that NLO volume effects do not anymore suffice to give a
good description of the observed proton cumulants, meaning that N2LO volume terms must be included.  This is demonstrated
in Fig.~\ref{fig:cumul02} which, for $y \in y_0\pm 0.2$, compares the effect of the volume correction at successive levels of
sophistication.  Shown are the reduced cumulants $\kappa_1, \kappa_2, \kappa_3$, and $\kappa_4$ as a function of $N_{\textrm{part}}$ when
using 5\% centrality bins: either not volume corrected (open triangles), or with only the leading order (LO) correction of Eq.~\eqref{eq:SkokovLO}
applied (open circles), or with the full N2LO correction applied (full squares).  To not clutter the pictures too much, the NLO corrected
points are not displayed explicitly but both fit curves are shown: NLO (dashed curve) done with Eq.~\eqref{eq:NL} and N2LO (solid curve)
done with Eq.~\eqref{eq:N2L}.  The corresponding statistical and systematic errors were obtained with the procedures described in
Sec.~\ref{Sec:ErrorBars}.  Figure~\ref{fig:cumul02} illustrates that the LO scheme proposed in \cite{Skokov2013,Rustamov2017a}
removes in our case only about 50 - 70\% of the volume fluctuations.  While using instead NLO corrections does improve the description,
it still does not lead to a fully satisfactory fit of the cumulants.  One can see that the linear fit of $\kappa_2$, in particular,
misses the data points which definitely display a substantial curvature.  When enlarging the accepted phase space further,
curvature terms become even more important, as shown in Fig.~\ref{fig:cumul04} which compares volume-corrected reduced proton
cumulants and fits in the two rapidity bins, $y \in y_0 \pm 0.2$ and $y \in y_0 \pm 0.4$.  Consequently, all results presented
in the following were obtained by consistently applying the full N2LO volume corrections.

Comparing furthermore the measured reduced proton cumulants of Fig.~\ref{fig:cumul04} with their transport calculation counterparts,
as shown in Fig.~\ref{fig:transport}, one can notice a qualitative agreement for the $y \in y_0 \pm 0.2$ rapidity bite.
Especially the IQMD model seems to capture the basic trends of $\kappa_n$ with $N_{\textrm{part}}$, including the presence of a curvature
in $\kappa_2$.  However, in our simulations, all three codes used (IQMD, UrQMD, and HSD) generally miss the absolute magnitudes
of $\kappa_n$, $\kappa_n'$, and $\kappa_n''$.  In the present study we refrained, however, from a more detailed comparison of our data
with model calculations.

\begin{figure*}[ht]
     \resizebox{1.0\linewidth}{!} {
       \includegraphics{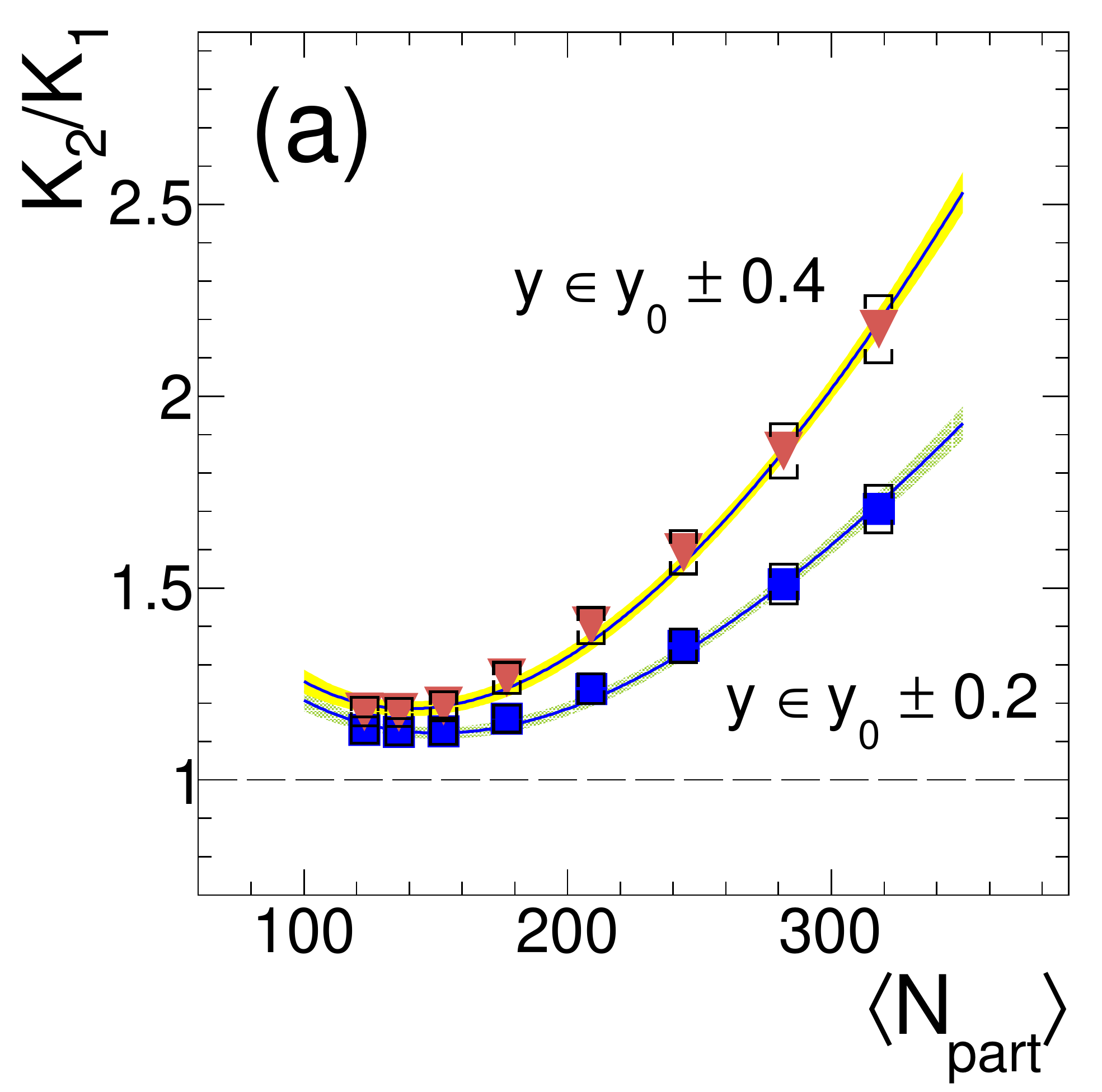}
       \includegraphics{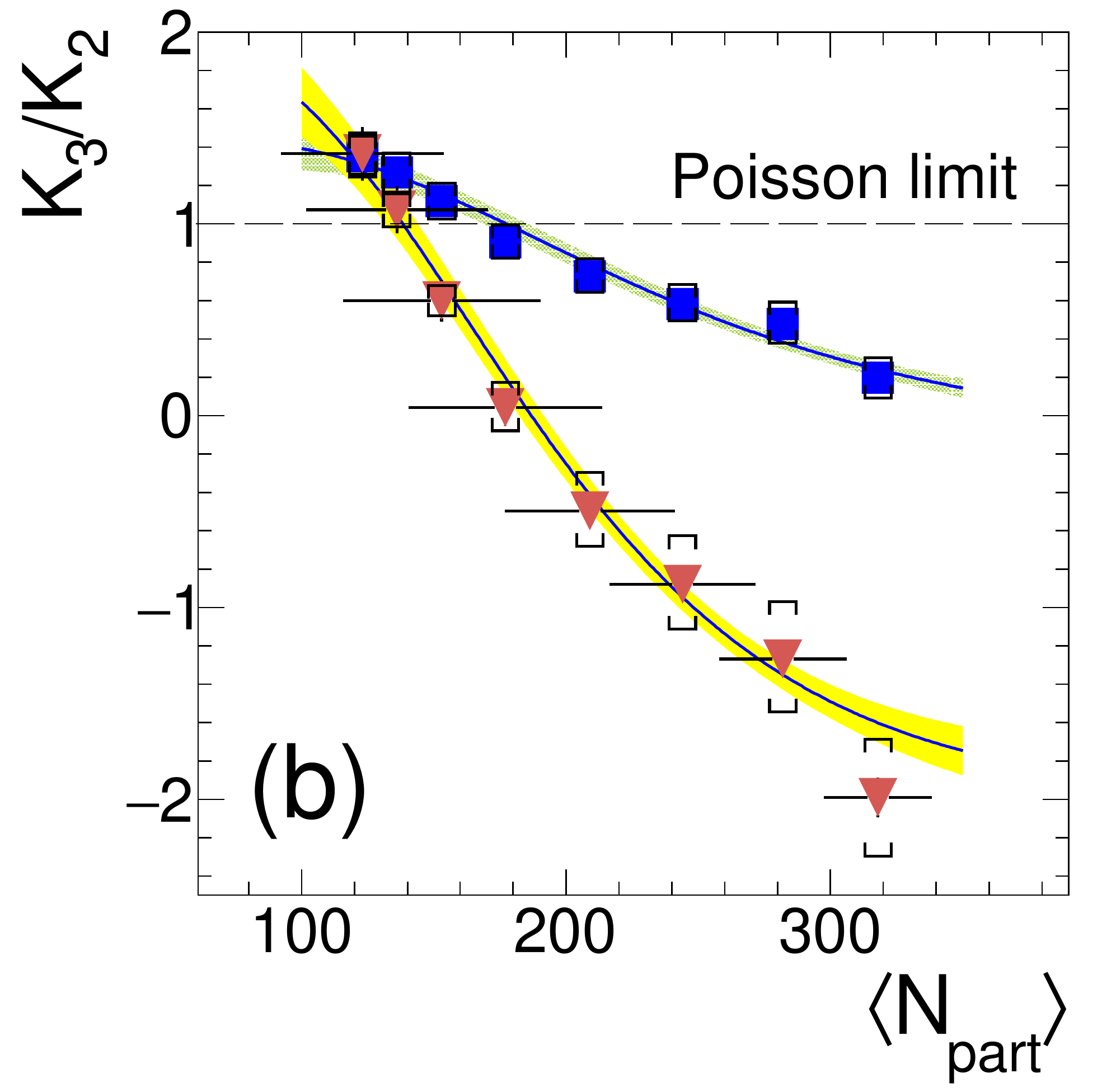}
       \includegraphics{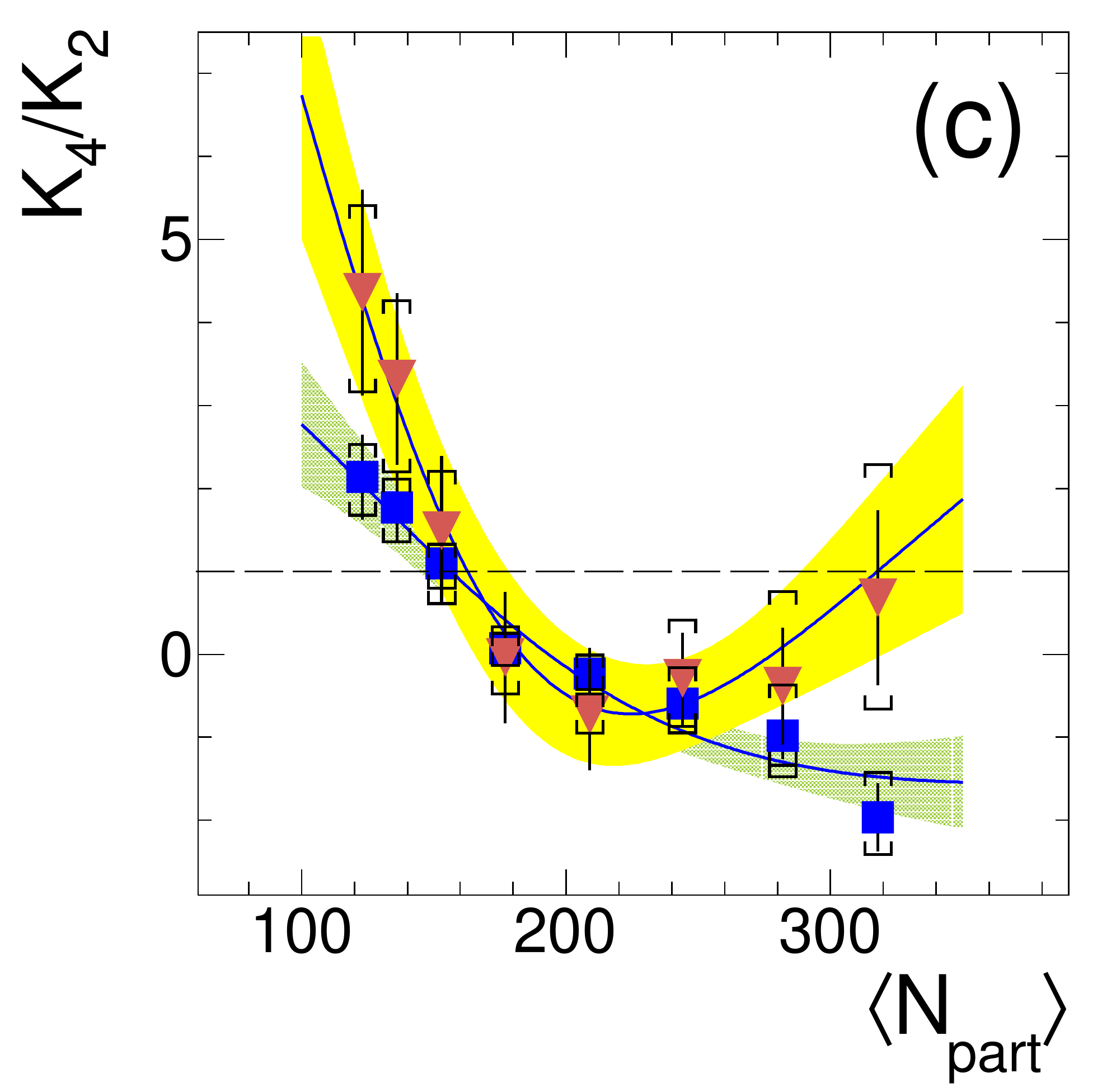}
     }
  \vspace*{-0.2cm}
  \caption[] {(Color online) Au+Au data: Efficiency and N2LO volume corrected proton cumulant ratios as a function of mean $N_{\textrm{part}}$,
               using 5\% centrality bins.  Two selections are shown: $y \in y_0 \pm 0.2$ (blue squares) and $y \in y_0 \pm 0.4$
               (red inverted triangles), both with $0.4 \leq p_t \leq 1.6$~\gevc.  Vertical error bars are statistical,
               cups demarcate full systematic errors;  horizontal bars, shown in (b) only, depict the width ($\pm 1$ s.d.)
               of the $N_{\textrm{part}}$ distribution within the given centrality bin.  Solid curves are N2LO fits, shaded bands
               represent the corresponding $\pm 1$ s.d. statistical fit errors.  The horizontal dashed lines correspond to
               the Poisson limit where $K_1 = K_2 = K_3 = K_4$.
             }
  \label{fig:KKratio}
\end{figure*}

From the reduced cumulants $\kappa_n$, the full proton cumulants $K_n = N_{\textrm{part}} \; \kappa_n$ as well as their ratios are readily
obtained.   Cumulant ratios are shown as a function of $N_{\textrm{part}}$ in Fig.~\ref{fig:KKratio} for rapidity bites $y \in y_0 \pm 0.2$
and $y \in y_0 \pm 0.4$.  In contrast to the narrow mid-rapidity bin $y \in y_0 \pm 0.05$ (cf. Fig.~\ref{fig:Poissonizer}),
the deviation from the Poisson limit -- where all $K_n$ would be equal -- is blatantly apparent: Except for the notable region around
$N_{\textrm{part}} = 150$, cumulant ratios at all orders differ strongly from unity and they display, overall, a highly non-trivial $N_{\textrm{part}}$
dependence.  Ratios of cumulants are intensive (although not strongly intensive) quantities, meaning that they do not depend on the
mean source volume.  They are therefore often favored when directly comparing data from different experiments, where e.g.\ the selected
centralities may differ.

\begin{figure}[!hbt]
  \begin{center}
     \resizebox{1.0\linewidth}{!} {
       \includegraphics{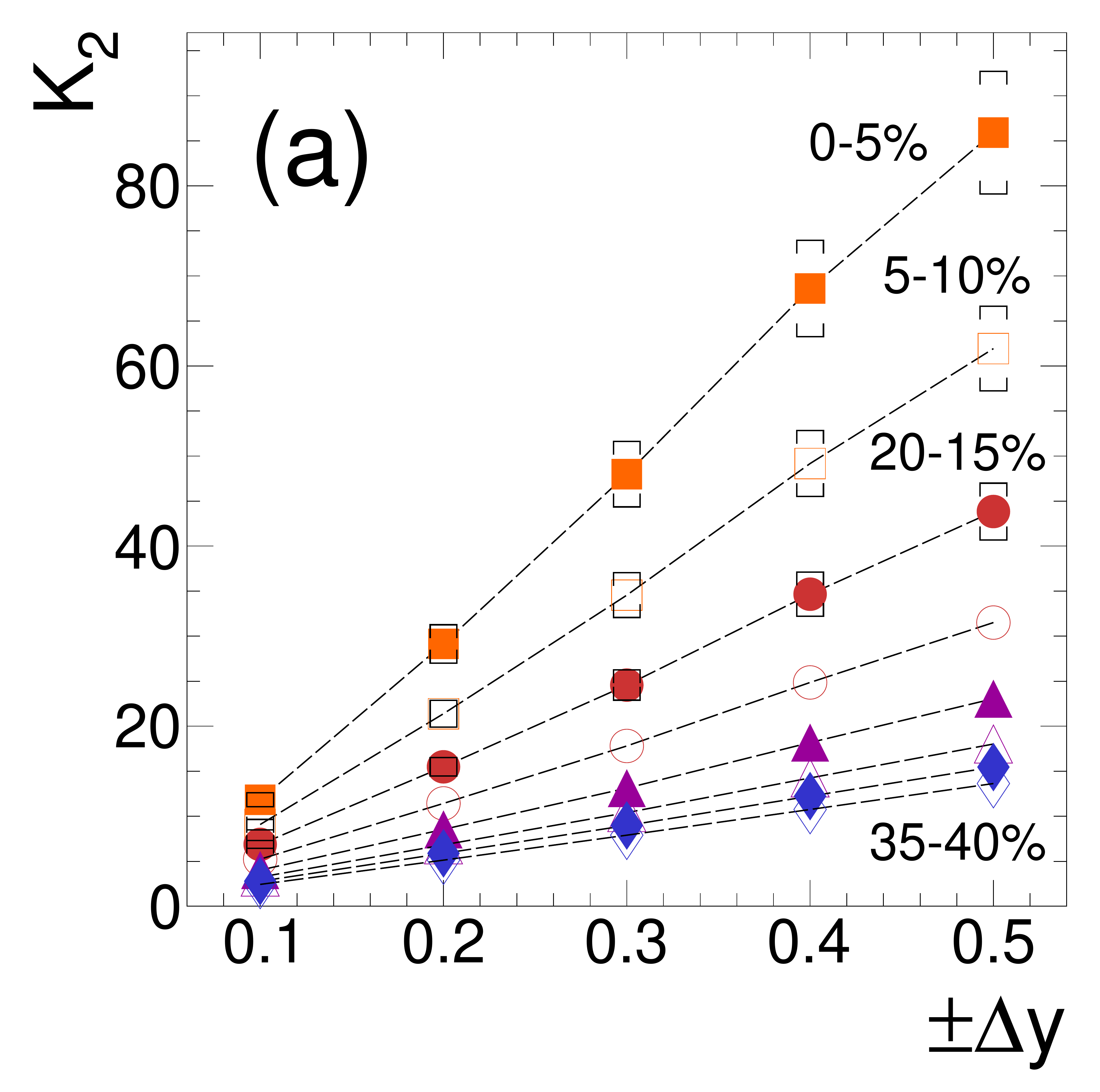}
       \includegraphics{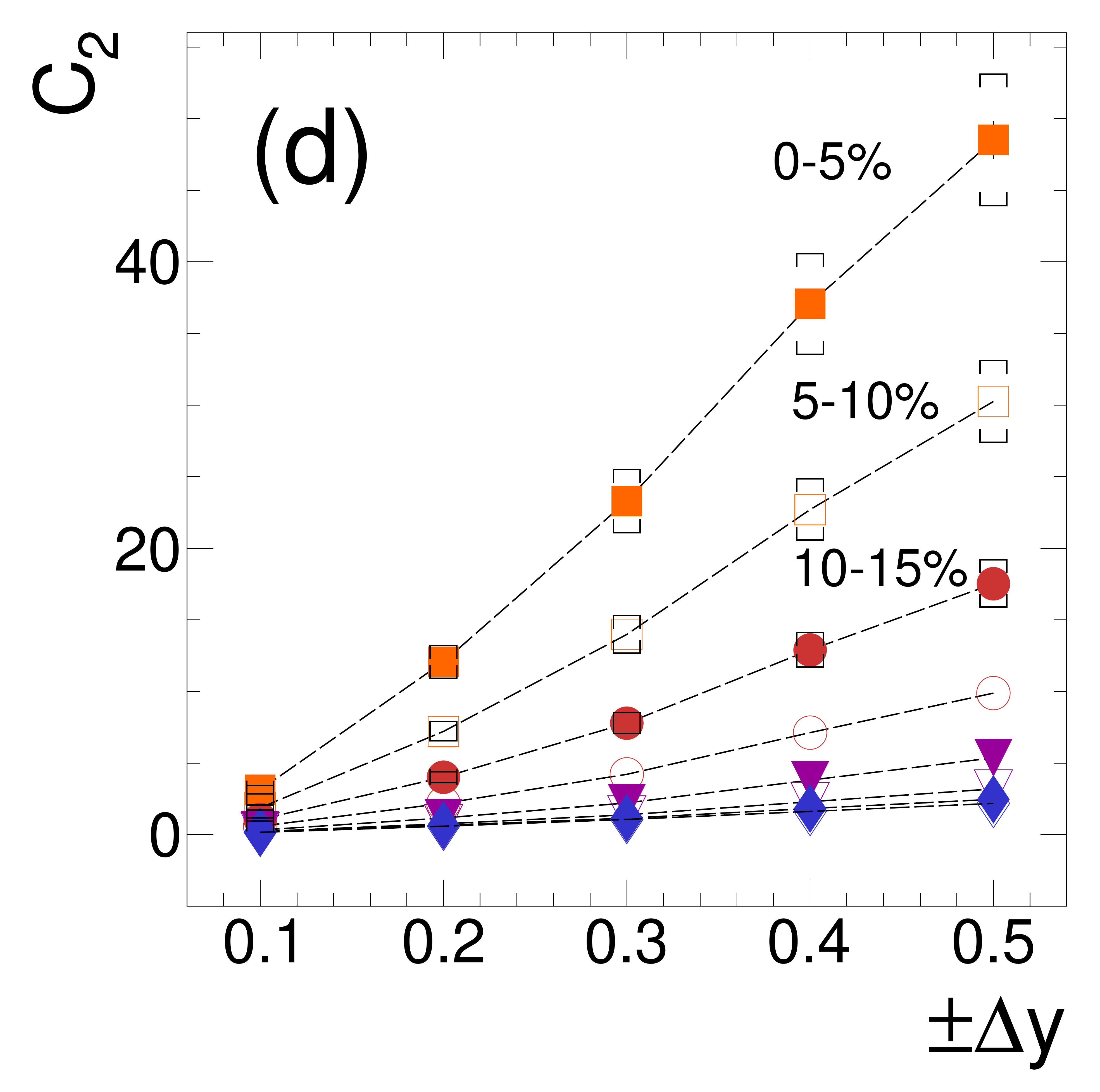}
     }
     \resizebox{1.0\linewidth}{!} {
       \includegraphics{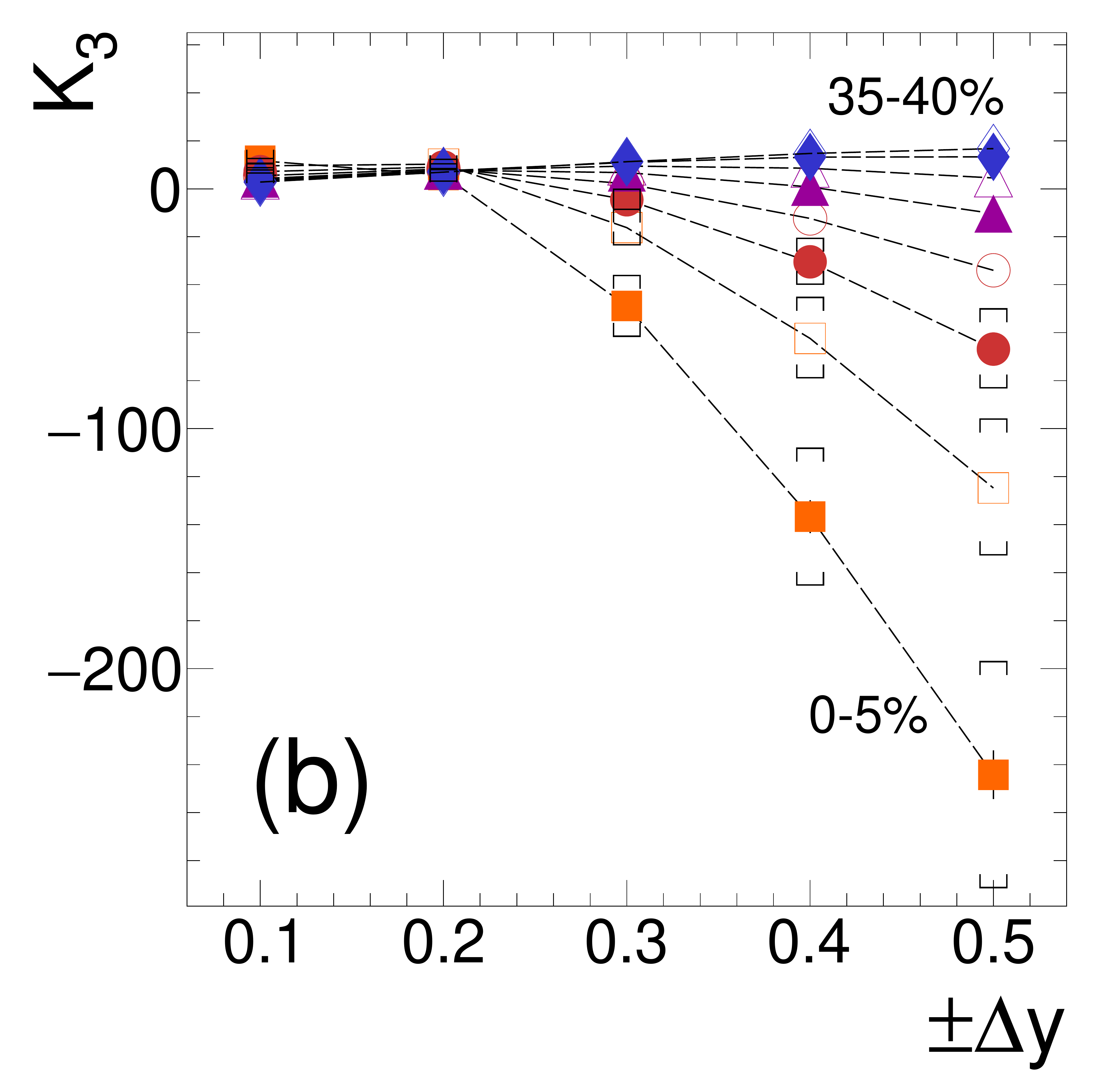}
       \includegraphics{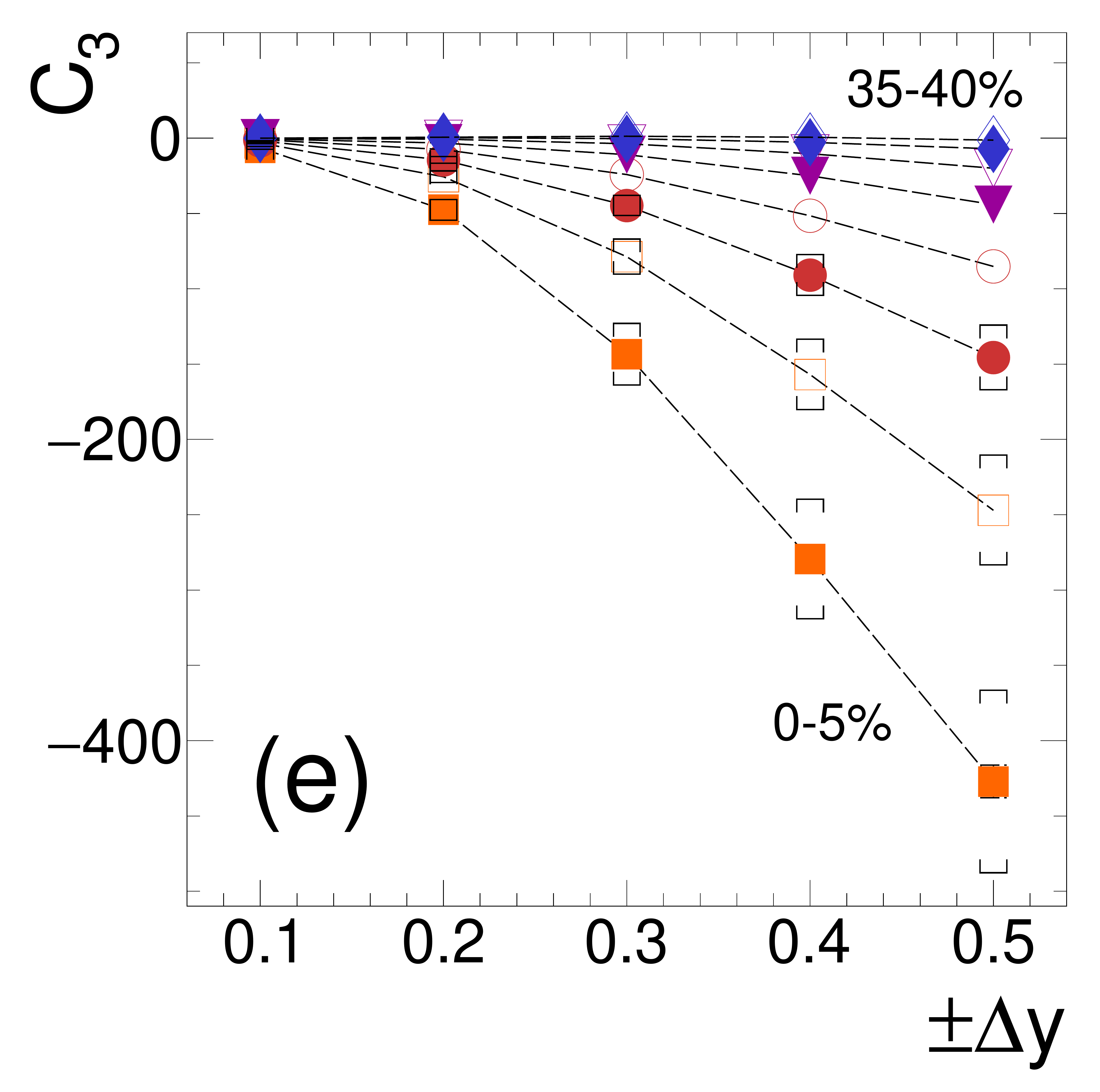}
     }
     \resizebox{1.0\linewidth}{!} {
       \includegraphics{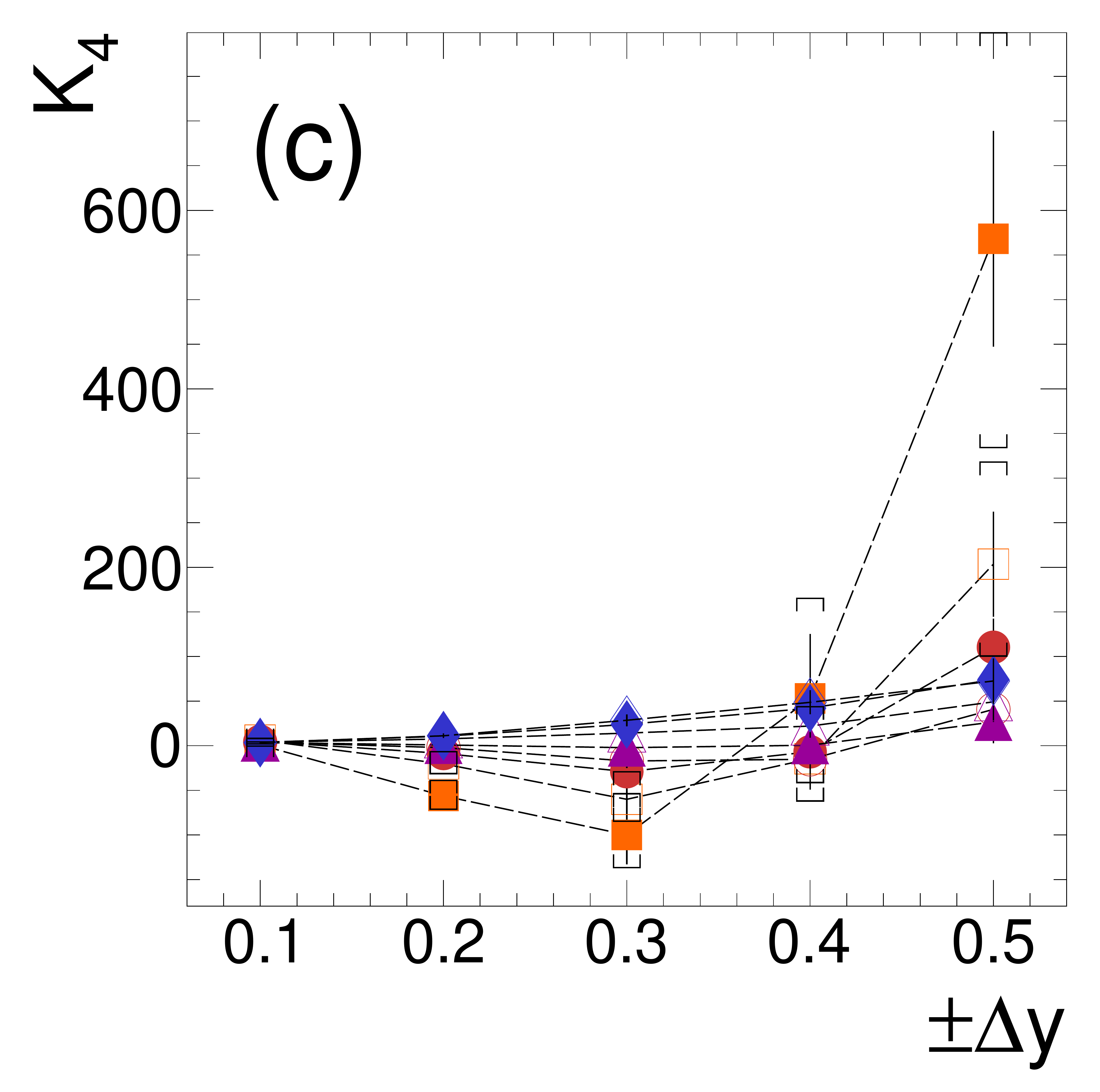}
       \includegraphics{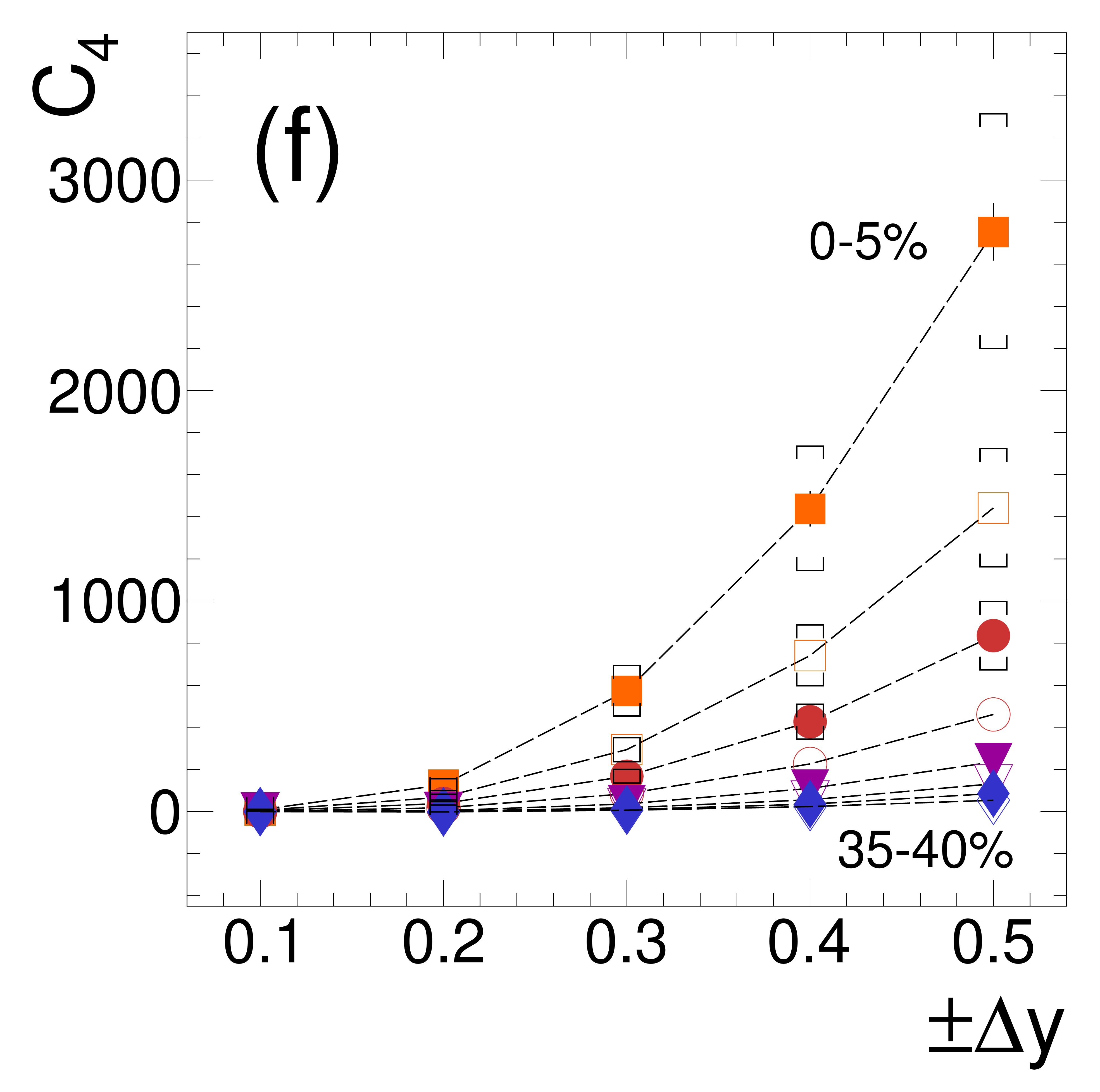}
     }
  \end{center}
  \vspace*{-0.2cm}
  \caption[] {(Color online) Au+Au data: Efficiency and N2LO volume corrected proton cumulants $K_n$ (a -- c)
              and correlators $C_n$ (d -- f), shown as a function of the rapidity
              bin $\pm \Delta y$, i.e.\ a phase space $y \in y_0 \pm \Delta y$ and $0.4 \leq p_t \leq 1.6$~\gevc.
              Dashed lines connect data points for guidance only, bars are statistical errors, and cups indicate
              full systematic errors.  To not clutter the graphs, the latter are shown only for a few of the
              centrality selections, i.e.\ 0 -- 5\%, 5 -- 10\%, and 10 -- 15\%.
             }
  \label{fig:KnCnvsDy}
\end{figure}


\subsection{Correlators}

As pointed out in Refs.~\cite{Ling2016,Bzdak2017str,Bzdak2017rap}, the essential information contained in particle number cumulants 
is related to the physics of multi-particle correlations, the underlying mechanism of which we hope to unravel.
Indeed, the cumulants of a given order $n$ contain contributions from multi-particle correlations of all orders up to $n$.
The $n$-particle correlators $C_n$ -- also called factorial cumulants or connected cumulants or sometimes correlation functions --
can be obtained straightforwardly from the cumulants $K_n$ via Eq.~\eqref{eq:momfacmom}.  Making use of this general expression,
we explicitly write down the correlators up to the 4\textsuperscript{th} order:

\begin{equation}
  \begin{split}
  C_1 &= K_1 \,,\\
  C_2 &= K_2 - K_1 \,,\\
  C_3 &= K_3 - 3 K_2 + 2 K_1 \,,\\
  C_4 &= K_4 -6 K_3 + 11 K_2 -6 K_1\,.
  \end{split}
\label{eq:correlators}
\end{equation}

\begin{figure}[bt]
  \begin{center}
     \resizebox{0.7\linewidth}{!} {
       \includegraphics{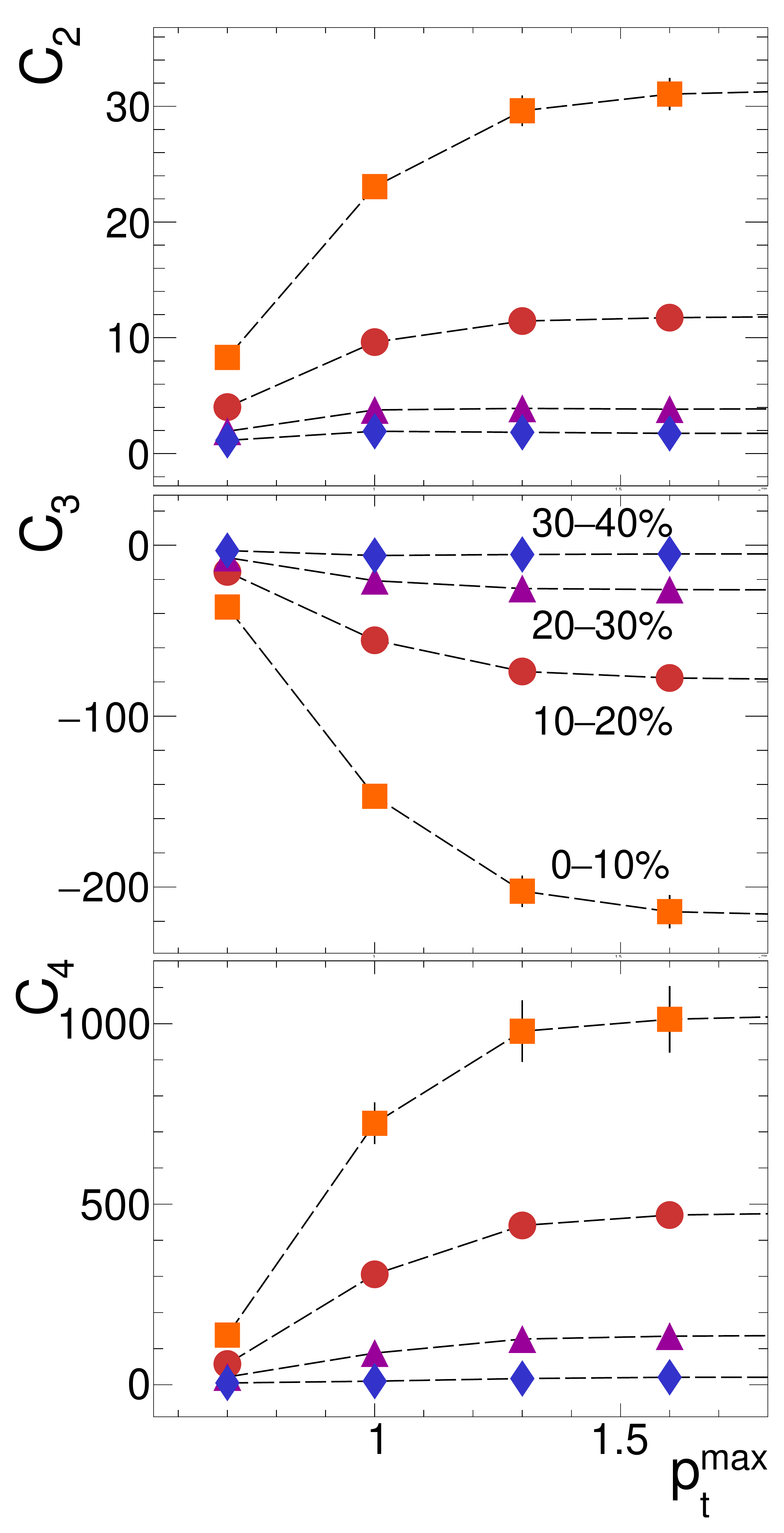}
     }
  \end{center}
  \vspace*{-0.2cm}
  \caption[] {(Color online) Au+Au data: Efficiency and N2LO volume corrected proton number correlators $C_n$ in the
              rapidity bin $y \in y_0 \pm 0.4$ as a function of the upper cut on transverse momentum $p_{t}^{\textrm{max}}$,
              shown here for 10\% centrality bins.  Error bars are statistical only.  Saturation of the correlators
              is reached for $p^{\textrm{max}}_t \cong 1.5$~\gevc.
             }
  \label{fig:CnvsPtmax}
\end{figure}

\noindent
To illustrate their differences, Fig.~\ref{fig:KnCnvsDy} displays side by side the full set of measured proton cumulants (left column)
and correlators (right column) as a function of the selected rapidity bin width $\Delta y$.  Also, in  Fig.~\ref{fig:CnvsPtmax} the dependence
of the correlators $C_n$ on the upper transverse momentum cut $p_t^{max}$ is shown, demonstrating the saturation of $C_n$ for a maximum
momentum of $p_t \simeq 1.5$~\gevc;  this is likely due to the proton yield fading quickly with increasing $p_t$.

\begin{figure*}[!hbt]
  \begin{center}
     \resizebox{1.0\linewidth}{!} {
       \includegraphics{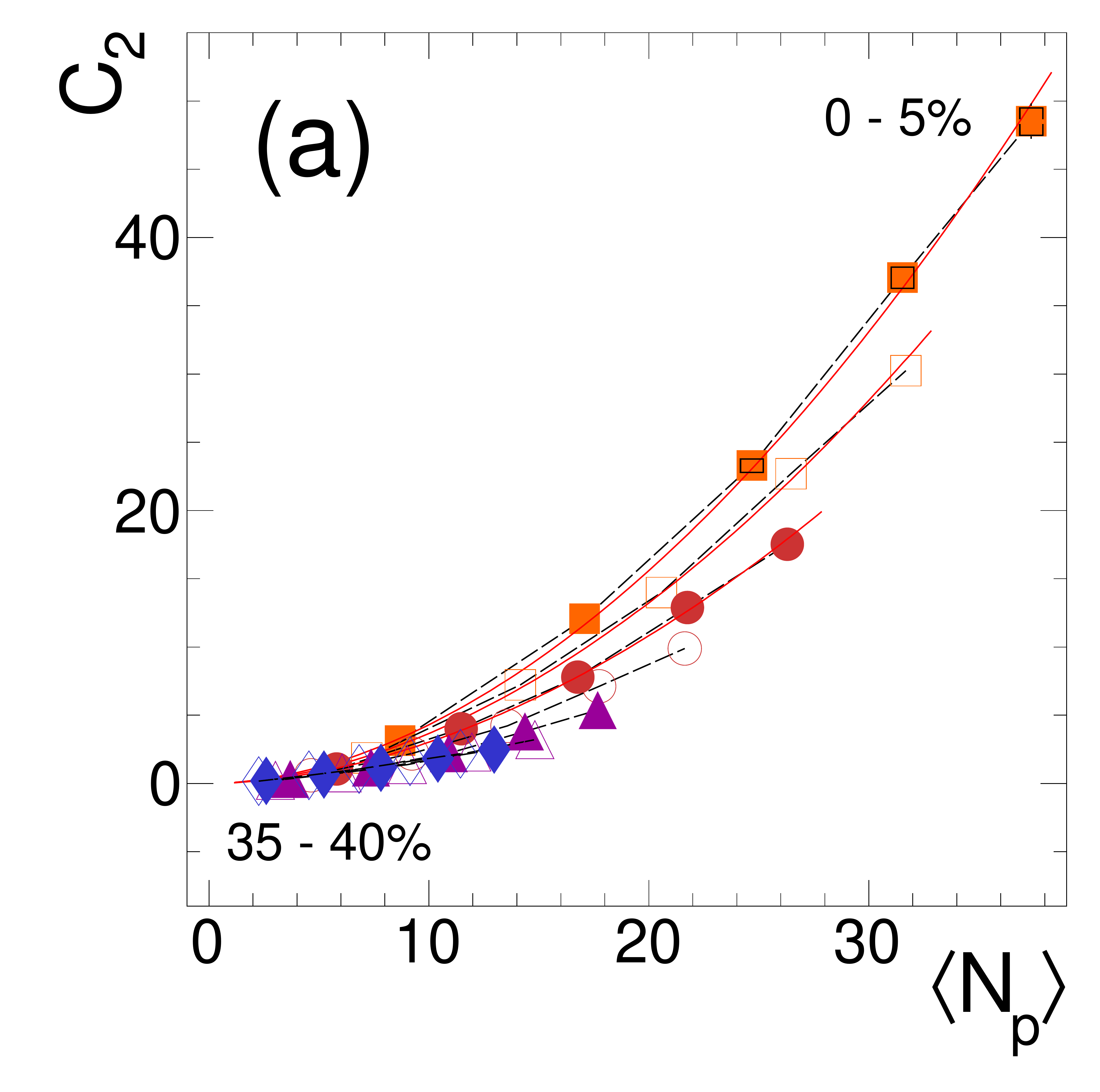}
       \includegraphics{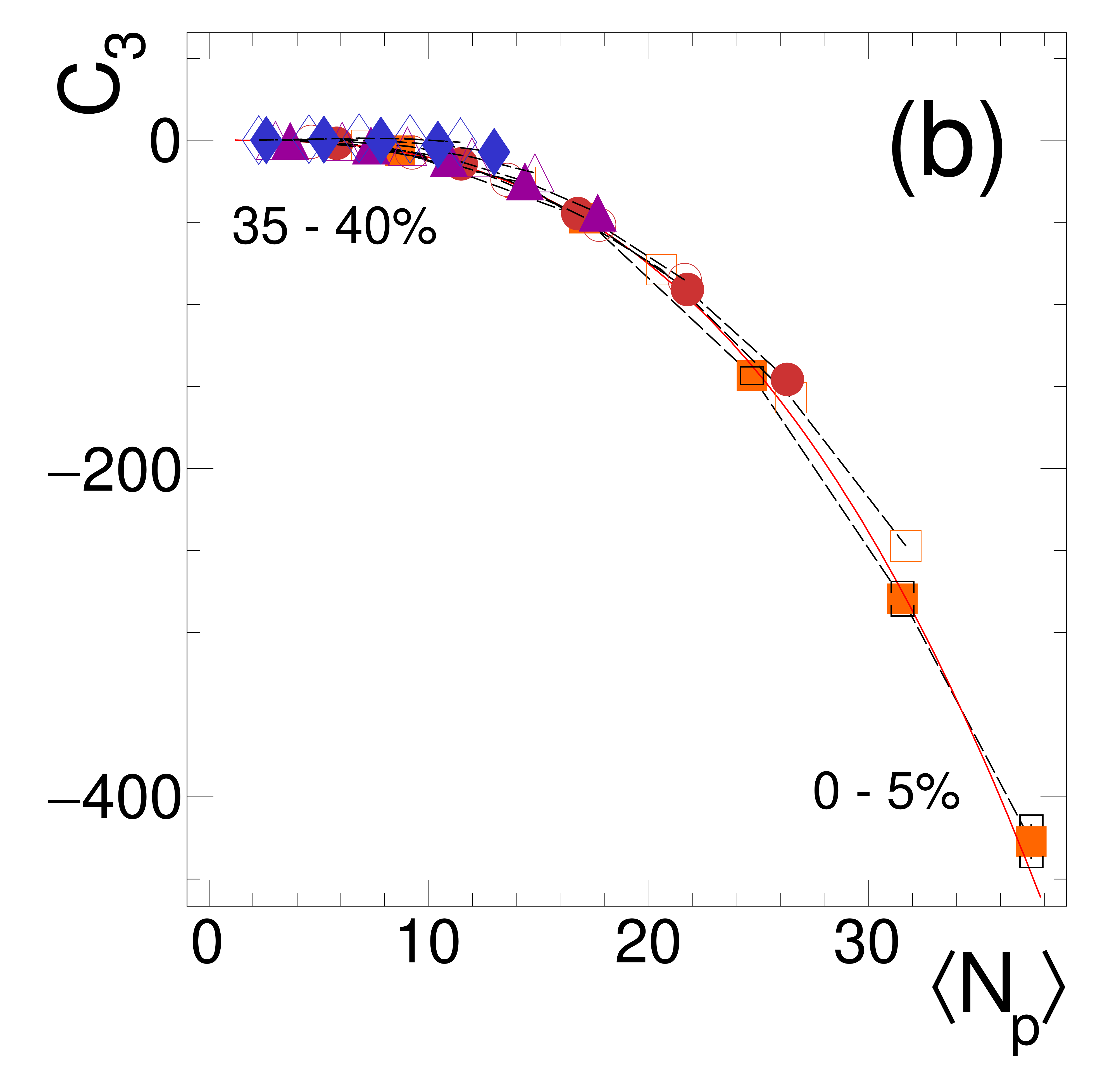}
       \includegraphics{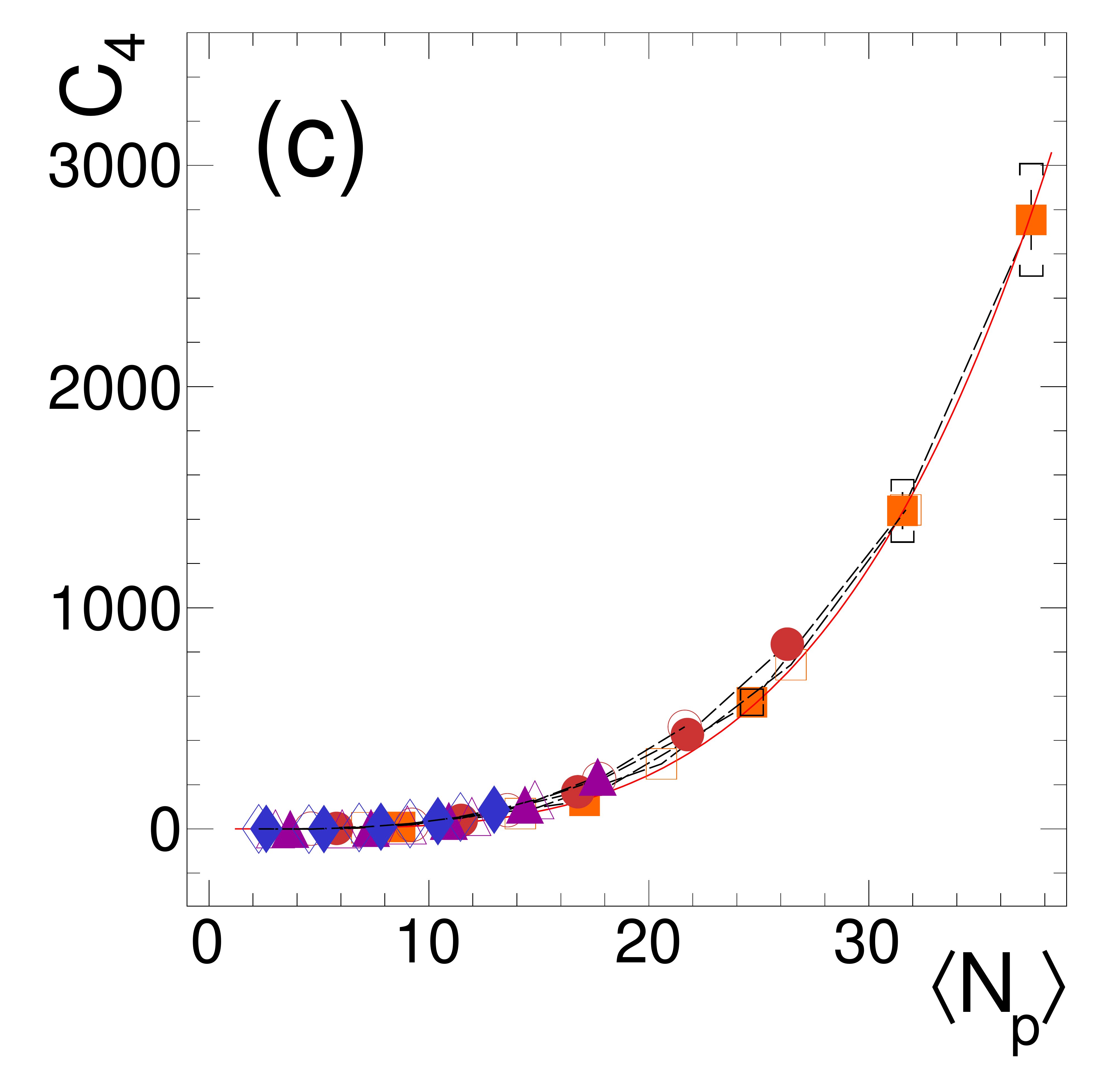}
     }
  \end{center}
  \vspace*{-0.2cm}
  \caption[] {(Color online) Au+Au data: Efficiency and N2LO volume corrected proton correlators $C_n$ as a function
              of the mean number of protons $\langle N_{p} \rangle$ within the selected phase-space bin, $y \in y_0 \pm \Delta y$
              ($\Delta y = 0.1, \ldots, 0.5$) and $0.4 \leq p_t \leq 1.6$~\gevc, and for eight centrality selections.
              Error bars on data are statistical, cups delimit systematic uncertainties (shown, for clarity, only on the 0 -- 5\% selection).
              Black dashed lines connect the data points in a given centrality selection and red solid curves are power-law fits
              $C_n \propto \langle N_{p} \rangle^{\alpha}$.  Only a few of the fit curves are actually presented,
              however all adjusted values of parameter $\alpha$ are listed in Table~\ref{tab:alpha}.
             }
  \label{fig:CnvsN}
\end{figure*}

General arguments regarding the nature of multi-particle correlations have been discussed in Refs.~\cite{Ling2016,Bzdak2017rap}.
In particular, in \cite{Ling2016} it was argued that the scaling of $C_n$ with the mean number of particles emitted into a
given phase-space bin $\Delta y$ depends on the range $\Delta y_{\textrm{corr}}$ in momentum space of these multi-particle correlations.
Two regimes were considered:  first, when very short-range correlations dominate, i.e.\ $\Delta y_{\textrm{corr}} \ll \Delta y$, one expects
a linear scaling $C_n \propto \Delta y$; second, when long-range correlations are important, that is $\Delta y_{\textrm{corr}} \gg \Delta y$,
one expects a $C_n \propto (\Delta y)^n$ scaling of the correlators.  However, the rationale underlying these considerations is given
by the regime of heavy-ion collisions at RHIC and LHC where the number of detected particles is proportional to $\Delta y$.
At low bombarding energies, the $dN/dy$ distribution is typically bell shaped and it is better to discuss the $C_n$ scalings directly
in terms of the number of particles emitted into $\Delta y$.  Hence, the two scaling regimes become $C_n \propto N$
for short-range correlations and $C_n \propto N^n$ for long-range.  This more adequate representation of the proton
correlators as a function of mean number of protons $\langle N_{p} \rangle$ is shown in Fig.~\ref{fig:CnvsN}, together with
power-law fits $C_n(N) = C_0 \; N^{\alpha}$, where the exponent $\alpha$ and the normalization constant $C_0$ are fit parameters.
The exponents resulting from these fits are listed with their error\footnote{This error is mostly statistical as $\alpha$ was found
to be very robust against systematic effects of types A, B and, C.} for all 5\% centrality selections in Table~\ref{tab:alpha}.
One can see that, for the most central events, the values of $\alpha$ are approaching $n$ for all $C_n$.  This suggests that
a setting close to the second scenario seems to be realized, that is, long-range correlations dominate the correlators
in Au+Au collisions at \sqrtsNN\ = 2.4~\gev.  In other words, $\Delta y_{\textrm{corr}}$ is of the same order or larger than the
accepted rapidity range of $y_0 \pm 0.5$, that is $\Delta y_{\textrm{corr}} \geq 1$.  \textcolor{black}{It is also interesting
to note that $C_2$, unlike the higher orders, displays a somewhat more intricate scaling with $\langle N \rangle$, i.e.\ a drop
of the exponent with the centrality of the collision.  The meaning of this behavior remains presently unclear to us.}

\begin{table}
\caption{Results of the power-law fits to the proton correlators shown in Fig.~\ref{fig:CnvsN} using
         $C_n \propto \langle N_p \rangle^{\alpha}$.  The fit parameter $\alpha$ and its statistical error
         are listed for 5\% centrality selections.  The few instances where the fit did not converge to
         a meaningful result are indicated by a dash.  Systematic errors on $\alpha$ are small, typically
         smaller than the statistical error listed. 
        }
 \label{tab:alpha}
 \vspace*{3mm}
 \setlength{\extrarowheight}{2pt}  
 \setlength{\tabcolsep}{6pt} 
 \begin{tabular}{c c c c}
  \hline \hline
  Centrality & $\alpha[C_2]$ & $\alpha[C_3]$ & $\alpha[C_4]$ \\
  \hline
  0 - 5\%   & $1.86 \pm 0.04$ & $2.84 \pm 0.05$ & $3.89 \pm 0.14$\\
  ~5 - 10\% & $1.85 \pm 0.04$ & $2.85 \pm 0.05$ & $3.75 \pm 0.13$\\
  10 - 15\% & $1.84 \pm 0.05$ & $2.80 \pm 0.06$ & $3.66 \pm 0.14$\\
  15 - 20\% & $1.82 \pm 0.07$ & $2.83 \pm 0.09$ & $3.72 \pm 0.22$\\
  20 - 25\% & $1.78 \pm 0.09$ & $2.95 \pm 0.15$ & $4.11 \pm 0.44$\\
  25 - 30\% & $1.67 \pm 0.10$ & $3.44 \pm 0.40$ & $4.46 \pm 0.82$\\
  30 - 35\% & $1.59 \pm 0.10$ & - & $4.76 \pm 1.36$\\
  35 - 40\% & $1.55 \pm 0.11$ & - & - \\
  \hline \hline
 \end{tabular}
\end{table}

The reason for these unexpectedly strong long-range correlations is not obvious.  We can only speculate that
they could be caused by collective phenomena (e.g.\ flow fluctuations) or, if they are of truly critical nature, they might signal
the close-by liquid-gas phase transition \cite{Siemens1983,Borderie2019}.  Indeed, in a recent model study using a hadron resonance gas
with minimal van der Waals interactions \cite{Vovchenko2017,Vovchenko2018,Poberezhnyuk2019b}, the fluctuation signal characterizing
the liquid-gas endpoint was found to persist over a surprisingly large part of the phase diagram.  A dynamic description is, however,
still missing and it remains open how exactly the correlations observed at freeze-out in momentum space do relate to the spatial
correlations build up in the initial state of the collision and/or during the expansion of the resulting fireball.  Note that some of
these issues have also been addressed by the authors of \cite{Ling2016,Asakawa2016} when discussing the observable signals of critical
fluctuations.  The authors of \cite{Bzdak2019} have recently demonstrated in more general terms that the observed behavior of the
correlators $C_n$ emerges naturally when the measured particle distribution results from the superposition of two event classes,
both uncorrelated -- Poisson or binomial -- but with distinct mean multiplicities.  Even a small contamination of order $10^{-3}$ of the
main event class would lead to large values of the combined $C_n$.  It is therefore very important to put limits on possible instrumental
origins of such a contaminant, as we have discussed in Sec.~\ref{Sec:Exp} (see in particular Table~\ref{tab:nuisance}).

Note also that the proton number cumulants as well as their ratios are expected to be affected by baryon-number
conservation effects \cite{Bzdak2013,BraunMunzinger2019,Pruneau2019,BraunMunzinger2020}.  Such effects have indeed been observed in LHC
data \cite{Rustamov2017a,BraunMunzinger2019} as deviations from the Skellam distribution expected in the grand canonical limit.
Introducing an acceptance factor $a = \langle N_p \rangle / \langle N_B \rangle$ as the ratio of the mean number of protons
accepted in a given phase-space bin and the corresponding total number of baryons, the authors of \cite{BraunMunzinger2019,BraunMunzinger2020}
could express the constraint of baryon-number conservation on the particle cumulants as function of $a$.  In the low-energy regime,
where the number of antibaryons drops to zero, the expected deviations of the cumulant ratios from the pure Poisson baseline, 
due to global baryon-number conservation, are then given by

\begin{equation*}
  \begin{split}
    K_2/K_1 =& 1 - a \;, \\
    K_3/K_2 =& 1 - 2 a \;, \\
    K_4/K_2 =& 1 - 6 a (1 - a) \;.
  \end{split}
\label{eq:baryonConservation}
\end{equation*}

\noindent
All three ratios are reduced by canonical suppression and our proton data, where $a$ before efficiency correction is in the
most central bin between 0.023 (for $y \in y_0 \pm 0.2$) and 0.051 (for $y \in y_0 \pm 0.5$), may be affected accordingly.
\textcolor{black}{Although $a$ turns out to be rather small in our case,}
baryon-number conservation as well as similar constraints due to electric charge conservation will have to be accounted for
in future model calculations.  The direct effect on the particle correlators $C_n$ has been discussed in \cite{Bzdak2017str}
where it was found to display a scaling behavior very different from the one observed in Fig.~\ref{fig:CnvsN}.

\begin{figure*}[!t]
  \begin{center}
     \resizebox{0.75\linewidth}{!} {
       \includegraphics{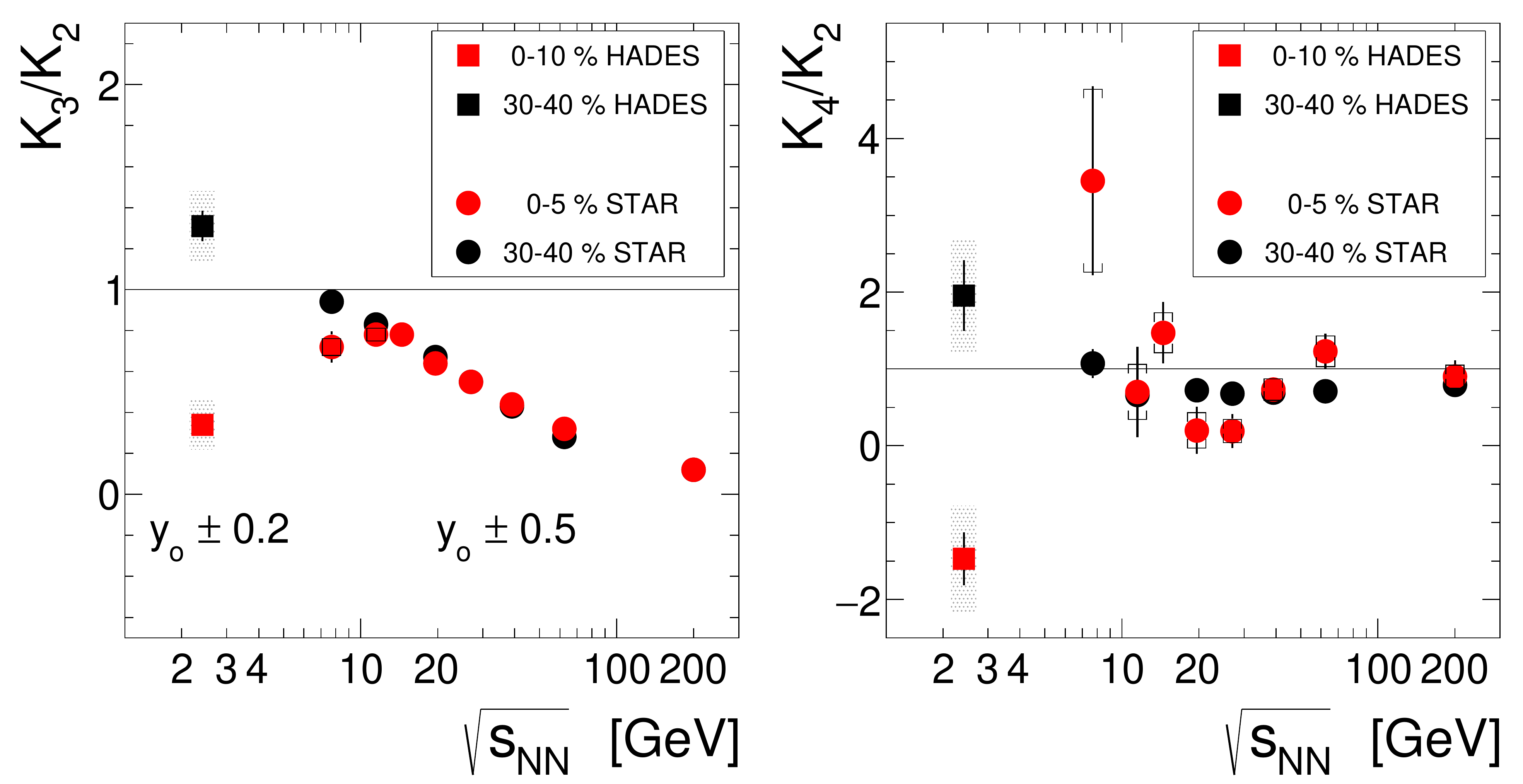}
     }
     \resizebox{0.75\linewidth}{!} {
       \includegraphics{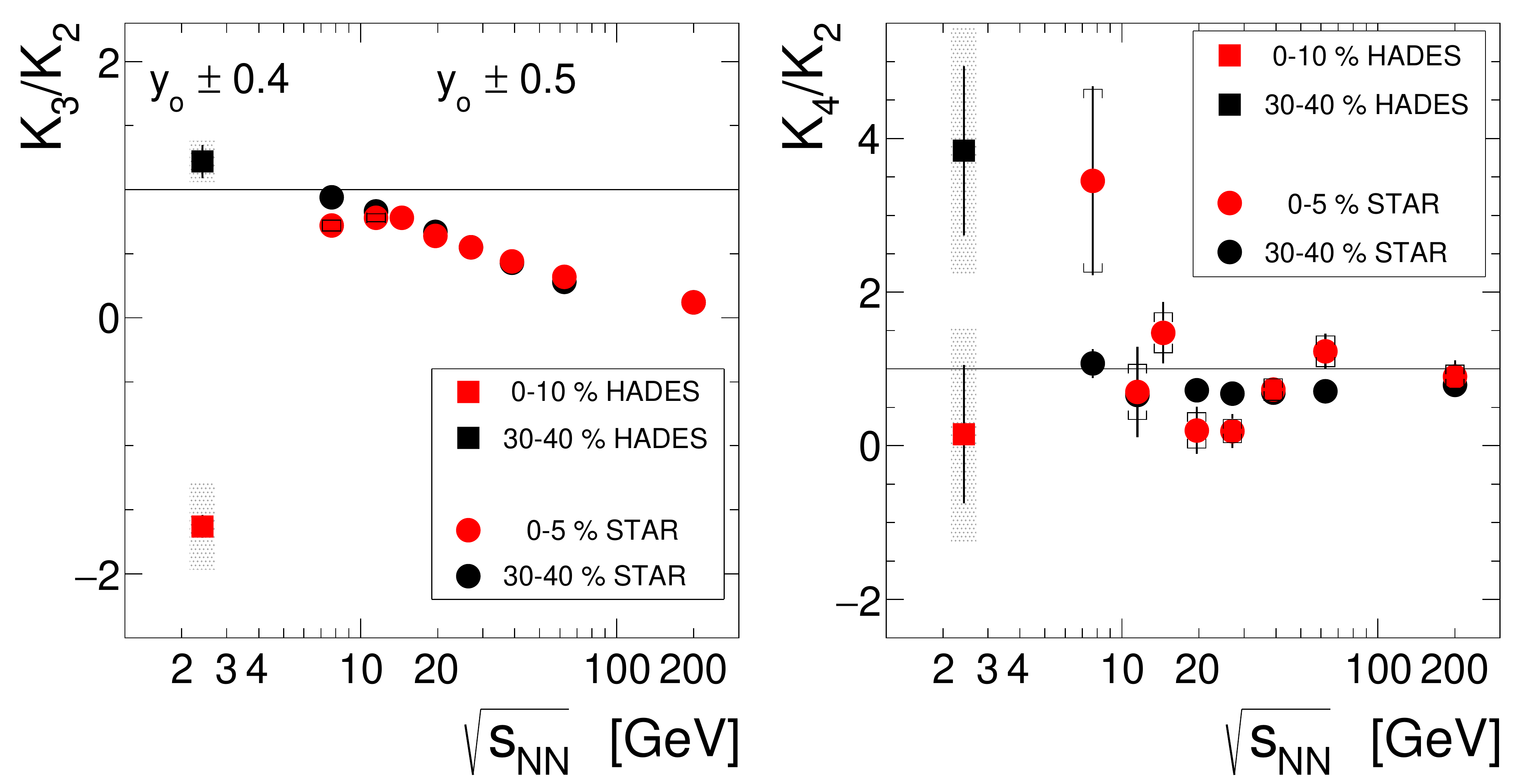}
     }
  \end{center}
  \vspace*{-0.2cm}
  \caption[] {(Color online) Au+Au data: Evolution of the scaled cumulants $K_n/K_2$ as a function of center-of-mass energy
              \sqrtsNN\ for two centrality bins (0 -- 10\% or 0 -- 5\%, red symbols, and 30 -- 40\%, black symbols) and shown as 
              $\gamma_1 \times \sigma$ (left column) and $\gamma_2 \times \sigma^2$ (right column).
              STAR data \cite{Luo2015b,Adam2020a} is shown for a $y_0 \pm 0.5$ phase-space bite,
              HADES data for $y_0 \pm 0.2$ (top row) and $y_0 \pm 0.4$ (bottom row), respectively.
              Vertical bars are statistical errors; full systematic errors are shown for the HADES data as shaded bars
              and for the STAR data points as cups.
             }
  \label{fig:HADESvsSTAR}
\end{figure*}

At higher energies a beam energy scan has been conducted by the STAR collaboration at RHIC for \sqrtsNN\ = 7.7 -- 200~\gev\
and net-proton number fluctuations have been analyzed and published \cite{Aggarwal2010,Adamczyk2014a,Adamczyk2014b,Luo2015b,Adam2020a}.
In Fig.~\ref{fig:HADESvsSTAR} we extend the STAR systematics of net-proton cumulant ratios $\gamma_1 \times \sigma$ and
$\gamma_2 \times \sigma^2$ with our low-energy point at \sqrtsNN\ = 2.4~\gev.  The STAR analysis was done for all beam energies in
the rapidity range covered by their TPC and time-of-flight detector, i.e.\ $y \in y_0 \pm 0.5$.  It is not at all clear how
the interplay between fluctuation signals from the central fireball and from spectator matter changes with energy, and how this affects
the measurements in rapidity intervals of a given size, in particular at low beam energies where the proton rapidity distribution is
more bell shaped and much narrower than at RHIC energies.  Therefore, we present the comparison with HADES data for two choices of the
rapidity bite: $y \in y_0 \pm 0.2$ and $y \in y_0 \pm 0.4$.  We prefer $\pm 0.4$ over the $\pm 0.5$ choice because of justified fears
that the latter, larger range contains sizable contributions from the abraded spectator matter.  \textcolor{black}{A naive estimate,
based on a Fermi gas nucleon momentum of $p_F = 0.27$~\gevc, leads indeed to a safe rapidity region of about 0.28 -- 1.20
(or $y_0 \pm 0.46$) for our bombarding energy;  this is also born out by transport calculations done with the IQMD, UrQMD, and HSD models.}
\textcolor{black}{As shown in Fig.~\ref{fig:HADESvsSTAR}, for both presented choices of the rapidity interval and for both centralities,
the HADES data smoothly extend the $K_3/K_2$ trend observed by STAR towards lower \sqrtsNN.  This seems to be true as well for the
$K_4/K_2$ trend in semi-peripheral events, whereas in the most central events, the HADES data suggest a sharp decrease of the
fourth-order ratio with respect to its value at the lowest STAR energy.  One has to keep in mind, however, that the sizable gap
remaining in the excitation function between \sqrtsNN\ = 2.4 and 7.7~\gev\ will have to be covered by experiment before firm
conclusions can be drawn.}

\section{Summary and outlook}

To summarize, we have investigated with HADES proton multiplicity fluctuations in \sqrtsNN = 2.4~\gev\ Au+Au collisions
up to 4\textsuperscript{th} order.  In this context, we have done an in-depth investigation and comparison of
various efficiency correction schemes, and we have in the end opted to apply a highly granular event-by-event correction to the
measured proton cumulants to account for both, phase-space interval dependent and track-density related, efficiency changes.  Furthermore,
guided by transport model calculations, we have extended the procedure proposed in the literature for stripping off volume fluctuations
by including the higher-order correction terms required in the low-energy regime where HADES operates.  The resulting fully corrected
proton cumulants and correlators are presented and discussed as a function of centrality and phase-space acceptance.  When only
a very narrow rapidity bin is selected, we find that the observed proton distributions are, as expected, close to Poisson.
However, this behavior changes dramatically when the acceptance opens up and multi-particle correlations set in. In particular,
from the dependence of the correlators on the number of emitted protons, we conclude that our results are dominated by rather
long-range ($\Delta y_{\textrm{corr}} \geq 1$) correlations, strongly positive in second and fourth order, but negative in third order.  Why and how
these correlations in momentum space at freeze-out build up from spatial correlations in the initial state and/or expansion phase
of the fireball, remains to be investigated by theory.  When joined with the STAR results \cite{Luo2015b,Adam2020a} obtained in the first
RHIC beam-energy scan, our data allow to extend the excitation function of net-proton cumulants in central Au+Au collisions to low energies.
While the present data show a rather smooth trend for $K_3/K_2$ with \sqrtsNN, they indicate a distinctive change of sign of
$K_4/K_2$ when moving from RHIC to SIS18 energies.  Again, the interpretation of these observations requires input from
advanced quantitative calculations, e.g.\ hydro or transport models including phase boundaries.

An interesting avenue to follow next is to evaluate fluctuations of bound protons by including nuclear cluster production
-- foremost deuteron, triton, and He isotopes -- into the analysis.  Indeed, in order to elucidate the origin of long-range
correlations observed for free protons, the role played by protons bound in clusters might turn out to be decisive, as the latter
represent at \sqrtsNN\ = 2.4~\gev\ about 40\% of the total number of protons emitted from the fireball \cite{Szala2019}.
Also, the HADES collaboration has recently done a high-statistics measurement in \agag\ collisions at the two bombarding energies
of 1.23 and 1.58\agev.  These data provide the opportunity to study both the system size and, to some extent, the energy
dependence of various fluctuation signals.  We are furthermore looking forward to see results from the 2\textsuperscript{nd}
beam-energy scan at RHIC \cite{Yang2017}, which will provide data at energies reaching down to \sqrtsNN\ = 3~\gev.
These, when combined with our measurements, will allow to map fluctuation signals across the QCD phase diagram.
Also, a dedicated beam-energy scan in the \sqrtsNN\ = 2 -- 2.5~\gev\ region, to be done at SIS18, could add vital
information with regard to the liquid-gas phase transition.  Further in the future, the heavy-ion experiments CBM at
FAIR and MPD at NICA will produce comprehensive data sets of extremely large statistics and will thus allow to extend
the present studies to much higher precision and to higher orders.\\


\begin{acknowledgements}
The HADES collaboration thanks Adam Bzdak, Xiaofeng Luo, Mazakiyo Kitazawa,
Volker Koch, and Nu Xu for elucidating discussions.
We also thank Yvonne Leifels for providing the clusterized IQMD events to us.
We gratefully acknowledge support by the following grants:
SIP JUC Cracow, Cracow (Poland), National Science Center, 2016/23/P/ST2/040 POLONEZ, 
2017/25/N/ST2/00580, 2017/26/M/ST2/00600; TU Darmstadt, Darmstadt (Germany), VH-NG-823, DFG GRK 2128,
DFG CRC-TR 211, BMBF:05P18RDFC1; Goethe-University, Frankfurt(Germany), HIC for FAIR (LOEWE), BMBF:06FY9100I,
BMBF:05P12RFGHJ, GSI F$\&$E; Goethe-University, Frankfurt (Germany) and TU Darmstadt, Darmstadt (Germany),
ExtreMe Matter Institute EMMI at GSI Darmstadt;
TU M\"unchen, Garching (Germany), MLL M\"unchen, DFG EClust 153, GSI TMLRG1316F, BMBF 05P15WOFCA, 
SFB 1258, DFG FAB898/2-2; NRNU MEPhI Moskow, Moskow (Russia), in framework of Russian
Academic Excellence Project No. 02.a03.21.0005 (27.08.2013), RFBR No. 18-02-00086, MSHE
of Russia No. 0723-2020-0041; JLU Giessen, Giessen (Germany), BMBF:05P12RGGHM; IPN Orsay, 
Orsay Cedex (France), CNRS/IN2P3; NPI CAS, Rez, Rez (Czech Republic), 
MSMT LM2015049, OP VVV CZ.02.1.01/0.0/0.0/16 013/0001677, LTT17003.
\end{acknowledgements}

\appendix  

\section{Non-binomial efficiencies: The occupancy model}
\label{Sec:Occupancy}

As discussed in section~\ref{Sec:Eff}, efficiency corrections to particle number cumulants are usually done \cite{Bzdak2015}
assuming the efficiency to be binomial, i.e.\ assuming that the detection processes of multiple particles in any given event are
independent.  This can be described best with the help of a dichromatic urn model, where successive draws from the urn stand for
particle detection processes: a white ball drawn is taken as 'particle detected' and a black one as 'particle not detected'.
The initial state of the urn, i.e.\ the initial content of white and black balls, is chosen such that the ratio of the number
of white balls to the total number of balls corresponds to the single-particle detection efficiency.  How the state of the urn
evolves with successive draws depends on the chosen addition rule.  In the binomial case, the balls are drawn with replacement,
meaning that every drawn ball is placed back into the urn which thus does not change its state.  Successive draws are hence
independent from each other, i.e.\ the urn does not have memory.  The probability of obtaining $p$ white balls in $m$ draws or,
equivalently, of detecting $p$ out of $m$ emitted particles in an event is described by

\begin{equation}
  P(p; m) = \binom{m}{p} \; \epsilon^p (1 - \epsilon)^{m-p} \;,
\label{eq:A1}
\end{equation}

\noindent
where $\epsilon$ is the probability to detect a given particle and $\binom{m}{p}$ are binomial coefficients. 
By expanding the $(1-\epsilon)$ term in Eq.~(\ref{eq:A1}) and averaging $P(p; m)$ over the distribution of emitted particles,
one finds a relationship between the average probability $P(p)$ to observe $p$ particles in the detector and the factorial
moments $F_n$ of the true particle distribution \cite{VDWerf1978}:

\begin{equation}
  P(p) = \sum^{\infty}_{m=p} \: (-1)^{m-p} \: \frac{F_m}{m!} \: \binom{m}{p} \: \epsilon^m  \,.
\label{eq:A2}
\end{equation}

\noindent
Using the definition of the observed factorial moments

\[ f_n = \sum^{\infty}_{p=n} \: P(p) \: p\, (p-1) \cdots (p-n+1) \]

\noindent
and inserting $P(p)$ from Eq.~\eqref{eq:A2}, one retrieves for the binomial efficiency model the well-known
relation between the measured $f_n$ and the true $F_n$ factorial moments \cite{Kirej2004,Bzdak2012,HeLuo2018}

\begin{equation}
  f_n = \epsilon^n \: F_n \,.
\label{eq:A3}
\end{equation}

\begin{figure}[!htb]
  \begin{center}
     \resizebox{0.9\linewidth}{!} {
       \includegraphics{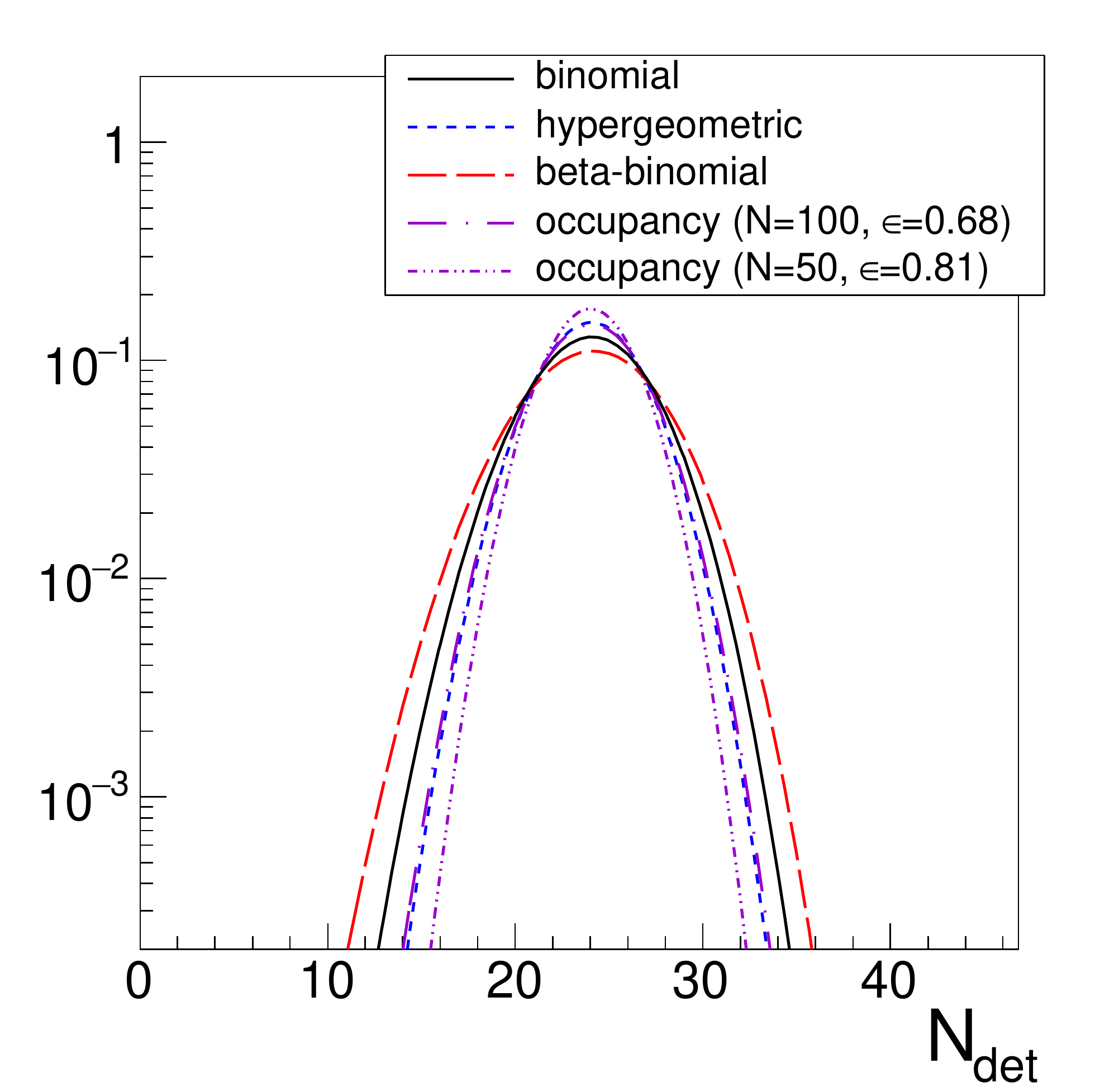} 
     }
  \end{center}
  \vspace*{-0.2cm}
  \caption[] {(Color online) Comparison of the normalized distribution of detected particles $N_{\textrm{det}}$ resulting from different
              efficiency models: binomial, hypergeometric, beta-binomial, and our occupancy model (see text).  In all cases,
              the input number of particles is $N_{\textrm{in}} = 40$ and the parameters of the models are adjusted to
              realize an average efficiency of $\langle \epsilon \rangle = 0.6$, resulting in $\langle N_{\textrm{det}} \rangle = 24$.
              Calculations done with the occupancy model are shown for two segmentations: $N = 100$, with a single-hit
              efficiency of $\epsilon = 0.68$, and $N = 50$, with $\epsilon = 0.81$.
             }
  \label{fig:A1}
\end{figure}

Real-life detectors are commonly designed with a finite occupancy; they can register only a limited number of particles
per given event and consequently their detection efficiency eventually decreases with ever increasing particle number.
This effect can be studied in simulations, and for HADES the efficiency drop was found to be of order 10\% - 15\% (see Sec.~\ref{Sec:Eff}).
Deviations from the binomial assumption have been discussed in \cite{Tang2013,Bzdak2016} and the authors of \cite{Bzdak2016} considered
in particular the hypergeometric and beta-binomial distributions.  Like the binomial distribution, these two distributions can also be
derived from a dichromatic urn model.  In the hypergeometric model, balls are drawn without replacement, i.e.\ they are not put back and
the urn state changes while it gets successively emptied of balls.  The resulting hypergeometric distribution of white balls obtained
per given number of draws is narrower than the binomial distribution (see Fig.~\ref{fig:A1}).  For the beta-binomial model, on
the other hand, balls are drawn with double replacement, meaning that for a white draw, two white balls are put back and for a
black draw, two black balls.  Again, the state of the urn changes, but the resulting beta-binomial distribution is now broader
than the binomial distribution (cf. Fig.~\ref{fig:A1}).  Although the properties of both of these adhoc models are well known,
their connection to physical phenomena playing a role in the actual detection process is not obvious.

Here we propose another urn model, known as the occupancy model \cite{Mahmoud2008}, that offers a more intuitive
connection with the behavior of detectors under particle bombardment.  In the occupancy model, balls are drawn with the
following replacement rule: A drawn black ball ('not detected') is just put back into the urn, but a drawn white ball ('detected')
is not put back, instead it is replaced by a back ball put to the urn.  The state of the urn changes as the number
of white balls gets depleted, realizing a decrease in efficiency.  In this picture, each white ball represents one
active detector module that turns inert or busy when hit by a particle and thus has to be replaced by a black ball to keep
constant the total solid angle covered by the device.  The occupancy model naturally applies to detectors segmented into a
finite number $N$ of modules of solid angle $\Omega$ each such that the total active solid angle covered is $\epsilon = N \Omega$.
Any given module can fire when hit by a particle, but only once, i.e.\ multiple hits of a module are not distinguishable from
single hits.  Following the approach presented in \cite{VDWerf1978, Bellia1984}, the binomial multi-hit detection probability
of Eq.~\eqref{eq:A1} transforms into

\begin{equation}
  P(p; m,N) = \binom{N}{p} \: \sum^p_{l=0} \: (-1)^{p-l} \: \binom{p}{l} \: \left(1 - \frac{N-l}{N} \: \epsilon \right)^{m} \;.
\label{eq:A4}
\end{equation}

\noindent
The resulting distribution of detected particles, shown in Fig.~\ref{fig:A1}, is wider than binomial.  Note that,
for a given $\epsilon$, the average number of detected particles decreases with respect to the binomial case, i.e.\ the
average efficiency is smaller than $\epsilon$.  In addition, the relation between factorial moments changes from Eq.~\eqref{eq:A3} into

\begin{equation}
\begin{split}
  f_n = \sum^{N}_{m=n} \: \frac{(-1)^m}{m!} \: F_m \: \sum^m_{p=n} \: (-1)^p \binom{N}{p} \: p \: (p-1) \cdots \\
     \cdots (p-n+1) \: \sum ^p_{l=0} (-1)^l \: \binom{p}{l} \: \left( \frac{N-l}{N} \: \epsilon \right)^m \,.
\end{split}
\label{eq:A5}
\end{equation}

\begin{figure}[thb]
  \begin{center}
     \resizebox{0.9\linewidth}{!} {
       \includegraphics{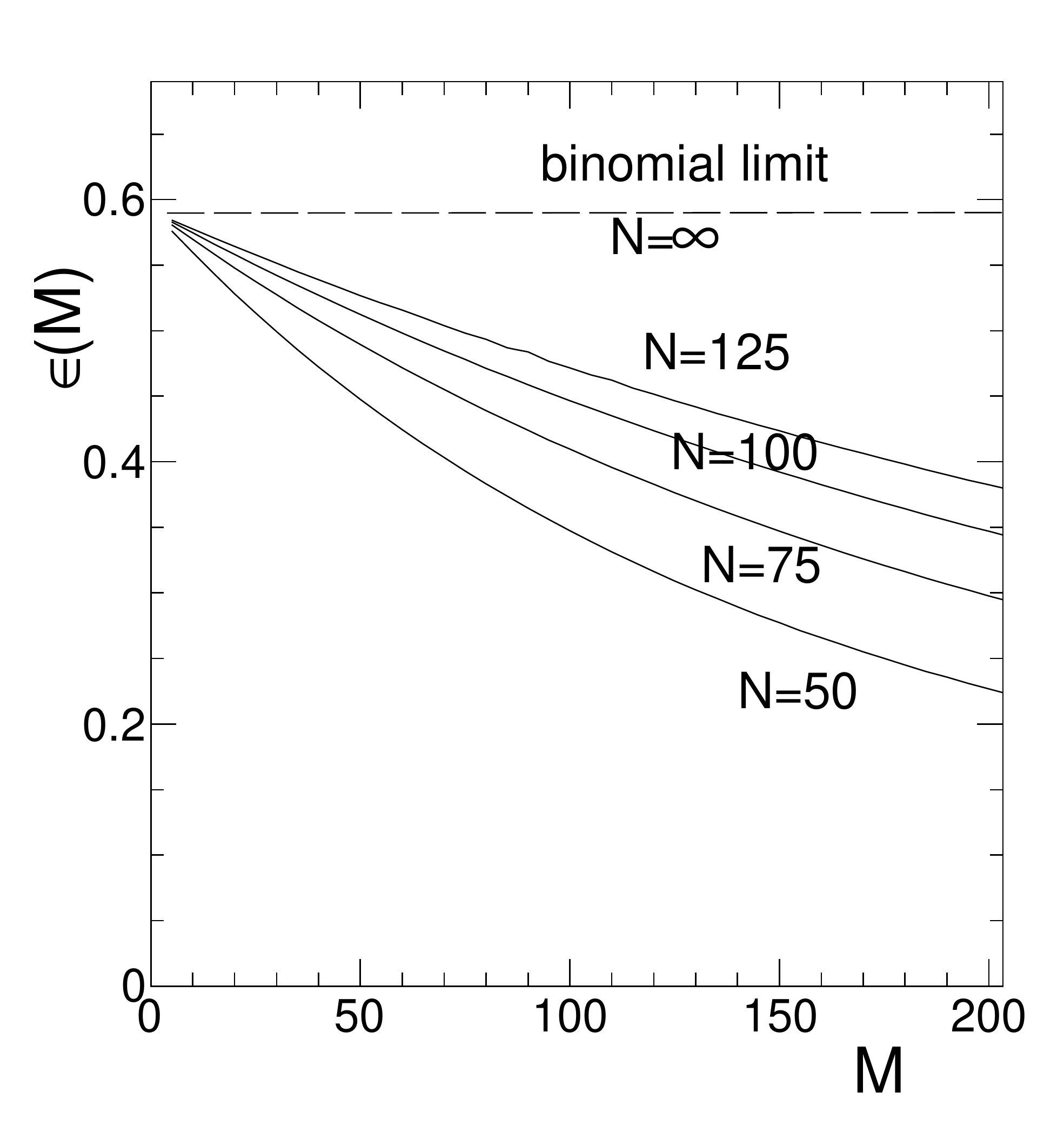} 
     }
  \end{center}
  \vspace{-0.7cm} 
  \caption{Non-binomial efficiencies of the occupancy model as a function of the particle multiplicity $m$ for
   $\epsilon \equiv N \Omega = 0.59$ and different values of $N$.
}
\label{fig:A2}
\end{figure}

\noindent
Unlike Eq.~\eqref{eq:A3},  Eq.~\eqref{eq:A5} is not anymore diagonal, i.e.\ an observed factorial moment of order $n$
depends on all true factorial moments $F_m$ of orders $n \leq m \leq N$.  Solving this system of equations for large $N$
requires eventually a truncation at some sufficiently high order.  We found, furthermore, that calculations
with Eq.~\eqref{eq:A5} need to be done with high numerical precision (at least 64 bit, if not 128 bit floating-point arithmetic)
in order to produce reliable results.  This had already been pointed out in \cite{Ockels1978} where also a faster converging
expansion of the hit probabilities in terms of cumulants was proposed.

Figure~\ref{fig:A2} illustrates the drop of the effective efficiency $\epsilon_{\textrm{eff}} = \epsilon(m)$ with particle multiplicity $m$
and its dependence on the segmentation $N$ of the detector;  for $N = \infty$, the binomial case is recovered.
For a ''continuous'' detector like HADES, a strict hardware segmentation into $N$ distinct modules is not realized,
but an effective segmentation $\tilde{N}$ can still be introduced, with $\tilde{N}$ and $\Omega$ adjusted to describe
the observed (or simulated) average efficiency behavior $\langle \epsilon_{\textrm{eff}} \rangle = \epsilon(\tilde{N}, \Omega)$.
Such an adjustment to IQMD simulated proton moments is also shown in Fig.~\ref{fig:effiqmddy02}.

\section{NLO volume fluctuation corrections}
\label{Sec:NL}

Here we list once more the reduced particle number cumulants $\tilde{\kappa}_n$ with all slope-related, i.e.\ NLO
volume fluctuation terms, including order $n = 4$ (see Sec.~\ref{Sec:Vol} for details).

\begin{widetext}
\label{eq:B1}
\begin{align}
  \tilde{\kappa}_1 &= \kappa_1 + v_2 \kappa_1' \,, \\
  \tilde{\kappa}_2 &= \kappa_2 + \kappa_1^{2} v_2 + \kappa_2' v_2 + 2 \kappa_1 \kappa_1' V_2 + 2 \kappa_1 \kappa_1' v_3 
                    + 2 \kappa_1'^2 v_2 V_2 + \kappa_1'^2 V_1 V_2 + 2 \kappa_1'^2 V_3 + \kappa_1'^2 v_4 \,, \\
  \tilde{\kappa}_3 &= \kappa_3 + \kappa_1^3 v_3 + 3 \kappa_1 \kappa_2 v_2
                    + 3 (\kappa_1 \kappa_2' + \kappa_1'  \kappa_2) v_3 + 6 \kappa_1' (\kappa_1^2 + \kappa_2') v_2 V_2
                    + 3 \kappa_1' (\kappa_1^2 + 2 \kappa_2') V_3 \\ \nonumber
                   &+ 3 \kappa_1' (\kappa_1^2 + \kappa_2') v_4 + 12 \kappa_1 \kappa_1'^2 V_2^2 + 3 \kappa_1 \kappa_1'^2 V_1 V_3
                    + 24 \kappa_1 \kappa_1'^2 v_2 V_3 + 6 \kappa_1 \kappa_1'^2 V_4 + 3 \kappa_1 \kappa_1'^2 v_5 + \kappa_3' v_2\\ \nonumber
                   &+ 3 (\kappa_1 \kappa_2' + \kappa_1' \kappa_2) V_2 + 8 \kappa_1'^3 v_2 V_2^2 + 6 \kappa_1'^3 V_1 V_2^2 
                    + 10 \kappa_1'^3 v_3 V_3 + \kappa_1'^3 V_1^2 V_3 + 24 V_2 V_3 \kappa_1'^3 \\ \nonumber 
                   &+ 3 \kappa_1'^3 V_1 V_4 + 12 \kappa_1'^3 v_2 V_4 + 3 \kappa_1'^3 V_5 + \kappa_1'^3 v_6
                    + 3 \kappa_1' \kappa_2' V_1 V_2 \,, \\  
  \tilde{\kappa}_4 &= \kappa_4 + \kappa_1^4 v_4 + 6 \kappa_1^2 \kappa_2 v_3 + (4 \kappa_1 \kappa_3 + 3 \kappa_2^2) v_2
                    + 24 (\kappa_1^3  \kappa_1' + 4 \kappa_1 \kappa_1' \kappa_2' + 2 \kappa_1'^2 \kappa_2) v_2 V_3 \\ \nonumber  
                   &+ 4 (\kappa_1^3 \kappa_1' + 6 \kappa_1 \kappa_1' \kappa_2' + 3 \kappa_1'^2 \kappa_2) V_4
                    + 2 (2 \kappa_1^3 \kappa_1' + 6 \kappa_1 \kappa_1' \kappa_2' + 3 \kappa_1'^2 \kappa_2) v_5 \\ \nonumber
                   &+ 48 ( \kappa_1^2 \kappa_1'^2 + \kappa_1'^2 \kappa_2') v_2 V_2^2
                    + 12 (4 \kappa_1^2 \kappa_1'^2 + 5 \kappa_1'^2 \kappa_2') v_3 V_3
                    + 72 (\kappa_1^2  \kappa_1'^2 + 2 \kappa_1'^2  \kappa_2') V_2 V_3 \\ \nonumber
                   &+ 6 (\kappa_1^2  \kappa_1'^2 + 3 \kappa_1'^2 \kappa_2') V_1 V_4 
                    + 72 (\kappa_1^2 \kappa_1'^2 + \kappa_1'^2 \kappa_2') v_2 V_4
                    + 6 (2 \kappa_1^2 \kappa_1'^2 + 3 \kappa_1'^2 \kappa_2') V_5 \\ \nonumber
                   &+ 6 (\kappa_1^2 \kappa_1'^2 + \kappa_1'^2 \kappa_2') v_6
                    + 2 (6 \kappa_1^2 \kappa_2' + 12 \kappa_1 \kappa_1' \kappa_2 + 4 \kappa_1' \kappa_3' + 3 \kappa_2'^2) v_2 V_2 \\ \nonumber
                   &+ 2 (3 \kappa_1^2 \kappa_2' + 6 \kappa_1 \kappa_1' \kappa_2 + 4 \kappa_1' \kappa_3' + 3 \kappa_2'^2) V_3  
                    + 2 (3 \kappa_1^2 \kappa_2 + 2 \kappa_1 \kappa_3' + 2 \kappa_1' \kappa_3 + 3 \kappa_2 \kappa_2') v_3\\ \nonumber
                   &+ (6 \kappa_1^2 \kappa_2' + 12 \kappa_1 \kappa_1' \kappa_2 + 4 \kappa_1' \kappa_3' + 3 \kappa_2'^2) v_4
                    + 96 \kappa_1 \kappa_1'^3 V_2^3 + 96 \kappa_1 \kappa_1'^3 V_3^2 + 288 \kappa_1 \kappa_1'^3 v_3 V_2^2\\ \nonumber
                   &+ 72 \kappa_1 \kappa_1'^3 V_1 V_2 V_3 + 4 \kappa_1 \kappa_1'^3 V_1^2 V_4 + 144 \kappa_1 \kappa_1'^3 V_2 V_4
                    + 128 \kappa_1 \kappa_1'^3 v_3 V_4 + 12 \kappa_1 \kappa_1'^3 V_1 V_5 \\ \nonumber
                   &+ 72 \kappa_1 \kappa_1'^3 v_2 V_5 + 12 \kappa_1 \kappa_1'^3 V_6 + 4 \kappa_1 \kappa_1'^3 v_7
                    + 24 (2 \kappa_1 \kappa_1' \kappa_2' + \kappa_1'^2 \kappa_2) V_2^2
                    + 6 (2 \kappa_1 \kappa_1' \kappa_2' + \kappa_1'^2 \kappa_2) V_1 V_3 \\ \nonumber
                   &+ 2 (2 \kappa_1 \kappa_3' + 2 \kappa_1' \kappa_3 + 3 \kappa_2 \kappa_2') V_2
                    + 48 \kappa_1'^4 v_2 V_2^3 + 48 \kappa_1'^4 V_1 V_2^3 + 48 \kappa_1'^4 V_1 V_3^2 + 240 \kappa_1'^4 v_2 V_3^2\\ \nonumber
                   &+ 32 \kappa_1'^4 v_4 V_4 + 288 \kappa_1'^4 V_2^2 V_3 + 24 \kappa_1'^4 V_1^2 V_2 V_3 + \kappa_1'^4 V_1^3 V_4 
                    + 144 \kappa_1'^4 v_4 V_2^2 + 72 \kappa_1'^4 V_1 V_2 V_4 \\ \nonumber 
                   &+ 128 \kappa_1'^4 V_3 V_4 + 4 \kappa_1'^4 V_1^2 V_5 + 72 \kappa_1'^4 V_2 V_5
                    + 56 \kappa_1'^4 v_3 V_5 + 6 \kappa_1'^4 V_1 V_6 + 24 V_2 V_6 \kappa_1'^4 v_2 V_6 + 4 \kappa_1'^4 V_7 \\ \nonumber
                   &+ \kappa_1'^4 v_8 + 36 \kappa_1'^2 \kappa_2' V_1 V_2^2
                    + 6 \kappa_1'^2 \kappa_2' V_1^2 V_3 + 4 \kappa_1' \kappa_3' V_1 V_2 + 3 \kappa_2'^2 V_1 V_2 + \kappa_4' v_2 \,.  \\ \nonumber
\end{align}
\end{widetext}

\section{The 4-parameter beta distribution}
\label{Sec:Beta}

The scheme proposed in Sec.~\ref{Sec:Vol} for volume fluctuation corrections of the observed particle number cumulants,
namely by subtraction of all volume terms, requires the $N_{\textrm{part}}$ distribution of the applied centrality selection.
While a distribution is in principle fully characterized by its moments or cumulants, a straightforward Taylor expansion
requires knowledge of all moments or at least of a sufficiently large number of them to keep the truncation error small.
More efficient schemes have however been proposed, for example the Poisson-Charlier expansion which approximates a given
distribution on the basis of its factorial cumulants by a sum of forward difference operators applied to the Poisson
distribution \cite{Bzdak2018,Aude2008}.  In Ref.~\cite{Bzdak2018} such an expansion has been used to model the STAR
proton distributions but, as also pointed out by the authors, the method can lead to unphysical results, e.g.\ a negative yield.
This is also our observation, since typically only the first few moments of a distribution are reliably known.  Hence we
took a different, more pragmatic approach based on the 4-parameter beta distribution \cite{Kendall2004} which is always
positive and, in all cases of interest to our analysis, is unimodal. 

Starting from the usual definition of the beta probability distribution on the support interval [0,1] one arrives at

\begin{equation}
  f(x;p,q) = \frac{x^{p-1} (1-x)^{q-1}}{B(p,q)} \,,
\label{eq:C1}
\end{equation}

\noindent
where $p$ and $q$ are dimension-less shape parameters fulfilling $p,q > 0$ and $B(p,q)$ is the beta function.
The latter one provides proper normalization to unity and it is defined with help of the gamma function $\Gamma(z)$

\begin{equation*}
  B(p,q) = \frac{\Gamma(p) \Gamma(q)}{\Gamma(p + q)} \,.
\end{equation*}

\noindent
The mean, variance, skewness $\gamma_1$, and excess kurtosis $\gamma_2$ of the beta distribution are given by \cite{Kendall2004}

\begin{equation}
 \begin{split}
  E[X] &= \frac{p}{p+q} \,,\\
  \textit{Var}[X] &= \frac{p \: q}{(p+q)^2 (p+q+1)} \,,\\
  \gamma_1[X] &= \frac{2 (p-q) \sqrt{p+q+1}}{(p+q+2) \sqrt{p \: q}} \,,\\
  \gamma_2[X] &= \frac{6 [(p-q)^2 (p+q+1) - p \: q \: (p+q+2)]}{p \: q \: (p+q+2) (p+q+3)} \,.
 \end{split}
\label{eq:C2}
\end{equation}

\noindent
Note that all moments are determined by the two shape parameters $p$ and $q$ which can be
mapped unambiguously onto skewness and excess kurtosis, with the restriction
that $\gamma_1^2 - 2 < \gamma_2 < \frac{3}{2} \gamma_1^2$.  This also imples that 
$\gamma_1$ and $\gamma_2$ can not both be zero.

In order to extend the range of the beta distribution beyond [0,1], we introduce two
additional parameters, a scaling $r$ and a shift $s$, via the linear transformation

\begin{equation*}
  x \longmapsto x' = r \: x + s \,.
\end{equation*}

\noindent
With this, Eq.~\eqref{eq:C1} becomes

\begin{equation}
  f(x';p,q,r,s) = \frac{(\frac{x'-s}{r})^{p-1} (1-\frac{x'-s}{r})^{q-1}}{r B(p,q)} \,,
\label{eq:C3}
\end{equation}

\noindent
now defined on the support interval $[s, s+r]$.
The resulting extended 4-parameter beta distribution $f(x';p,q,r,s)$ can be used to approximate unimodal distributions for
which the first four moments are given: $p$ and $q$ are obtained directly from the dimensionless skewness and kurtosis,
the scaling $r$ is determined by the variance (or width $\sigma' = r \:\sigma$), and lastly, the shift parameter $s$
is fixed by the mean ($\langle x' \rangle = r \: \langle x \rangle + s$).  Indeed, introducing the variable $\nu$ as

\begin{equation*}
  \nu = p + q = \frac{3 (\gamma_2 - \gamma_1^2 + 2)}{\frac{3}{2} \gamma_1^2 - \gamma_2} \,,
\label{eq:C4}
\end{equation*}

\noindent
parameters $p$ and $q$ can be obtained from skewness and kurtosis with

\begin{equation*} \label{eq:C5}
  \begin{split} 
  \gamma_1 = 0:&  \hspace{7mm} p = q = \frac{\nu}{2} = \frac{\frac{3}{2} \gamma_2 + 3}{-\gamma_2}\,,\\
  \gamma_1 \neq 0:& \hspace{7mm} p, q = \frac{\nu}{2} [1 \pm (1 + \frac{16 (\nu + 1)}{(\nu + 2)^2 \, \gamma_1^2})^{-1}] \,.
  \end{split}
\end{equation*}

\noindent
Next, the scaling parameter $r$ follows from the width

\begin{equation*} \label{eq:C6}
  r = \frac{\sigma}{2} \, \sqrt{(\nu + 2)^2 \, \gamma_1^2 + 16 (\nu + 1)} \,.
\end{equation*}  

\noindent
And last, from the mean, the shift parameter $s$ is obtained with

\begin{equation*} \label{eq:C7}
  s = \langle x' \rangle - \frac{p \, r}{\nu} \,.
\end{equation*}  

\begin{figure}[!ht]
  \begin{center}
     \resizebox{0.7\linewidth}{!} {
       \includegraphics{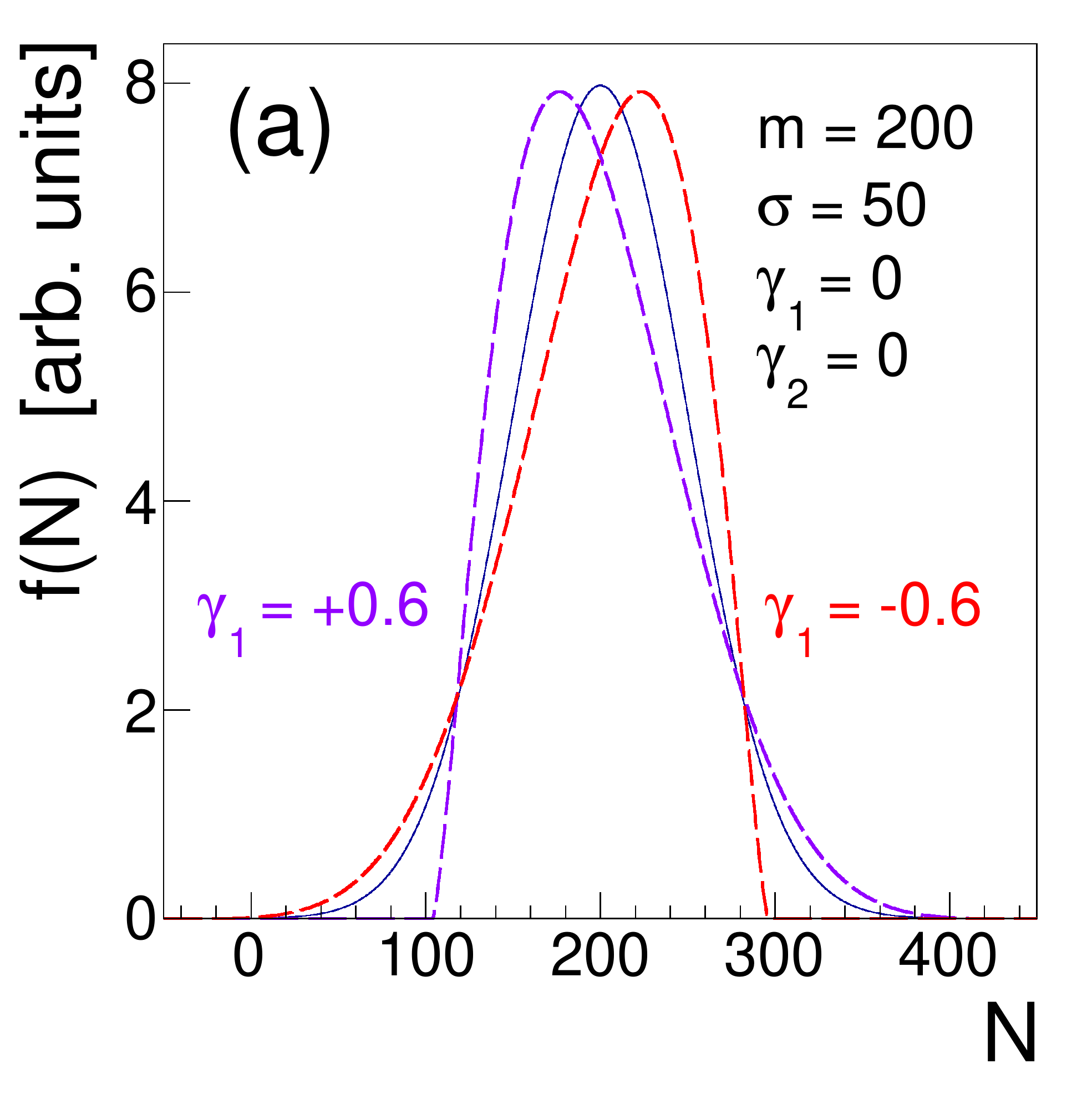}
     }
     \resizebox{0.7\linewidth}{!} {
       \includegraphics{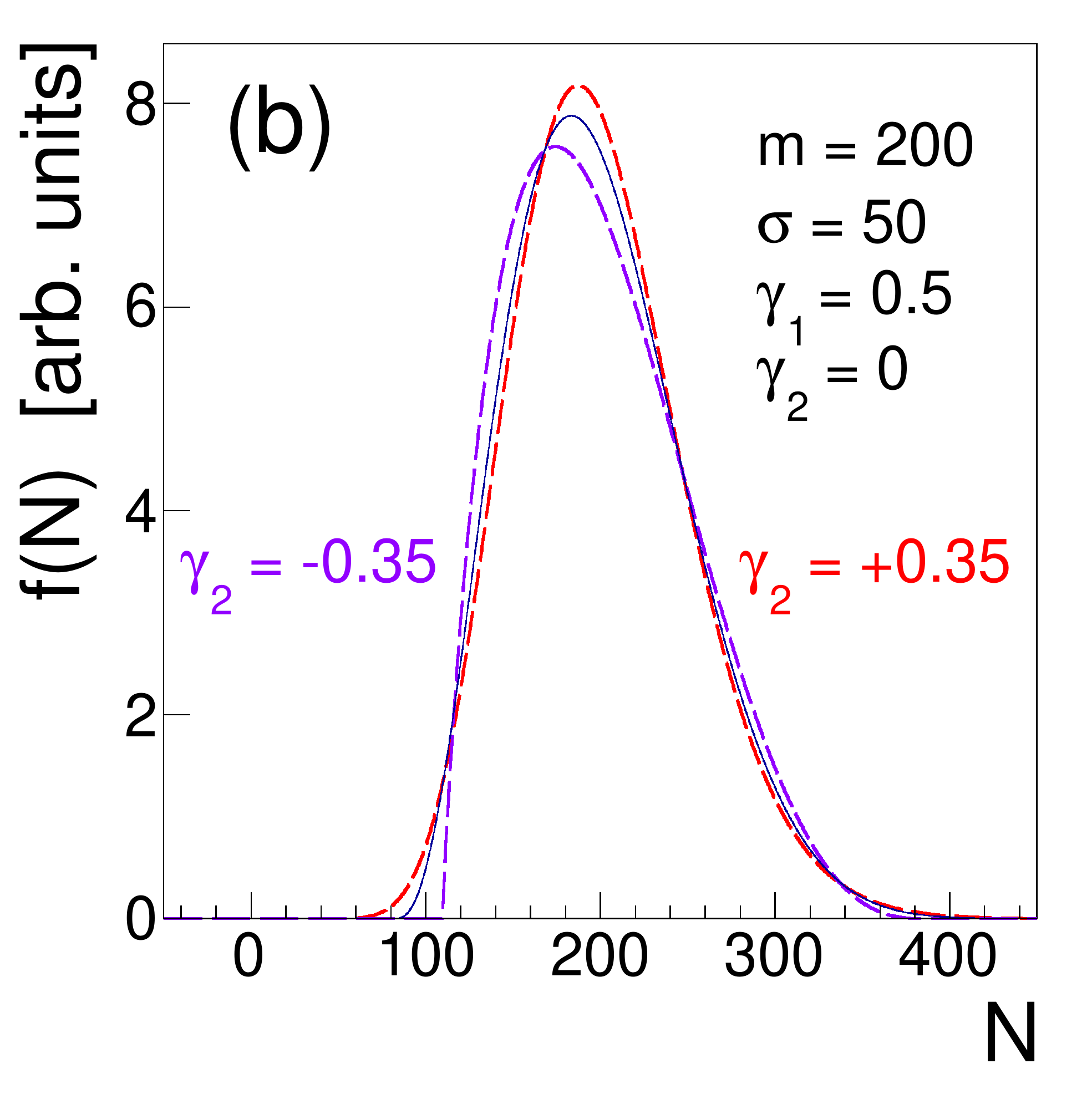} 
     }
  \end{center}
  \vspace{-0.7cm} 
  \caption{Illustration of the 4-parameter beta distribution (Eq.~\eqref{eq:C3}, not normalized).  The effect of non-zero skewness (a)
           and non-zero kurtosis (b) on the distribution shape is shown.  Note that the limits $\gamma_1 \simeq 0$ and $\gamma_2 \simeq 0$
           approximate the $N(200, 50)$ normal distribution also plotted in (a). 
}
\label{fig:A3}
\end{figure}

\noindent
The ability of Eq.~\eqref{eq:C3} to render various unimodal function shapes is illustrated in Fig.~\ref{fig:A3} with a set of
distributions having same mean and width, but different values of the higher-order shape parameters $\gamma_1$ and $\gamma_2$. 

\begin{figure}[!th]
  \begin{center}
     \resizebox{1.0\linewidth}{!} {
       \includegraphics{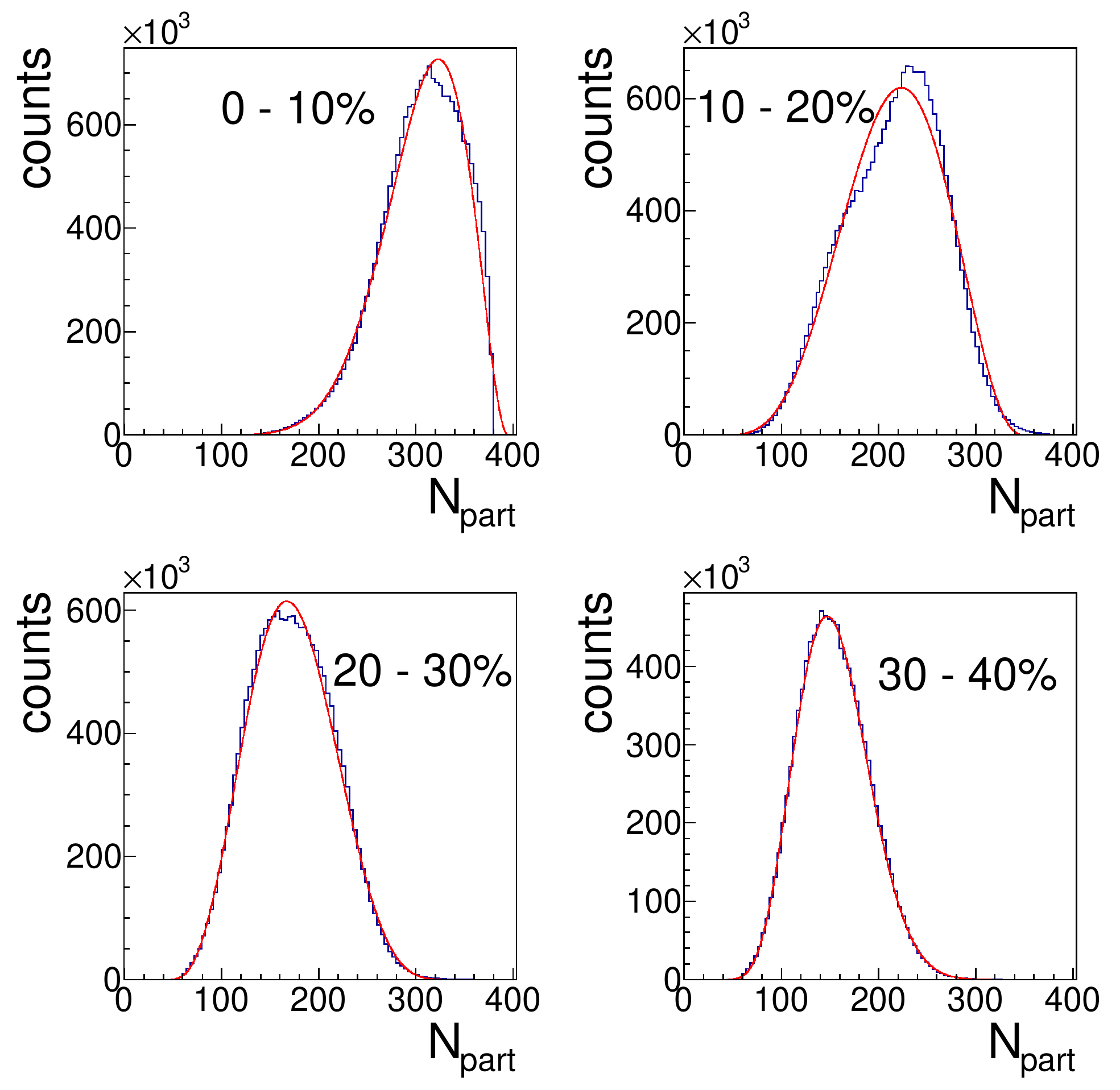} 
     }
  \end{center}
  \vspace{-0.7cm} 
  \caption{(Color online) Comparison of the 4-parameter beta distribution (red line) with the true $N_{\textrm{part}}$
           distributions (dark blue histogram) obtained from IQMD for four different centrality selections.  For each selection,
           the two distributions are normalized to the same number of counts.
          }
\label{fig:A4}
\end{figure}

The extent to which unimodal distributions can be approximated by Eq.~\eqref{eq:C3} is demonstrated in Fig.~\ref{fig:A4},
where the 4-parameter beta distribution is compared to $N_{\textrm{part}}$ distributions obtained from IQMD transport model calculations.
These were done for different centrality selections on the HADES forward wall sum of charges signal, as discussed in
Sec.~\ref{Sec:Npart} (see also Fig.~\ref{fig:iqmdNpart}).  In all cases, the first four moments agree by construction and any
visible residual deviations are caused by moments of order higher than~4.  In the fluctuation analysis of our Au+Au data we
have also used Eq.~\eqref{eq:C3} to visualize the experimental $N_{\textrm{part}}$ distributions (see Fig.~\ref{fig:dataNpart8bins}).
To improve the quality of this modeling, moments beyond 4\textsuperscript{th} order would have to be included and hence a
more sophisticated expansion, like Gram-Charlier \cite{Kendall2004} or Poisson-Charlier \cite{Barbour1987}, would have to
be employed.

\newpage
\bibliography{flucsHades}

\end{document}